# An Ultraviolet Spectroscopic Atlas of Local Starbursts and Star-Forming Galaxies: The Legacy of FOS and GHRS

Claus Leitherer

*Space Telescope Science Institute[1], 3700 San Martin Drive, Baltimore, MD 21218*

[leitherer@stsci.edu](leitherer@stsci.edu)

Christy A. Tremonti

*Univ. of Wisconsin-Madison, Dept. of Astronomy, 475 N. Charter St., Madison, WI 53706*

[tremonti@astro.wisc.edu](tremonti@astro.wisc.edu)

Timothy M. Heckman

*John Hopkins Univ., Dept. of Physics & Astronomy, Homewood Campus, Baltimore, MD 21218*

[heckman@pha.jhu.edu](heckman@pha.jhu.edu)

and

Daniela Calzetti

*Univ. of Massachusetts, Dept. of Astronomy, LGRT-B 524, Amherst, MA 01003*

[calzetti@astro.umass.edu](calzetti@astro.umass.edu)

# Abstract

We present 46 rest-frame ultraviolet (UV) spectra of 28 local starburst and star-forming galaxies which were observed with the Faint Object Spectrograph (FOS) and the Goddard High Resolution Spectrograph (GHRS) of the Hubble Space Telescope (HST) at a spectral resolution of a few 100 km s$^{-1}$. We compare the HST spectra with lower resolution International Ultraviolet Explorer (IUE) spectra of the same galaxies and find systematic differences: the bright star clusters targeted in HST's ~1″ apertures provide about 15% of the starburst luminosity traced by IUE's 10″ × 20″ aperture; they are bluer and have stronger stellar-wind features suggesting that the HST apertures have preferentially been placed on the youngest areas of the burst. In contrast, lines arising from the interstellar medium (ISM) show similar equivalent widths in both the large and small aperture observations, suggesting similar ISM properties from larger to smaller scales. In order to quantify the UV spectral morphology of star-forming galaxies, we created a set of UV line indices similar to the standard optical Lick indices. We discuss the relation between the UV spectral morphology and the properties of the galaxy host. We present our atlas of FOS and GHRS spectra both in print and electronically. The data set is useful as a baseline for comparisons with observations of the rest-frame UV spectra of star-forming galaxies at high redshift.

---

# 1. Motivation

While our understanding of galaxy formation and evolution is in many ways incomplete, it has become increasingly evident that much of the stellar mass in the universe was assembled in violent and short-lived episodes of star formation at high redshift (e.g., Papovich, Dickinson, & Ferguson 2001; Shapley et al. 2001; Pérez-González et al. 2008). In the latter half of the Hubble time, the overall star formation rate of the universe has declined precipitously (Lilly et al. 1996; Cowie, Songaila, & Barger 1999; Hopkins 2004). Nevertheless, there remain present-day examples of galaxies undergoing intense bursts of star formation. Indeed, while few in number, these starburst galaxies are responsible for ~20% of the entire high-mass star formation in the local universe (Brinchmann et al. 2004). Local starburst galaxies provide us with an opportunity to investigate in detail a phenomenon with profound cosmic consequences.

In this study we examine the UV spectral properties of 28 local starburst galaxies which were observed with the FOS and the GHRS instruments on board HST. The rest-frame UV is of particular interest because it is the window onto the high-redshift universe provided by large ground-based telescopes working in the visible. Our study complements the earlier work of Kinney et al. (1993) and Heckman et al. (1998; hereafter H98) based on spectra obtained with the IUE satellite. It also provides guidance for population synthesis work that attempts to model the UV spectra of star forming galaxies theoretically (Leitherer et al. 2010). While the majority of our sample galaxies were included in the earlier IUE analysis, substantial differences between the HST and IUE observations motivate our present undertaking. The velocity resolution of our FOS and GHRS data (a few100 km s$^{-1}$) represents nearly a factor of 10 improvement over that of IUE in low-dispersion mode, enabling us to make use of the spectral *line profile* information and allowing us to study the kinematics of the ISM. In addition, the HST apertures employed in our sample observations have projected sizes more than a factor of 10 smaller than that of IUE, often encompassing only a few bright star clusters. Thus, our study provides a small scale probe of the ISM and stellar populations, which we compare to the more global view offered by IUE.

In this paper we present a UV spectral atlas of HST observations and consider the differences inherent in the HST and IUE spectra due to the different spatial scales sampled. We take advantage of the superior resolution and signal-to-noise of the HST spectra and of the availability of multi-wavelength data for our local sample to perform a thorough analysis of the connection between a starburst's UV spectral morphology and its luminosity and dust content and the mass and metallicity of its host galaxy.

In Section 2 we review the origin of the prominent spectral features in the UV and the conclusions of previous studies of UV starburst spectra based on IUE data. The 46 HST galaxy spectra which constitute our sample are presented in Section 3, as are the details of the data compilation and reduction procedures. A key element in unlocking the diagnostic power of the rest-frame UV is the availability of multi-wavelength data for our local starburst sample. In Section 4 we describe the UV, optical, infrared (IR), and H I observations which we use to infer the starburst luminosity and dust content, as well as the host galaxy mass and metallicity. We compare the IUE and HST spectra in Section 5. A new set of standard UV indices is introduced in Section 6. In Section 7 we discuss the correlation of the UV features with the physical properties of the galaxies. Finally, Section 8 presents our conclusions.

## 2. Background

Data obtained early in the IUE mission demonstrated the existence of a rich absorption spectrum in star forming galaxies (Weedman et al. 1981; Huchra et al. 1983). The identification of prominent UV spectral features with lines seen in massive stars provided some of the first *direct* evidence for the presence of large numbers of massive stars in actively star forming galaxies. The vastly better sensitivity and spectral resolution of HST's instruments brought subsequent advances in detecting and classifying the stellar and interstellar spectral features (e.g., Heckman & Leitherer 1997; Vázquez et al. 2004).

The space-UV lines observed in starburst galaxies are essentially of four types: stellar wind features, stellar photospheric absorption lines, interstellar absorption lines, and nebular emission lines. In Table 1 we provide a comprehensive compilation of the four types of lines over the wavelength range 1150 to 3000 Å. Many of the listed lines may not be observable in typical UV spectra of a starburst galaxy. However, our goal is to provide the community with a database that is as complete as possible for comparison with a wide variety of data. The table lists the atom or ion (column 1), and is arranged by the laboratory vacuum wavelength (column 2). Columns 3 and 4 give the ionization ($E_{i\text{on}}$) and excitation ($E_{\text{exc}}$) energies for the preceding ionization stage and the lower excitation level, respectively. These are the relevant quantities for the formation of almost all observed features, except for recombination lines like He II λ1640, for which the upper $E_{\text{ion}} = 54.4$ eV matters. Crosses in columns 5, 6, 7, and 8 indicate whether the lines predominantly form in the wind, the stellar photosphere, the ISM, and/or have a nebular origin, respectively. The distinction between these categories is sometimes not unique, as low-ionization resonance lines could be both interstellar and stellar. Therefore we used *typical* FOS and GHRS starburst spectra to guide our categorization scheme. The references are listed in column 9.

The most prominent features are typically the N V λ 1240, Si IV λ1400, and C IV λ1550 stellar-wind lines, which are produced in the expanding atmospheres of O and B stars. They are broad (~2000 km s$^{-1}$), blueshifted (by ~1000 km s$^{-1}$), and often show a P-Cygni profile. Because these lines are produced in a radiatively driven wind which must overcome the gravitational well of the mass-losing star, the line strength and profile shape have a mass dependence and can be used as a tracer of the initial mass function (IMF) and star-formation history. However, the line profile is also metallicity sensitive since both stellar evolution and mass loss have a strong metallicity dependence (Kudritzki & Puls 2000; Puls, Vink, & Najarro 2008).

Stellar photospheric features abound shortward of ~2000 Å, where O and B-star spectra exhibit heavy line-blanketing, primarily by transitions of highly ionized iron and nickel (Dean & Bruhweiler 1985; Brandt et al. 1998). Among the multitude of photospheric lines, there are a handful of well developed features (e.g., C III λ1176, O IV λ1342, Fe V λ1363, Si III λ1417, C III λλ1426/28, S V λ1502), many of which are comprised of groups of neighboring lines. However, these features are typically quite weak in contrast with the stellar-wind lines. Because their equivalent widths (EW) range from <1 to a few Å (as compared to 5 – 15 Å for the wind lines), high signal-to-noise (S/N) moderate resolution spectra are required. Nonetheless, the photospheric features are of interest because they provide another tracer of the young stellar population. Rix et al. (2004) used theoretical models to calibrate several weak photospheric features as metallicity indicators. This method was used by Halliday et al. (2008) to determine the stellar metallicity of star-forming galaxies at redshift $z \approx 2$.

The ISM also leaves a pronounced imprint on the space-UV spectra of galaxies. Strong interstellar absorption lines are produced by the resonance transitions of the abundant ionic species: H I, C II, C IV, N I, N V, O I, Al II, Al III, Si II, Si III, Si IV, Mg I, Mg II, and Fe II. These lines appear optically thick (on the flat part of the curve of growth), and therefore their strength is primarily determined by the velocity dispersion of the starburst and the covering factor of the ISM. The ISM absorption lines offer a unique probe of the kinematics of the gas over a broad range in ionization.

Nebular emission lines are also present in the UV, although their EWs are typically small in all but the lowest metallicity galaxies. The most prominent lines are C III] $\lambda\lambda$1907/09 and Si III] $\lambda\lambda$1883/92. The C III] line can be used in concert with some of the weaker lines – ideally O III] $\lambda\lambda$1661/66 – to derive the C/O abundance ratio (Kobulnicky & Skillman 1998; Garnett et al. 1999).

While the UV spectra of star forming galaxies are rich in stellar absorption lines which are sensitive to age and metallicity, the low-order shape of the continuum is surprisingly robust to these parameters (quite unlike the case in the optical). The shape of the spectrum can be parameterized by a power law ($F_\lambda \propto \lambda^\beta$) between 1200 and 3000 Å. When stars younger than about 10 Myr are present, the value of β exhibits a narrow range of values between −2.7 and −2.5 (Meurer et al. 1997; Calzetti 2001). The relative insensitivity of the UV continuum to the particulars of the stellar population turns out to be fortunate happenstance because this part of the spectrum is extremely sensitive to dust. Even a modest extinction of $A_V = 1$ implies a factor of 10 dimming around 2000 Å.

However, starburst galaxies appear to obey a well defined wavelength dependent reddening law which accounts for the net effects of dust/star geometry, scattering, and total absorption (Calzetti 2001). Thus, if the presence of massive stars can be ascertained from the stellar absorption lines, the dust content can be estimated by comparing the observed power law index of the continuum to the expected intrinsic one.

The sheer diversity of the spectral features in the space UV makes the diagnostic power of this spectral region unparalleled. However, at present few empirical tools exist to help mine the information. In this paper we focus on developing automated techniques for extracting the maximum information from UV spectra. Our sample of nearby starbursts is ideal for calibrating UV diagnostics of galaxy physical properties since their properties are generally already well known from detailed observations in a number of wavebands.

An important milestone in the study of star forming galaxies was the creation of a catalog of UV spectra by Kinney et al. (1993). The sample comprised 143 nearby ($\lesssim$100 Mpc) spiral, irregular, blue compact, Seyfert 2, LINER, and starburst galaxies with spectra available in the IUE archives. IUE was ideal for characterizing the UV properties of these nearby galaxies because its $10'' \times 20''$ aperture was large enough to encompass the central few kiloparsecs of most objects. Besides laying the ground work for many more detailed studies, Kinney et al. presented the first qualitative picture of the UV spectra of galaxies of different morphological types and activity classes.

H98 undertook a more quantitative study of a subset of 45 starburst galaxies in the sample of Kinney et al. (1993). This work explored the connection between the UV spectral morphology of the starbursts and the mass, metallicity, luminosity, and dust content of their host galaxies. While the majority of the strong UV spectral features where well known by the time of H98, the modest spectral resolution and S/N of the IUE spectra precluded detailed analysis of individual lines. Figure 1 shows two examples of the very best quality IUE spectra and their corresponding FOS and GHRS spectra. IUE's 6 Å (~1200 km s$^{-1}$) resolution is barely sufficient to distinguish between the broad stellar-wind lines and the much narrower stellar photospheric and interstellar features. Some features which show up clearly in the HST spectrum, for example Si II λ1526, become blended with neighboring lines – including foreground Milky Way lines – in the IUE spectrum (see NGC 1705 in Figure 1, in particular). Because of this, and the generally modest S/N of the IUE spectra, H98 chose to characterize the UV spectral morphology of their starbursts with three parameters: β, the slope of the UV continuum between 1250 and 1850 Å; $W_{IS}$, the average EW of the strongest low-ionization interstellar features Si II λ1260, O I λ1302 + Si II λ1304, C II λ1335; and $W_W$, the average of the Si IV λ1400 and C IV λ1550 lines, which contain potentially large contributions from stellar winds.

The simple parameterization of the UV spectral features adopted by H98 proved adequate to elucidate fundamental trends in UV spectral morphology with galaxy physical parameters. The spectral parameters β, $W_{IS}$, and $W_W$ were found to be correlated with the strength and dustiness of the starburst, and the luminosity, mass, and metallicity of the host galaxy. Starbursts were shown to occupy a very small part of the available parameter space: more luminous starbursts tended to occur in more massive, metal-rich hosts, and consequently their spectra exhibited more dust reddening and stronger UV lines. This finding was a major leap forward in our understanding of the physical parameter space occupied by starbursts and their host galaxies, and it provided a framework for interpreting rest frame UV observations of high-redshift galaxies.

Our present undertaking is motivated by goals similar to that of H98, namely the development of tools for analyzing spectra of high-redshift galaxies. However, the ensuing years have brought major improvements in the data available at both low and high redshift, thanks to the availability of UV spectrographs on board HST and the improved light gathering power of ground-based telescopes. This has enabled us to explore the UV spectral morphology of galaxies with more detail and sophistication. We therefore revisit some of the correlations outlined by H98 with an improved parameterization of the UV spectral features.

The FOS and GHRS were two of the four original axial instruments aboard HST. Both instruments were removed from HST and replaced by the Space Telescope Imaging Spectrograph (STIS) during Servicing Mission 2 in February 1997. STIS operates in the UV as well and is currently still operational. It is complemented by the Cosmic Origins Spectrograph (COS), which was installed during Servicing Mission 4 in May 2009. The two legacy instruments FOS and GHRS have left behind a rich and self-contained data set whose calibration is complete and final. The spatial sizes of their apertures $(1 - 2'')$ are an order of magnitude *smaller* than that of IUE, yet they are an order of magnitude *larger* than those of the STIS apertures typically employed $(0.2'')$. Ongoing STIS observations, by virtue of their spatial sampling, often focus on individual star clusters in starburst galaxies, whereas FOS and GHRS spectra encompass small starburst regions having sizes of ~100 pc (corresponding to $2''$ at 10 Mpc). This makes the FOS and GHRS data set the sample of choice for our study.

# 3. Observations

*3.1 The sample*

Our sample is composed of 28 local starburst and star-forming galaxies with high quality UV spectra available from the HST Archive. Because of the small size of the HST apertures (1 to a few arcsec), our spectral atlas includes multiple pointings within many of the sample galaxies, for a total of 46 unique spectra[2]. Of these, 29 were obtained with the FOS and 17 with the GHRS. The characteristics of the FOS and GHRS instruments are described in Section 3.2 and 3.3, respectively. Details of the individual observations which constitute our sample are given in Table 2. The entries in this table are as follows: column1: spectrum identifier used in this paper; column 2: host galaxy name; column 3: name of the starburst region; columns 4 and 5: right ascension and declination of the region (note that the absolute pointing of HST has typical systematic uncertainties of ~1″); column 6: instrument used[3]; column 7: grating used; column 8: aperture used; column 9: HST program ID; column 10: wavelength region covered; column 11: measured S/N per pixel. HST images of our sample galaxies with the spectroscopic apertures overlaid are shown in the Appendix.

Our sample was culled from all of the FOS and GHRS spectra available in the HST archive. We included all spectra of star forming or Seyfert 2 galaxies observed both pre- and post-COSTAR which met the following criteria:

- wavelength coverage of at least 250 Å between 1150 and 3200 Å in the galaxy rest frame;

---

[2] We count multiple wavelength settings at the same pointing as one spectrum.
[3] See http://www.stsci.edu/documents/dhb/pdf/FOS.pdf and http://www.stsci.edu/instrument-news/handbooks/ghrs/GHRS_1.html for descriptions of the FOS and GHRS, respectively.

- spectral resolution of 0.5 − 3 Å;

- S/N of at least 2 in the continuum in the combined spectrum;

- clear evidence of spectral features caused by star formation. (This criterion eliminates all but a handful of Seyfert 2 galaxies.)

Despite the unsystematic selection inherent in this archival sample, the galaxies span a fairly broad swath of parameter space. They include a wide range of morphological types (irregulars, nuclear starbursts, blue compact dwarfs) and activity classes (HII galaxies, LINERS, Seyfert 2's), and encompass a broad range of global galaxy parameters such as metallicity and absolute blue magnitude (see Section 4 for details). Since both the FOS and GHRS instruments have completed their mission lifetimes, this sample represents their legacy.

*3.2 FOS*

The FOS (Harms et al. 1979) operated on board HST for six years beginning with the telescope's launch in 1990. The FOS had two independent optical channels which focused spectral images on the photo-cathodes of photon-counting Digicon detectors. The detector on the red side was sensitive from 1620 to 8500 Å, while the detector on the blue side was sensitive from 1150 to 5400 Å. Dispersers were available for both high ($\lambda/\Delta\lambda = 1300$) and low ($\lambda/\Delta\lambda = 250$) resolution modes. The data that met our selection criterion were obtained at high resolution using the G130H, GH190H, or G270H gratings, on either the red or blue sides. The G130H grating provided a dispersion of ~1.00 Å per diode in the 1140 − 1606 Å regime; the G190H, ~1.45 Å per diode from 1590 − 2312 Å; and the G270H, ~2.05 Å per diode from 2222 − 3277 Å.

The spectral resolution depends on the dispersion of the grating, the aperture used, and the degree to which the source fills the aperture. Table 2 lists the configuration employed for each observation. The apertures range from 0.26″ to 1.0″, with the 0.86″ circular aperture being the most common. (Note that the introduction of the COSTAR optics caused a 16% scale change, such that the original 1″ aperture was reduced to 0.86″.) The degree to which our spectroscopic targets − generally bright star clusters − resemble point or extended sources is difficult to estimate due to the redshift range of our sample and the highly inhomogeneous nature of starburst galaxies. A spectral line from an object uniformly filling the 0.86″ circular aperture would have a full width at half maximum (FWHM) of 2.29 Å, whereas a point source would yield a FWHM of ~1 Å. The resolutions of the FOS spectra in our sample will fall in between these extremes.

The FOS target acquisition could achieve a pointing accuracy of 0.1″ − 0.2″ for a point source (There will still be the systematic HST pointing uncertainty of ~1″.) However, the inhomogeneous nature of our starburst galaxies induces greater uncertainty in the process. We note here that the object coordinates listed in Table 2 and shown in Appendix Figure 1 are the input coordinates; the final pointing of HST could differ by a few arcseconds.

We re-calibrated our FOS spectra by running the Post Operational Archive version of the FOS calibration pipeline (POA_CALFOS v1.2.1). POA_CALFOS performs all of the tasks of the original CALFOS pipeline with some enhancements. Most significantly, the accuracy of the wavelength zero point on the FOS/BLUE side is restored to its original specification of ±1 pixel. The standard pipeline reduction begins with a correction for dead diodes, geomagnetic induced image motion, and saturation in the detector electronics. The data are then dark subtracted, scattered light corrected, flat fielded, and corrected for aperture throughput and time-dependent instrumental sensitivity variations. Finally, a dispersion solution is constructed, and the spectra are converted into absolute flux units of erg s$^{-1}$ cm$^{-2}$ Å$^{-1}$.

After pipeline reprocessing, we boxcar smoothed the spectra by five pixels to recover the original sampling of the diodes. We combined multiple observations acquired at the same pointing with the same grating by averaging the spectra weighted by their exposure times with the IRAF task `spcombine`.

*3.3 GHRS*

The GHRS (Brandt et al. 1994) was also one of the first-generation science instruments on board HST. It was designed to complement FOS by providing higher resolution and throughput in the UV, at the expense of more limited wavelength coverage. Like FOS, GHRS had two channels with different photocathode sensitivities: Side 1, which was optimized in the far-UV (~1050 – 1700 Å) and Side 2, which was optimized in the near-UV (~1700 – 3200 Å). The GHRS spectra in our sample were obtained with the G140L grating which provided a dispersion of ~0.57 Å per pixel over a 285 Å variable bandpass in the 1100 – 1900 Å range. Two science apertures were available: the 2.0″ square Large Science Aperture (LSA; 1.74″ post-COSTAR) and the 0.25″ Small Science Aperture (SSA; 0.22″ post-COSTAR). All of our sample observations, with the exception of those of NGC 1705 where made through the Large Science Aperture. The resolution of a point source observed with the G140L grating through the LSA is ~0.9 Å, which corresponds to ~250 km s$^{-1}$ at 1100 Å and ~140 km s$^{-1}$ at 1900 Å. The resolution of a fully extended source observed through the LSA is degraded by a factor of 4 since four GHRS diodes are encompassed by the LSA. Our spectra have resolutions in between these extremes, generally closer to the point source value.

The target acquisition generally succeeded in peaking up on the bright knots of star formation targeted in our sample galaxies. However, due to the nature of the process, the acquisition input coordinates recorded in Table 2 and shown in Appendix Figure 1 should not be considered accurate to more than a few arcseconds.

We recalibrated our GHRS data by re-running the standard CALHRS reduction pipeline with an updated set of reference files. The standard reduction begins by computing the count rate per pixel and then correcting for non-uniform diode sensitivity and for the finite response time of the detector electronics. The spectrum's location on the photocathode is then mapped out and a correction is applied for vignetting. In order to fully sample the HST point spread function, the diodes were sub-sampled by moving the spectrum on the detector by a fraction of a diode and storing the spectra separately. These sub-sampled spectra are recombined by the pipeline. The spectrum's position on the photocathode is then converted into a wavelength, and the wavelength zero point is adjusted for the appropriate aperture, and corrected to the heliocentric system. The detector background is subtracted off and a sensitivity vector is applied to convert the spectrum from counts per second per diode to absolute flux units of erg $s^{-1}$ $cm^{-2}$ $Å^{-1}$.

After pipeline re-processing, we recomputed the wavelength zero point of the spectrum if supporting "SPYBAL" observations were available with the STSDAS IRAF task `waveoff`. For NGC 1068, sub-integrations acquired in FP-SPLIT mode were aligned and combined using the STSDAS IRAF tasks `poffsets` and `specalign`. In cases where multiple observations were acquired at the same pointing, we averaged the spectra weighted by their exposure times using the IRAF task `scombine`.

*3.4 Correcting the wavelength zero points*

The wavelength calibration of the FOS and GHRS spectra are expected to be accurate to ±1 diode, or about ~1.5 and ~0.5 Å, respectively. To improve the wavelength calibration, we cross-correlated each spectrum with a zero-velocity template. We constructed the templates individually for each spectrum. When far-UV spectra were available, we compared them with Starburst99 synthetic spectra (Leitherer & Chen 2009; Leitherer et al. 2010) for a range of ages and selected the model which produced the best fit (for details see, e.g., Tremonti et al. 2001). We redshifted this template to the heliocentric velocity of the galaxy using redshifts obtained from the NASA/IPAC Extragalactic Database (NED). We then added the foreground Milky Way absorption lines listed in Table 3 to the template assuming a velocity width of 200 km s$^{-1}$. The entries in this table are the median and standard deviation of the velocity EWs along various Milky Way sightlines as reported by Savage et al. (2000). See Section 3.5 for further information about the Milky Way lines. Unfortunately the strongest features in the galaxy spectra − the stellar-wind lines and the interstellar lines − are sometimes blueshifted relative to the stellar photospheric lines, and hence are not useful for wavelength calibration. We masked these features and derived the wavelength zero points from the remaining pixels by doing a cross-correlation in real space (rather than Fourier space). The zero point is thus set by a combination of photospheric lines and Milky Way lines. The latter are used only when they are not blended with the intrinsic ISM lines ($v_{helio} > 800$ km s$^{-1}$). We assumed zero velocity for the Milky Way lines, i.e., a negligible contribution by Galactic high-velocity clouds. Each cross-correlation was examined by eye. We estimate the remaining error in the far-UV zero points to be of order ±0.3 Å.

The FOS spectra obtained in the near-UV necessitated a different procedure. The wavelength range covered by the G190H grating (1600 – 2300 Å) does not contain any strong Milky Way lines, however it does contain some prominent photospheric features and the C III] $\lambda\lambda$1907/09 emission line. Unfortunately, the empirical stellar libraries used by Starburst99 do not extend beyond 1850 Å. To extend the wavelength coverage of the models out to 2200 Å, we used some fully theoretical OB star spectra (Leitherer et al. 2010). We synthesized template spectra at ages of 1 and 5 Myr, and made a by-eye assessment of which model provided the best fit to each spectrum. To this template we added a C III] emission line in cases where it was apparent in the data. Because C III] is not a resonance line, we assumed it to be free of the complex radiative transfer effects and at rest with respect to the stars. These templates provided reasonably good fits to most of the data with the exception of NGC 3690, NGC 4670, and NGC 7552, whose data exhibited strong photospheric features not evident in the models. For these spectra we used a near-UV STIS spectrum of He 2-10 (Chandar et al. (2003) which shows similar photospheric features as a template. In a few cases where the S/N was not adequate no correction was made. We estimate the remaining errors in the near-UV wavelength zero points to be of order ±0.5 Å.

In the wavelength range covered by the FOS G270H grating (2300 – 3200 Å), few strong photospheric features exist, but there are a large number of strong interstellar lines. For these spectra we constructed templates using the Milky Way lines and the intrinsic interstellar absorption lines. Because many of the interstellar lines are strong and saturated, the Milky Way and intrinsic features are often blended. Where possible, we masked the intrinsic lines and used only the Milky Way lines.

After correcting the wavelength zero points, we rebinned the spectra to a common wavelength grid with a linear dispersion of 1.0 Å for the FOS spectra and 0.57 Å for the GHRS spectra. We merged FOS observations taken with different gratings weighting each segment by the standard deviation in the overlap region (1590 – 1606 Å and 2220 – 2310 Å). The agreement of the data in the overlap regions is generally quite good. However, the decline in grating throughput in these regions causes the spectra to be fairly noisy, particularly at the ~1600 Å junction.

*3.5 Correcting for Galactic contamination*

To obtain accurate measurements the UV spectral properties of galaxies it is desirable to remove any contribution to the spectrum from airglow or the Milky Way foreground absorption. The ISM of the Milky Way causes both discrete and continuous absorption in the UV − line absorption due to resonance transitions of ions in the gas, and continuum absorption due to dust.

A detailed analysis of Milky Way absorption lines has been carried out as part of HST Quasar Absorption Line Key Project (Savage et al. 2000). Using 83 FOS spectra of quasars, Savage et al. measured the velocity EWs ($W_v = cW_\lambda/\lambda$) of 29 ionic species in the range 1150 – 3200 Å. This study showed the Milky Way halo to be a mixed ionization system, with the low and intermediate ions having high EWs ($W_v = 180$ km s$^{-1}$) and the high ionization ions (Si IV and C IV) having weaker absorption ($W_v = 60$ km s$^{-1}$). Most of the low ionization Milky Way lines are saturated, and therefore their EWs reflect the kinematical complexity of the sight line. Variations from sight line to sight line were ~30%.

The pattern of ISM absorption lines caused by the Milky Way is very similar to that seen in more actively star forming galaxies. This has unfortunate consequences for the spectroscopy of low redshift galaxies because the foreground Galactic lines blend with the intrinsic ISM lines. This contamination affects 9 galaxies (18 spectra) in our sample which have velocities in the range ±600 km s$^{-1}$. In the remaining galaxies, the foreground lines are also problematic, because they mar regions of good continuum adjacent to the ISM lines. We have therefore attempted to remove the Milky Way lines from all of our spectra. Due to the limited positional information available on the Milky Way ISM lines, we chose to correct our spectra using the median of the EWs reported by Savage et al. (2000). We performed this correction by subtracting off Gaussians centered at $v = 0$ whose widths are set to the approximate instrumental resolution. The Milky Way absorption lines that we correct for are listed in Table 3.

Figure 2 shows the spectrum of NGC 1705 before and after correction for Milky Way absorption. In this high resolution (~0.6 Å) spectrum the Milky Way absorption lines are clearly visible just slightly blueward of the starburst ISM lines. The agreement between our estimates of the foreground absorption and that seen in the spectrum (top panel) is remarkably good. This level of concordance is very typical of our spectra, although there are relatively few cases where this can be assessed reliably because the Milky Way lines tend to be heavily blended with features in the rich starburst spectrum. The importance of the foreground absorption correction is illustrated by comparing the upper and lower panels of Figure 2. At the lower resolution typical of most of our spectra (~2 Å), the Milky Way lines would clearly blend with the starburst lines, and account for nearly half of the line width. In the weakest-lined galaxies in our sample the foreground absorption correction reduces the interstellar line flux by nearly 100%. The correction to the high ionization stellar wind lines is typically much smaller.

Each galaxy's spectral energy distribution is also modified by Galactic foreground reddening. We used the $E(B-V)$ value derived from the dust maps of Schlegel, Finkbeiner, & Davis (1998) and the Galactic extinction curve of Fitzpatrick (1999) to deredden the starburst spectra. The reddening corrections are relatively modest for all of our galaxies ($E(B-V) \lesssim 0.1$) with the exception of NGC 1569 ($E(B-V) \approx 0.5$). Nevertheless, a few spectra show hints of a residual in the 2175 Å dust feature, which is most likely a consequence of a slight error in the foreground reddening correction. For NGC 1569 and I Zw 18 we adopted improved $E(B-V)$ values derived from detailed analysis of the stellar populations with HST (Angeretti et al. 2005; Aloisi et al. 2007).

Besides the Milky Way absorption, our starburst spectra were also contaminated by airglow emission from the Earth's atmosphere, namely Ly-α at 1216 Å and O I at 1304 Å (and occasionally [O II] λ2470, and two unidentified lines at λ2137, λ2802). Unfortunately the strength of the airglow features is not a simple function of position on the sky, but depends sensitively on the solar angle, the column of Earth atmosphere along the line of sight, and the solar activity cycle. We therefore made no attempt to remove these features. The Ly-α airglow has a sufficiently strong equivalent EW (100 – 1000 Å) that it precludes analysis of the underlying starburst Ly-α emission and absorption in most cases. The O I emission is about a factor of 100 fainter, however, so we made measurements of starburst absorption features in the vicinity (typically the O I and Si II lines), but consider the airglow contamination as a possible (large) source of error.

*3.6 The spectral atlas*

The final corrected spectra are presented in Figure 3 – Figure 6. The spectral coverage for the majority of the starburst region is incomplete and in many cases confined to shorter wavelengths. In Figure 3 we reproduced those galaxies with available spectra between 1150 and 1750 Å. These spectra were taken the G130H and G140L gratings of FOS and GHRS, respectively. Those galaxies with spectra in the 1700 – 2300 Å region are shown in Figure 4. The FOS and its G190H grating were used in all cases. Relatively few galaxies were observed at the longest wavelengths in the 2300 to 3200 Å region (Figure 5). This wavelength region contains few *stellar* lines in young starbursts and has received less attention than the line-rich regime at shorter wavelengths.

We generated panchromatic spectra covering the entire 1150 – 3200 Å for the seven starburst regions with full spectral coverage. The spectra for these galaxies were obtained with the G130H, G190H, and G270H gratings of FOS. They are reproduced in Figure 6. Finally we constructed a composite starburst spectrum by taking the median of all available spectra at each wavelength. A continuum normalized version of this spectrum is shown in Figure 7. This spectrum is particularly useful for comparison with spectra of star forming galaxies. It provides guidance for the expected stellar and interstellar lines over the wavelength region 1175 – 3000 Å and can be used, e.g., as a template for isolating the massive star component in a galaxy with mixed populations. In order to study the influence of metallicity on the spectra, we constructed separate high-and low-metallicity templates. The high-metallicity template was restricted to galaxies with $12 + \log(O/H) > 8.7$, and the low-metallicity template to $12 + \log(O/H) < 8.2$. These two templates are contrasted in Figure 8. As expected, the spectral features are generally weaker at lower metallicity. The stellar-wind lines (e.g., C IV $\lambda$1550) are not only weaker at lower metallicity but also show a smaller blueshift. This is the result of stellar winds having lower terminal velocities at lower metal abundance.

All spectral templates are available in electronic format from http://www.stsci.edu/science/starburst/templ.html. We provide two versions of the fully processed spectra: (i) spectra in the observed frame and (ii) spectra in the galaxy rest frame with corrections applied for the dust obscuration and ISM line absorption caused by the Milky Way.

## 4. Multi-Wavelength Galaxy Properties

The principal advantage of using a sample composed of well studied nearby galaxies is the existence of a large body of multi-wavelength data in the literature. We use these data to obtain measures of the galaxy physical properties such as mass, metallicity, and starburst luminosity. In Table 4 we record a number of relevant quantities. Except for the properties noted further below, these quantities were obtained from the NASA/IPAC Extragalactic Database (NED): galaxy morphology (column 2), heliocentric velocity $v_{helio}$ (column 3), blue magnitude $m_B$ (column 4), optical diameter $D_{optical}$ (column 5), inclination $i$ (column 6), foreground reddening $E(B-V)$ (column 7), and a list of alternate names (column 8). The blue magnitudes have been taken preferentially from the RC3.9 catalog (de Vaucouleurs et. al. (1991) and alternately from the ESOLV catalog (Lauberts

& Valentijn 1989). Except in a minority of cases, the RC3 magnitudes are $B_T$ magnitudes, which have been not been *k*-corrected or corrected for Galactic extinction. The galaxy inclinations are from Huchtmeier & Richter (1989). The values of *E(B−V)* tabulated come from Schlegel et al. (1998), with the exception of the values listed for NGC 1569 and I Zw 18, which come from HST photometry (Angeretti et al. 2005; Aloisi et al. 2007).

In Table 5 we collate infrared (IR), UV, and H I data for our sample. The IRAS 60 and 100 μm fluxes (columns 2 and 3 of this table) were taken preferentially from the catalog of IRAS observations of large optical galaxies by Rice et al. (1988), then from the IRAS Bright Galaxy Sample (Soifer et al. 1989) and from the IRAS Faint Source Catalog (Moshir et al. 1990). The IRAS fluxes for He 2-10 are from the Point Source Catalog (IRAS PSC 1988).

We measured UV fluxes directly from the IUE spectra in two 100 Å band passes defined by Kinney et al. (1993) beginning at 1432 and 1863 Å. Reduced spectra were obtained in electronic form from the Kinney et al. atlas where possible. The remaining spectra were downloaded in processed form from the IUE archive. We combined all available IUE archive spectra with reasonable S/N obtained through the large (10″ × 20″) aperture at the same pointing, including data from the Short Wavelength camera (SWP; 1150 – 2000 Å; $\Delta\lambda \approx 5$ Å) and the Long Wavelength cameras (LWR and LWP; 1900 – 3300 Å; $\Delta\lambda \approx 8$ Å). Before measuring the fluxes, the spectra were corrected for foreground Galactic extinction and deredshifted. The fluxes recorded in columns 4 and 5 of Table 5 are the average fluxes in each band pass. For each of the IUE spectra we also measured the spectral slope, $\beta_{IUE}$ (column 6), using a method similar to that of Calzetti et al. (1994).

H I data were used to derive the parameter $\Delta v_{20}$, which represents the measured width of the galaxy's H I $\lambda 21$ cm emission line at 20% of the peak intensity (column 7). When corrected for inclination, $\Delta v_{20}$ is approximately twice the galaxy rotation speed and hence is a good proxy for mass. *The values in column 7 are not corrected for inclination. The corrected values are in Table 6 (see below).* The H I data were taken from the sources listed in column 8. In cases where multiple measurements were listed, we chose the observation with the largest beam size to obtain the fullest sampling of the galaxy rotation curve.

Column 9 of Table 5 gives the *K* magnitudes ($m_K$) for the galaxy sample. The values were taken from the sources in column 10. The GALEX far-UV magnitudes ($m_{FUV}$) are in column 11. The GALEX far-UV channel covers the wavelength region from ~1350 to 1750 Å, with peak sensitivity around 1500 Å. The magnitudes were taken from Gil de Paz et al. (2007). $\beta_{GALEX}$ in column 12 is the UV spectral slope calculated from the GALEX FUV and NUV fluxes listed by Gil de Paz et al.

In Table 6 we list a variety of physical parameters of the galaxy sample. The galaxy distances *d* are in column 2. We used direct distances where available from the HST key project and from various HST programs where the tip of the red giant branch was measured. The remaining distances are taken from the NASA/IPAC Extragalactic Database (NED). They were computed from the galaxy redshift assuming $H_0 = 73$ km s$^{-1}$, $\Omega_{matter} = 0.27$, $\Omega_{vacuum} = 0.73$ (Spergel et al. 2007) and using a model for the local Hubble flow that includes the Virgo Cluster, the Great Attractor, and the Shapley Supercluster (Mould et al. 2000). The reference for each galaxy is in column 3.

We derived absolute *B* magnitudes ($M_B$; column 4) from the apparent magnitudes in Table 4 and the distances given in column 2 of Table 6. We corrected for foreground extinction, but not for any extinction internal to the galaxy. We have not applied *k*-corrections since these would be negligible, given the low redshift of our sample. Likewise, we derived absolute *K* magnitudes ($M_K$; column 5) from the apparent magnitudes in Table 5. In the analysis that follows, we will use $M_K$ to characterize the starburst host galaxy. One concern is that the light from the starburst itself might contribute significantly to $M_K$. However, this was shown not to be the case for $M_B$ in the H98 sample, which is similar in character to the present one. H98 calculated that the median contribution of the starburst to the total *B*-band galaxy light was only 20% in the absence of dust extinction. We will assume that the same applies to the *K* band and that the *K*-band luminosity is dominated by the luminosity of the host galaxy.

The UV and IR luminosities $L_{UV}$ and $L_{IR}$ reported in columns 6 and 7 of Table 6, respectively, are expected to arise primarily from the young stellar population in the starbursts. We derived an estimate of the total UV luminosity between 912 and 3650 Å using the relation

$$\log L_{UV} = \log L_{1500} + 3.2,$$

where $L_{UV}$ is in erg s$^{-1}$ and $L_{1500}$ is in erg s$^{-1}$ Å$^{-1}$ (Leitherer et al. 2002). The UV luminosity we tabulate is corrected for Galactic foreground extinction, but not for extinction internal to the starburst. For the case of an unreddened starburst, the adopted definition of $L_{UV}$ would include about 75% of the bolometric luminosity of the starburst, assuming no underlying old population (Leitherer et al.). We approximated $L_{1500}$ using the GALEX far-UV fluxes (column 11 of Table 5) when available. For those galaxies without GALEX data we used the average IUE fluxes in the 1432 – 1532 Å interval (recorded in column 4 of Table 5). Since our aim is to tabulate the global properties of the galaxies, the GALEX flux is a better measure of the starburst luminosity than the IUE flux because the GALEX field of view encompasses the entire galaxy. One practical concern is that a fraction of the UV flux will be missing in those cases without GALEX data. We can estimate the fraction of the UV flux not included in the IUE data by comparing $L_{1500}$ for those galaxies with both IUE and GALEX data available. On average, $L_{1500}$ derived from IUE is ~25% the GALEX value. While this is significant, our main conclusions are not primarily dependent on $L_{1500}$, but rather on the total bolometric luminosity, which is the sum of $L_{UV} + L_{IR}$ as discussed in the following paragraph. In most cases, the $L_{IR}$ is the dominant term. Moreover, the relations shown in the figures extend over 2 – 3 orders of magnitude in luminosity, which is much larger than a potential 0.6 dex offset in $L_{1500}$.

We calculated the total far-IR flux $F_{IR}$ from the IRAS 60 and 100 μm fluxes according to the prescription of Helou et al. (1988):

$$F_{IR} = 1.26\ (2.58\ f_{60} + f_{100})\ 10^{-11}\ \text{erg s}^{-1}\ \text{cm}^{-2}.$$

The $L_{IR}$ values listed in column 7 of Table 6 were obtained from $F_{IR}$ and the distances in column 2. The sum of the UV and IR luminosities approximates the intrinsic UV luminosity since the IR emission is primarily powered by UV photons which have been absorbed and re-radiated by dust grains (Meurer, Heckman, & Calzetti 1999; Leitherer et al. 2002; Calzetti 2008). Consequently the ratio of $L_{UV}/L_{IR}$ is a good measure of the dust content.

The width of the H I $\lambda$21 cm line (column 8 of Table 6) characterizes the host galaxy (rather than the starburst). In the simple case of circular rotation, this value is related to the line-of-sight rotation speed. The values reported in the table have been corrected for galaxy inclination (see Table 4) where possible. Under the assumption that the H I line width traces the rotation speed, $\Delta v_{20,i}$ is a good proxy for galaxy mass. However it is likely that turbulent motions contribute or even dominate for galaxies which have experienced recent mergers (e.g., IRAS 0833+6517) and for some of the dwarfs.

The nebular oxygen abundances of the galaxies are listed in column 9 of Table 6. In case of multiple pointings with available oxygen abundances, we list the values for all pointings but give the other properties for the first pointing only. The solar oxygen abundance in these units is 8.70 (Meléndez & Asplund 2008); LMC and SMC abundances are 8.4 and 8.0, respectively (Garnett 1999). The oxygen abundances quoted were obtained from a large number of sources. When several measurements were available, the most recent one was, in general, adopted. For a few of the most metal-poor systems the oxygen abundance was calculated directly after measuring the electron temperature from the ratio of [O III] $\lambda$4363/[O III] $\lambda\lambda$4959,5007. However, for the majority of the galaxies, the abundance was computed using "strong-line methods", which rely on empirical calibrations of various nebular line ratios such as, e.g., ([O II] + [O III])/H$\beta$. The errors in this process can be large due to the necessity of removing the underlying Balmer absorption and correcting for reddening. Further systematic errors result from the disagreement among strong-line calibrations (Kewley & Ellison 2008). Moustakas et al. (2010) compared the oxygen abundances resulting from theoretical and empirical (strong-line) calibrations and found differences of a factor of ~5. The sample of Moustakas et al. includes three galaxies studied in this work: NGC 1705, NGC 2403, and NGC 5055. Their quoted theoretical and empirical oxygen abundances for these three galaxies are 8.3 and 8.0, 8.9 and 8.4, 9.2 and 8.5, respectively. These values bracket our derived abundances listed in Table 6, and suggest that the errors in our tabulated abundance measurements are on the order of ±0.3 dex. This is consistent with the spread in values seen when multiple measurements are available. H98 demonstrated that the metallicity of a starburst region has a pronounced impact on its UV spectrum. We have therefore obtained the metallicity closest to the actual HST pointing when possible. This is particularly important for spiral galaxies where there can be strong radial metallicity gradients of the order of 0.4 dex kpc$^{-1}$ (Van Zee et al. 1998). Fortunately the bright star clusters that HST observed where typically associated with bright giant H II regions where nebular metallicities were most likely to be obtained. Column 10 of Table 6 gives the references the oxygen abundances.

## 5. Starbursts versus Star Clusters: comparing HST and IUE

An important goal of this work is to determine how much can be reliably deduced about a galaxy's physical properties based on its UV spectrum alone. H98 demonstrated that the strength of the stellar and interstellar lines and the spectral slope β measured in IUE data were strongly correlated with the global properties of the host galaxy. These correlations can be investigated in more detail with our higher quality HST data set, with the caveat that the HST apertures sample only a small part of the starburst. Since this is rather different from the situation at high redshift where the integrated light of the whole galaxy is observed, it is extremely important to investigate whether HST spectra are representative of the starburst properties as a whole.

The availability of IUE data for nearly all of our sample galaxies makes it possible to compare the UV properties of starbursts on two very interesting spatial scales. In columns 11 and 12 of Table 6 we list the projected aperture sizes $A_{IUE}$ and $A_{HST}$ for each data set, respectively. The $10'' \times 20''$ IUE aperture is a good match to the circumnuclear sizes (a few kpc) of most of the starbursts in our sample and the sizes of bright star-forming complexes (a few 100 pc) in the disks of nearby spirals. In contrast, the HST apertures (typically the $1.74''$ Large Science Aperture for GHRS and the $0.86''$ circular aperture for the FOS) generally encompass only a few bright star clusters. However, in starburst galaxies, one or a few bright clusters can account for a substantial portion of the UV light.

In Figure 9 we contrast the average UV fluxes of the IUE and HST spectra in the two band passes defined by Kinney et al. (1993). The median flux ratio is 15%, however, a few galaxies have over half of their UV flux in the HST aperture. This suggests that individual clusters can be important contributors to the total starburst luminosity but that star formation is typically more extended.

In the upper left panel of Figure 10 we compare the UV spectral slopes $\beta_{IUE}$ and $\beta_{HST}$ of the IUE and HST spectra, respectively. The values of $\beta_{IUE}$ and $\beta_{HST}$ have been measured using a method similar to that of Calzetti et al (1994). Because the extent of the wavelength coverage can have a fairly strong effect on the measured spectral slope, we measure $\beta_{IUE}$ and $\beta_{HST}$ only over the wavelength region in common between the IUE and HST spectra to facilitate the comparison. (Note that because of the rather patchy nature of the HST wavelength coverage, the common wavelength region varies significantly from spectrum to spectrum.)

Figure 10 suggests that the IUE spectra are redder on average than the HST spectra, with the median difference between $\beta_{IUE}$ and $\beta_{HST}$ being 0.6. This trend is likely to be partly a selection effect: in choosing targets for HST preference was usually given to the regions of the starburst with the highest UV surface brightness, thereby favoring clusters which were young but not heavily dust enshrouded. Another factor which could contribute to the color difference is the presence of underlying old stellar populations, which probably account for a larger fraction of the light in the IUE aperture, especially in cases where the starburst is fairly compact. It is of interest to determine if the difference in the UV spectral slopes in the HST and IUE spectra can be accounted for purely by stellar population differences or whether some patchiness of the ISM is required. This has major implications for the "foreground screen" models of the dust distribution (Calzetti 2001) and for the leakage of ionizing photons into the intergalactic medium. Additional relevant clues can be provided by a comparison of the strong stellar and interstellar absorption lines as we will demonstrate in the following. Theoretical energy distributions predict $\beta \gtrsim -3$ (Leitherer et al. 1999). A few galaxies in Figure 10 fall outside this relation. However, we caution from over-interpreting this apparent discrepancy since all the outliers have limited spectral coverage and the measurement errors are large.

The UV spectral slopes in Figure 10 can also be compared to the slopes derived from the GALEX FUV and NUV fluxes $\beta_{GALEX}$ (column 12 in Table 5). Ten galaxies have available measurements of both $\beta_{GALEX}$ and $\beta_{IUE}$. On average the two sets of UV spectral slopes agree: $\beta_{GALEX} - \beta_{IUE} = 0.02 \pm 0.57$. However, this agreement is significantly biased by several nearby, large galaxies such as NGC 2403 where aperture size effect come into play. Most galaxies whose star formation is concentrated in nuclear star clusters have bluer IUE than GALEX slopes. This is consistent with the trend between HST and IUE suggested by Figure 10.

The upper right panel of Figure 10 compares the EW of the average of the three strongest interstellar absorption lines, Si II $\lambda1260$, O I $\lambda1303$ + S II $\lambda1304$, and C II $\lambda1335$ measured in the IUE and HST spectra. For consistency, we have smoothed the HST data to the 6 Å resolution of IUE and used the windows defined by H98 to measure the EWs. ISM line EWs range from about 1 to 6 Å and appear to be in relatively good agreement in the IUE and HST spectra. Given that the apertures employed differ by roughly two orders of magnitude, this agreement is rather striking. Taken at face value, the results of this figure may suggest a constant covering factor when going from the small HST to the larger IUE apertures.

Although we cannot make use of the line profile information due to the low resolution of the IUE data (~1000 km s$^{-1}$), we can draw some inferences based on the ISM line EWs on the assumption that the lines are saturated. In this case, the EWs do not depend strongly on column density but reflect the velocity distribution and the covering factor of the gas. If rotation were the dominant factor affecting the line width, then pronounced differences would be expected between the small aperture HST observations and the large aperture IUE observations. The lack of such systematic differences in the measured EWs suggests a turbulent origin for the line widths. This is plausible given that starbursts deposit mechanical energy into the ISM by via stellar winds and supernovae.

It is also illuminating to compare the EWs of the strong stellar wind lines Si IV λ1400 and C IV λ1550. (Note that there is an interstellar contribution to the total EW as well.) We measured the stellar-wind lines in the same manner as the interstellar lines using the windows of H98. The lower two panels of Figure 10 show that the EW of Si IV is very similar in the IUE and HST spectra, while the EW of C IV is larger in the HST spectra by an average of 2.5 Å. The agreement of the Si IV lines and the disagreement of the C IV lines seems hard to reconcile at first since both features are a blend of interstellar absorption lines and stellar-wind P-Cygni features. One clue is the fact that the contribution of the wind lines to the absorption is greater at C IV than at Si IV. Using our same line measurement techniques on the Starburst99 models (Leitherer et al. 1999), we found that the difference in the EW of C IV at 5 Myr (when the wind lines are maximal) and at 50 Myrs (when the wind lines are absent and only interstellar lines contribute) is ~3 Å, whereas the difference for Si IV is only ~1 Å. The similarity between these numbers and the differences manifested in Figure 10 suggests this as the likely explanation, provided that the clusters observed by HST are younger than the surrounding starburst.

## 6. New Measures of UV Spectral Features

In order to quantify the UV spectral morphology of galaxies it is desirable to develop a repeatable method for measuring the various UV absorption features. The major challenge in this undertaking is fitting the UV continuum. The continuum is difficult to

pinpoint – even by eye – due to the myriad weak stellar features, the strong absorption by Ly-α, and the effects of reddening on the continuum shape. Most prior works utilized hand-tuned polynomial or power-law fits. We experimented with automated versions of these techniques with little success: power laws seemed to have trouble fitting the small-scale structure, while polynomials tended to fit too much structure and perform poorly at the endpoints. These techniques were also extremely sensitive to the wavelength coverage of the spectrum. It proved to be considerably more robust to define a local continuum for each spectral feature of interest.

To promote standardization, we have created a set of 12 UV line indices similar in form to the well known Lick indices (Worthey1994; Schiavon 2007). These 12 indices are presented in Table 7. Column 1 of this table gives the name of the index. SiIV_1400 and CIV_1550 are the only indices which are typically dominated by stellar-wind lines. These indices are useful for determining ages, metallicities, and initial mass functions of stellar populations in stars-forming galaxies (e.g., Leitherer 2010), in particular if the observational data have insufficient quality to permit detailed line profile fits. AlIII_1860 is a blend of wind and interstellar lines, and the remaining 9 indices are mostly interstellar. As demonstrated by H98, the interstellar lines in starburst galaxies, even if saturated, are sensitive to the metallicity of the host galaxy and are therefore a useful abundance indicator. For each index we define a central band pass and two flanking continuum band passes (columns 2, 3, and 4, respectively). We define the continuum at the midpoint of the flanking bands to be the median value of the flux within the band pass. We use median rather than the average because it is more robust at low S/N levels and stable against spectral artifacts, such as the presence of a weak intervening absorber.

The continuum in the central index band pass is defined by a straight line connecting the midpoints of the two continuum bands. The most important stellar absorption features included in the band pass of the index are in column 5 of Table 7.

We have taken considerable care in placing both the central band pass and the continuum bands. We have endeavored to make the central band passes broad enough to accommodate a range of velocity dispersions, but no broader than necessary. In several instances, we have grouped together several neighboring lines (e.g., Fe II_2370 = Fe II $\lambda$2344, $\lambda$2374, $\lambda$2383.) The continuum bands have been carefully chosen to avoid regions of stellar line-blanketing, which show up clearly only in the most metal-rich systems. The indices are illustrated in Figure 11 and Figure 12. The spectrum shown in Figure 11. is a high metallicity composite made from the median of our sample galaxies with abundances greater than $12 + \log(O/H) = 8.7$ (see also Figure 8). The composite is illustrative because of its superior S/N and because the weak stellar features are strongest at high-metallicity. For contrast we show a relatively low S/N spectrum of NGC 5253 in Figure 12. This spectrum demonstrates that the indices do an adequate job of constraining the continuum even at modest S/N.

Following these definitions, we measured the line indices in all galaxy spectra where feasible. No measurements were performed in cases of AGN activity leading to an emission contribution to the interstellar lines. The measurements are summarized in Table 8. The table gives the line index measurements (in Å) in columns 2 – 13, as well as the mean velocity of the nine interstellar indices with respect to the galaxy rest frame velocity (column 14).

We attempted to estimate the error in our index measurements even though we have not propagated the error array output by the HST pipeline through to our final calibrations. We define the error in the continuum fit as the standard deviation of the data points in the blue and red band passes about the line defined as the continuum. This is appropriately propagated into the error in the EW of the index. A by-eye examination of the indices for our galaxies suggests that the predominant error source is the continuum fit. The method we have adopted of approximating the error in the line index appears to do a very good job of representing the general quality of our index measurements. We will use this method for the error estimates presented in the following section.

## 7. UV Diagnostics of Galaxy Physical Properties

Having established that our HST spectra correlate with large-aperture UV spectra in predictable ways, we now turn to examining how the various spectral features in our HST data correlate with the physical properties of galaxies.

*7.1 Correlations between starburst and host galaxy properties*

We are interested both in the properties of the starburst itself and those of the host galaxy. We characterize the starburst by its intrinsic UV luminosity $L_{UV} + L_{IR}$ and its dustiness – measured by the ratio of the emitted versus reprocessed flux $L_{UV}/L_{IR}$. The salient properties of the host galaxy are its luminosity ($M_K$), its mass (as traced by $\Delta v_{20,i}$), and its metallicity ($12 + \log(O/H)$). As pointed out by H98 and many other authors, very strong correlations exist among the physical properties themselves: the most massive galaxies are the most optically luminous and metal-rich, and they host the dustiest and most UV-luminous starbursts. These correlations are shown for our sample in Figure 13. In this figure we show how galaxy mass, metallicity, starburst strength, and dust obscuration correlate with the galaxy luminosity as expressed by $M_K$. Since mass, metallicity, dustiness, and host and starburst luminosity are all correlated with one another, we do not plot all possible combinations of parameters, but instead focus on the trends that are the most physically motivated.

We have included in some of the figures the well-studied lensed Lyman-break galaxies MS 1512-cB58 (Pettini et al. 2000), the "Cosmic Horseshoe" (Quider et al. 2009), and the "8 O'clock Arc" (Dessauges-Zavadsky et al. 2010). The oxygen abundance of $\log(O/H) + 12 = 8.4$ of MS 1512-cB58 has been derived by Teplitz et al. (2000) based on IR observations of the common optical nebular emission lines and by Rix et al. (2004) from modeling weak photospheric absorption lines. Hainline et al. (2009) and Finkelstein et al. (2009) derived oxygen abundances of 8.4 and 8.6 using the N2 index for the Cosmic Horseshoe and the 8 O'clock Arc, respectively.

We have calculated the sum of the far-UV and far-IR luminosities for MS1512-cB58 using the rest-frame UV flux measured by Ellingson et al. (1996) and the rest-frame far-IR flux by Siana et al. (2008) assuming a gravitational lens amplification of 30 (Seitz et al. (1998). $M_K$ was derived from the rest frame 8 μm data in Siana et al. (2008). MS 1512-cB58 clearly ranks among the most luminous starbursts and has correspondingly strong ISM features. There are no corresponding luminosity measurements for the other two Lyman-break galaxies.

*7.2 Interstellar lines: correlations between EW and galaxy parameters*

We first discuss the interstellar lines. The UV interstellar lines hold great promise as diagnostics of high-redshift galaxy properties. The Si II λ1260, O I λ1302 + Si II λ1304, and C II λ1335 lines are some of the strongest features observed in UV starburst galaxy spectra, as shown by our composite spectrum in Figure 7. Previous studies of the ISM lines in starbursts have yielded important information about the chemical enrichment and gas kinematics of galaxies at both low and high redshift (Heckman & Leitherer 1997; Sahu & 1998; Kunth et al. 1998; Pettini et al. 2002; Savaglio et al. 2004; Chen, Prochaska, & Bloom 2007; Grimes et al. 2009). However, these works made use of high-resolution spectra which resolved the velocity structure of the ISM lines. Since such observations are costly in terms of telescope time, it is of interest to determine how much information can be reliably deduced from observations with moderate resolution.

We begin our analysis with a simple examination of the correlation of line EWs with the starburst luminosity and metallicity. We measured the interstellar absorption lines using the UV line indices defined in Section 6 and Table 7. The indices have been measured after subtracting off the foreground Milky Way ISM lines, as described in Section 3. The error in this correction has not been included in errors quoted for the line indices. Since we have endeavored to select continuum windows which are free of other features, our index measurements approximate true EWs, and should therefore always be positive in the case of absorption (unlike the Lick indices which may take on positive or negative values due to changes in the continuum bands.)

The EW of a given interstellar line depends on several factors: the column of gas along the line of sight, the covering factor of the gas, the abundance of the element, the fraction of the atoms in the appropriate ionization state, the transition probability (oscillator strength), and the velocity distribution of the gas. However, when the lines become strongly saturated, the EW depends mainly on the velocity distribution and the covering factor of the gas. Unfortunately the instrumental resolution of our spectra (a few 100 km s$^{-1}$) is insufficient to determine if the lines are saturated directly from their line profiles. However, because the strong Si II λ1260 λ and λ1526 lines have oscillator strengths that differ by a factor of ~8, the ratio of their EWs provides a quick test. On the linear part of the curve of growth EW(Si II λ1260)/EW(Si II λ1526) ≈ 5.2, whereas our spectra have ratios ranging from ~0.5 – 2, as shown in Figure 14. This suggests that Si II transitions are generally close to being optically thick. The same is likely to be true for the other strong low-ionization lines.

In Figure 15 we plot the line EWs of our nine new interstellar line indices versus metallicity. We also included in this figure the EWs measured for the three lensed Lyman-break galaxies MS 1512-cB58, the Cosmic Horseshoe, and the 8 O'clock Arc. Only data points with errors of less than 1.5 Å are shown. For comparison we indicate the median EWs of the Milky Way ISM lines as reported by Savage et al. (2000). Given the complex factors governing the line EWs, correlations with metallicity are not a priori expected. However, strong trends are evident in the SiII_1260, OI_SiII_1303, and CII_1335 indices, in agreement with the findings of H98 for the sum of these features. An additional line whose EW appears to be well correlated with metallicity is Si II λ1526. The AlII_1670 index is generally too noisy to be very useful. While this line is nearly as strong as the Si and C features, the continuum estimation in this heavily line-blanked region of the spectrum is rarely satisfactory. The weaker Fe II λ1608 line also suffers from these problems to some degree. Unfortunately the spread in metallicity of the galaxies for which near-UV spectra were obtained is not large, so no reliable conclusions can be drawn about the usefulness of the FeII_2370, FeII_2600, and MgII_2800 indices.

In Figure 15 we included data points from the $z = 2.73$ lensed Lyman-break galaxy MS 1512-cB58. We measured the line indices of MS 1512-cB58 from a spectrum kindly provided by Max Pettini (Pettini et al. 2000). Also included in this figure are the results for the Lyman-break template of ……..Shapley et al. (2003) where we measured the indices in the same way and assume its average metal abundance to be identical to that of MS 1512-cB58. Good agreement is seen between the high-redshift galaxies and our low-redshift sample, with a slight tendency for MS 1512-cB58 to have stronger UV lines than local galaxies of the same metallicity.

The origin of the correlation between the ISM line EW and metallicity is somewhat perplexing given the very different star forming environments spanned by our sample. If the ISM lines were unsaturated (i.e., on the linear part of the curve of growth), the trend with metallicity would imply similar gas columns in the starbursts. This seems fairly implausible given that our sample includes SA galaxies like NGC 5055 as well as unevolved gas-rich dwarfs such as I Zw 18. A more plausible explanation is that the ISM lines tend to be saturated (as suggested by the Si II $\lambda1260$/Si II $\lambda1526$ ratio) and that the correlation with metallicity is a secondary effect. As discussed previously, more metal-rich galaxies tend to host more powerful starbursts which are likely to deposit more energy into the ISM, thereby increasing the turbulent velocity dispersion.

In Figure 16 we show the relation between ISM line equivalent width and the intrinsic UV luminosity of the starburst, $L_{UV} + L_{IR}$. The SiII_1260, OI_SiII_1303, CII_1335 and SiII_1526 indices all display good correlations. The FeII_1608 index shows a hint of an anti-correlation. We note, once again, the excellent agreement of the Lyman-break galaxies with the trends defined by the local galaxies. While an examination of trends in the ISM line EWs is illuminating, it does not take advantage of the superior resolution of the HST data. At HST's resolution, kinematical studies of the gas become possible: bulk inflows or outflows of the gas can be measured relative to the photospheric features, and the turbulent velocity of the ISM can be inferred from the line widths. While the ISM line profiles are not fully resolved in the HST data, many of our sample spectra exhibit line widths well in excess of the instrumental resolution. Also, because we sample transitions with different oscillator strengths, some information about the Doppler broadening can in principle be recovered. We have therefore attempted to model the ISM absorption in each spectrum using a very simple prescription.

*7.3 Line profiles and velocity measurements*

We fitted the line profiles of the ISM lines between 1200 and 1700 Å using a model with three free parameters: the total column density of H I, the turbulent velocity (Doppler *b* parameter) and the velocity offset of the lines. This made the implicit assumption that gas at different ionization stages has the same velocity profile. However, we mitigated this by only performing the actual fit of the model over spectral regions encompassing the first ions – Si II λ1260, Si II λ1304, C II λ1335, Si II λ1526, Fe II λ1608, Al II λ1670. The other simplifying assumption we make is that the relative abundances of the elements in the gas phase are the same as the solar values reported by Meléndez & Asplund (2008). The oxygen abundance of the gas was then scaled to match the measured nebular metallicity. The implicit assumptions built into this approach are that the α-enhancement is similar to the solar pattern (although is only affects Fe II λ1608), that the refractory nature of the elements does not change strongly with metallicity, and that ionization corrections are not required. While these assumptions are unlikely to hold in detail, they are reasonable approximations given the quality of the data. The final step in constructing the models was to convolve the synthetic spectra to the appropriate instrumental resolution. However, as discussed in Section 3, the instrumental resolution is not well known for our spectra. We assumed a resolution of 250 km s$^{-1}$ for the FOS spectra and 160 km s$^{-1}$ for the GHRS spectra, except for the NGC 1705 data where we use 120 km s$^{-1}$. This assumption is likely to be a source of error in the derived $b$ parameter.

We now turn to examining how the measured velocity dispersions and velocity offsets correlate with galaxy physical properties. In this analysis we include only data where the model fit converged and where a by-eye inspection suggested that the fit was reasonable. We note that at our velocity resolution there is a certain amount of degeneracy expected between the derived gas column and the Doppler $b$ parameter. However, the velocity offsets are completely robust to this degeneracy.

In Figure 17 we show how the velocity offsets relate to various host galaxy parameters. The velocity offsets are measured relative to the weak photospheric features as described in Section 3. There appears to be no strong trend in the velocity offset with rotation speed or starburst luminosity. However, it is interesting to note that galaxies with large absolute values of the velocity offsets (>100 km s$^{-1}$) have starburst luminosities greater than $10^{43}$ erg s$^{-1}$ (corresponding to star-formation rates larger than 0.2 M$_\odot$ yr$^{-1}$). This could be interpreted as evidence of a threshold star-formation rate needed to drive a galactic wind. A correlation between outflow velocity as measured with the optical Na I D lines and the global star-formation rate in starburst galaxies was suggested by Martin (2005) and Rupke et al. (2005). Chen et al. (2010) found a similar but weaker trend in a sample of star-forming galaxies drawn from Sloan Digital Sky Survey Data Release 7. This is consistent with theoretical expectations for the velocity of a swept-up shell of gas at the point where it blows out of the disk. However, more data are clearly necessary for a definitive conclusion.

*7.4 Stellar-wind lines*

Next we examine how the properties of the stellar-wind lines depend upon the starburst luminosity and metallicity. We utilize the SiIV_1400 and CIV_1550 line indices defined in Table 7 to characterize the *absorption* component of the Si IV and C IV stellar-wind lines (see Figure 11 for an illustration). We have not used the N V $\lambda 1240$ line, due to the difficulty of measuring the continuum in such close proximity to the Ly-α absorption line. We note that the measured Si IV and C IV features are not always stellar in origin but include large – sometimes dominant – contributions from interstellar absorption at Si IV $\lambda\lambda 1394,1402$ and C IV $\lambda\lambda 1548,1551$. In Figure 18 and Figure 19 we show the correlation of the EWs of the Si IV and C IV features with several galaxy properties. As demonstrated by H98, there is a clear increase in the EW of the stellar-wind lines with starburst luminosity and metallicity.

It is clear that the properties of hot-star winds must depend upon metallicity because the winds are driven via photon momentum transfer through metal-line absorption. The strength of the N V λ1240, Si IV λ1400, and C IV λ1550 P-Cygni features in individual O and B stars is set by the mass-loss rate, which theoretical and observational results suggest scales as $\dot{M} \propto Z^{0.5-1.0}$ (Vink, de Koter, & Lamers 2001; Mokiem et al. 2007). However, the situation in composite stellar populations is complicated by the fact that the stellar-wind line profiles also depend strongly on the effective temperature and luminosity class of the ionizing star. The wind features in galaxy spectra therefore reflect the make-up of the stellar population (age, IMF) as well as the metallicity (Leitherer et al. 2001). Unfortunately there is a certain amount of degeneracy between these effects since both metal-poor and old stellar populations will show weak stellar-wind features.

The age-metallicity degeneracy problem is of particular relevance to our present study because the HST spectra are primarily dominated by the light of individual star clusters which are likely to contain stellar populations with discrete ages. Given this fact, the existence of a trend in the Si IV and C IV line strength with metallicity is remarkable. There are several reasons why this might occur. First, the HST targets selected where generally those with the highest UV-brightness. This would tend to preferentially select very young clusters since their 1500 Å luminosities peak at about 3 Myr and decline rapidly thereafter. Young ages are clearly suggested for most of the spectra shown in Section 3 by the existence of strong C IV P-Cygni emission. The second possibility is that the interstellar components of Si IV and C IV are driving the observed trend since a correlation between interstellar line equivalent width and metallicity was demonstrated in Figure 15.

Once again, the line profile information available with HST's superior resolution provides insight. Figure 20 shows the Si IV and C IV line profiles of template spectra constructed by averaging spectra of similar metallicities. The range in metallicity is from 1/5 solar to nearly 3 times solar. While it is difficult to gauge the exact interstellar contribution to the Si IV and C IV lines, the asymmetric blueshifted portion of the lines can clearly be attributed to stellar winds. Notably, the stellar and interstellar contributions appear to scale roughly in lock-step. Even at the lowest metallicities, very broad blueshifted components are evident. Visual inspection of the profiles suggests that the interstellar component could plausibly account for ~2/3 of the Si IV EW and ~1/3 of the C IV EW. Clearly, both the stellar and interstellar components contribute to the observed scaling of the line equivalent widths with metallicity.

Besides mitigating the effects of the stellar population age on the observed Si IV and C IV features, another positive side effect of the blend of stellar and interstellar features is that the line strengths of the combined features are nearly double that of the pure interstellar features. Thus, the Si IV and C IV lines can serve as valuable metallicity diagnostics even in spectra with relatively poor S/N. This fact has already been taken advantage of to map the chemical enrichment history of the universe. Mehlert et al. (2002) used the C IV EW measured from 57 VLT spectra of galaxies at redshifts from 1.37 to 3.4 to infer the evolution of the mean cosmic metallicity at these epochs. However, some caution is in order due to the composite nature of the lines. In Figure 21 we compare the spectrum of the lensed Lyman-break galaxy MS 1512-cB58 with the metallicity templates shown in Figure 20. The stellar-wind features match moderately well with the two low-metallicity templates (12 + log(O/H) = 8.0 and 8.2), but the interstellar features appear much stronger. The results in this figure suggest care when interpreting absorption features containing both interstellar and stellar contributions. Crowther et al. (2006) found that for the C IV profile to give reliable abundances at high redshift, its usage should be limited to data of sufficiently high spectral resolution (>300 km s$^{-1}$) in order to separate the stellar and interstellar CIV components. Oxygen abundances will be systematically overestimated if the local calibration is applied to spectra of high-redshift galaxies obtained at lower resolution, as is typical for many current wide-field surveys.

Ironically, the well-studied, archetypal Lyman-break galaxy MS 1512-cB58 may have rather atypical interstellar line strengths for its general properties. Heckman et al. (2001) measured residual intensities in the line cores of the UV interstellar lines of a few percent. The corresponding EWs are about a factor of 2 higher than the average of the Lyman-break sample discussed by Steidel, Pettini, & Adelberger (2001). The reason for the difference is not understood and may hint at the diversity of the ISM in high-redshift galaxies.

## 8. Conclusions

We have presented 46 UV spectra of local starburst and star-forming galaxies obtained with HST's legacy instruments FOS and GHRS through ~1 – 2 arcsec apertures with a spectral resolution of a few 100 km s$^{-1}$. These observations provide a unique data set to study local starbursts due to the specific spatial scales that they probe. Earlier observations through the large IUE aperture sample starburst sizes of order ~1 kpc. In contrast, STIS, one of HST's current UV spectrographs probes scales as small as a few pc. FOS and GHRS bridge this gap by observing multi-cluster regions extending over tens of pc. When compared at the same resolution, HST and IUE spectra show some substantial differences. However, these differences are not random, but systematic, and fully consistent with our expectation based on the aperture differences. The star clusters sampled by the HST spectra provide about 15% of the UV luminosity measured by IUE. Their spectra are bluer and have stronger stellar-wind features than the starburst as a whole. This suggests either that the clusters observed by HST are the youngest pieces of the burst, or that a significant amount of light from an older stellar generation contributes to the IUE spectrum. Interestingly, the interstellar lines appear to be generally consistent in the IUE and HST spectra, which is evidence of similar ISM properties along most sightlines. Under the assumption that the ISM lines are saturated, the lack of any systematic differences implies a turbulent rather than rotational origin for the line widths. Altogether, the nature and degree of differences between the small-aperture FOS and GHRS spectra and the large-aperture IUE spectra indicate that the HST spectra are suitable for our stated purpose of learning how UV spectral morphology correlates with galaxy physical properties.

The spectra presented in this work can be an invaluable tool for the interpretation of star-forming galaxies at any redshift, and they can test and complement theoretical spectral models. Efforts to interpret the UV spectra of galaxies tend to fall into three categories: comparison with observed galaxy templates, generation of synthetic spectra with observed stellar templates, and a fully theoretical approach using stellar model atmospheres. Each approach has its distinct pros and cons, and taken together they can identify shortcomings and provide guidance for improvement. The galaxy templates presented here cover the hitherto relatively unexplored spectral region between 2000 and 3000 Å. While this region is devoid of strong stellar-wind features, it contains numerous interstellar lines and a plethora of photospheric blends which can be used to constrain chemical abundances and other properties. The full wavelength region of our templates from 1150 to 3000 Å provides insight into the behavior of the four principal spectral-line types found: stellar-wind P-Cygni lines, photospheric absorption blends, narrow interstellar absorption, and nebular emission lines.

Numerous authors have highlighted the similarity between local starbursts and Lyman-break galaxies (Meurer et al. 1997; H98; Chandar, Leitherer, & Tremonti 2004; Grimes et al. 2006; Schwartz et al. 2006; Grimes et al. 2007; Pettini 2008). Our atlas is suitable for comparisons with the most recent generation of high-redshift observations, whose wavelength coverage, S/N, and spectral resolution are close matches to the FOS and GHRS data. The spectra presented here are available in electronic form for download at http://www.stsci.edu/science/starburst/templ.html.

*Acknowledgments.* We thank Miroslova Dessauges-Zavadsky and Max Pettini for kindly providing electronic versions of the rest frame UV spectra of the "Eight O'clock Arc" and the "Cosmic Horseshoe", respectively. Support for this work has been provided by NASA through grant number AR-08348.01-97 from the Space Telescope Science Institute, which is operated by AURA, Inc., under NASA contract NAS5-26555. This work has made extensive use of the NASA/IPAC extragalactic database (NED), which is operated by the Jet Propulsion Laboratory, Caltech, under contract with the National Aeronautics and Space Administration. We also utilized the Atomic Line List (v2.04) maintained by Peter van Hoof at http://www.pa.uky.edu/~peter/atomic/. We thank Ed Jenkins for clarifying discussions on the importance of Galactic high-velocity clouds.

# Figures

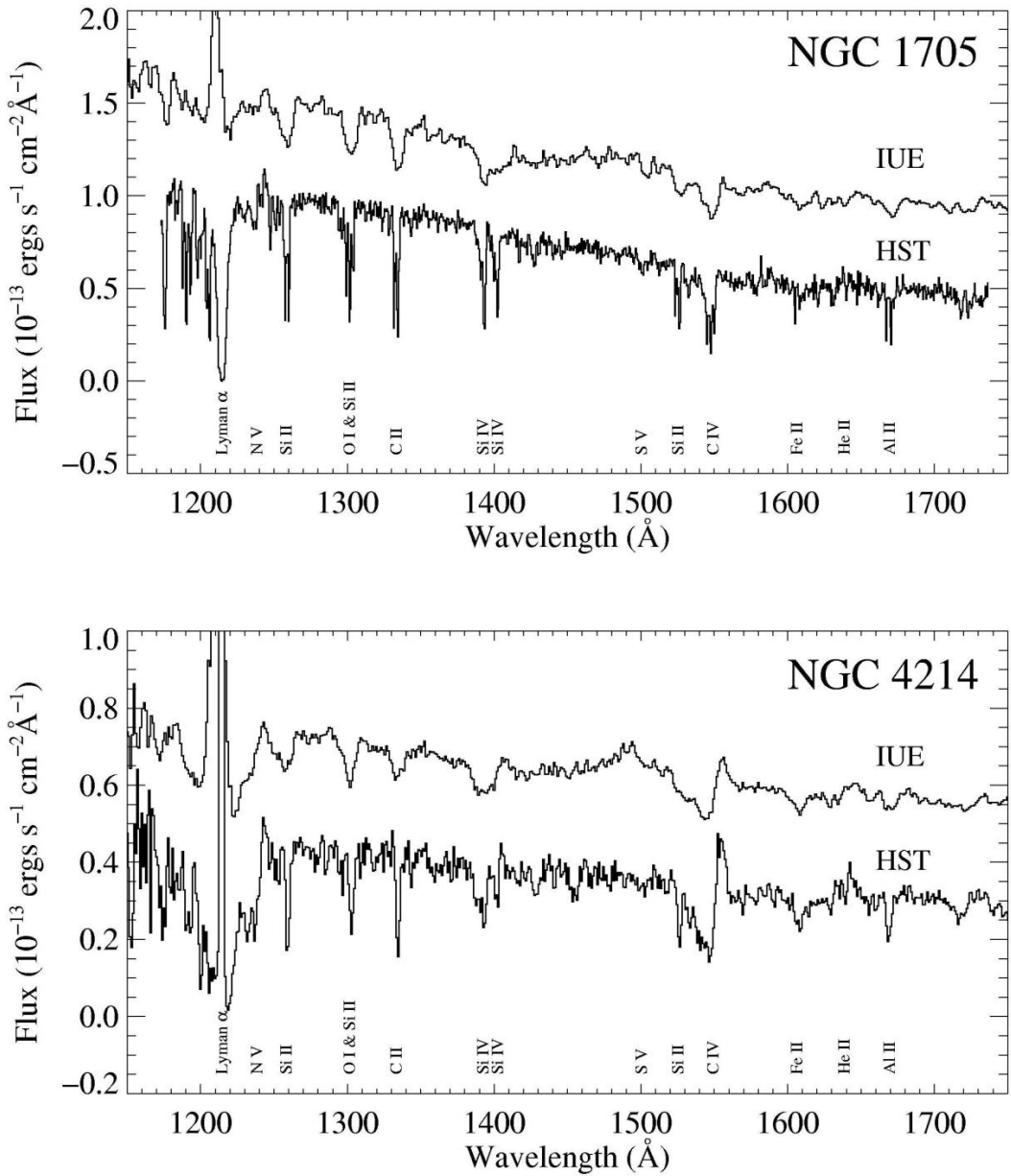

Figure 1. – Comparison of IUE and HST spectra of NGC 1705 (upper) and NGC 4214 (lower). The IUE spectrum has been scaled so that the flux at 1500 Å matches the HST spectrum and then offset by 0.5 (top) and 0.3 (bottom) for clarity. The slightly blue-shifted lines visible in the HST spectrum of NGC 1705 are due to interstellar absorption from the Milky Way.

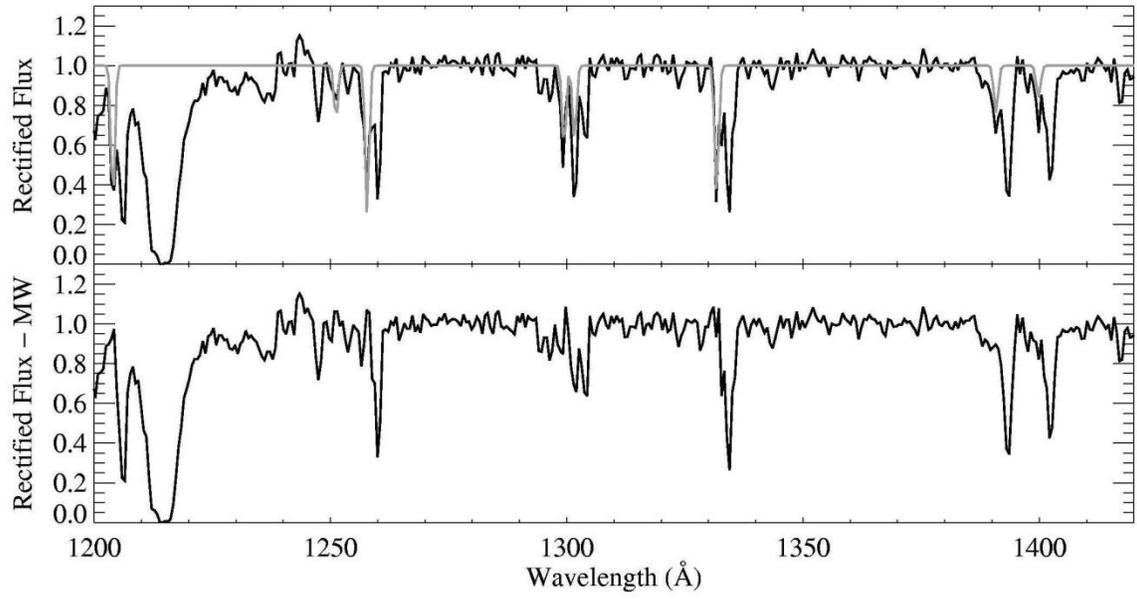

Figure 2. – Spectrum of NGC 1705. The upper panel illustrates the correction that we make for the foreground Milky Way absorption (light gray line) and the lower panel shows the spectrum after subtraction of the Milky Way lines.

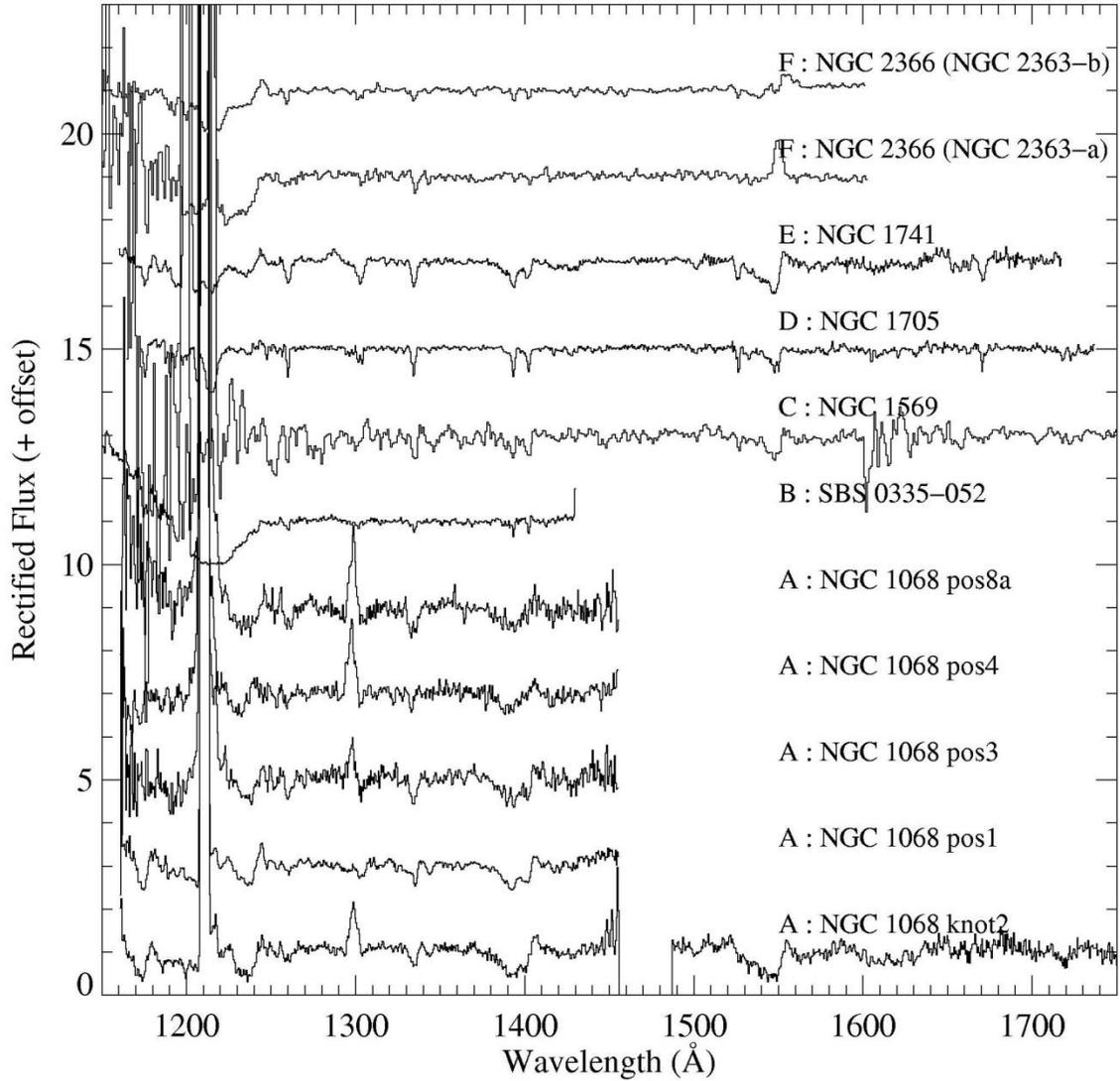

Figure 3. – Galaxies with spectra in the 1150 – 1750 Å region arranged by right ascension. Spectra are labeled with a capital letter followed by the name of the galaxy. The capital letter preceding the galaxy name is the identifier introduced in Table 2. The intervening Milky Way ISM lines have been subtracted from the spectra as described in Section 3.5. The Ly-α and O I λ1304 airglow lines have not been removed from the spectra. Details about the individual observations can be found in Table 2.

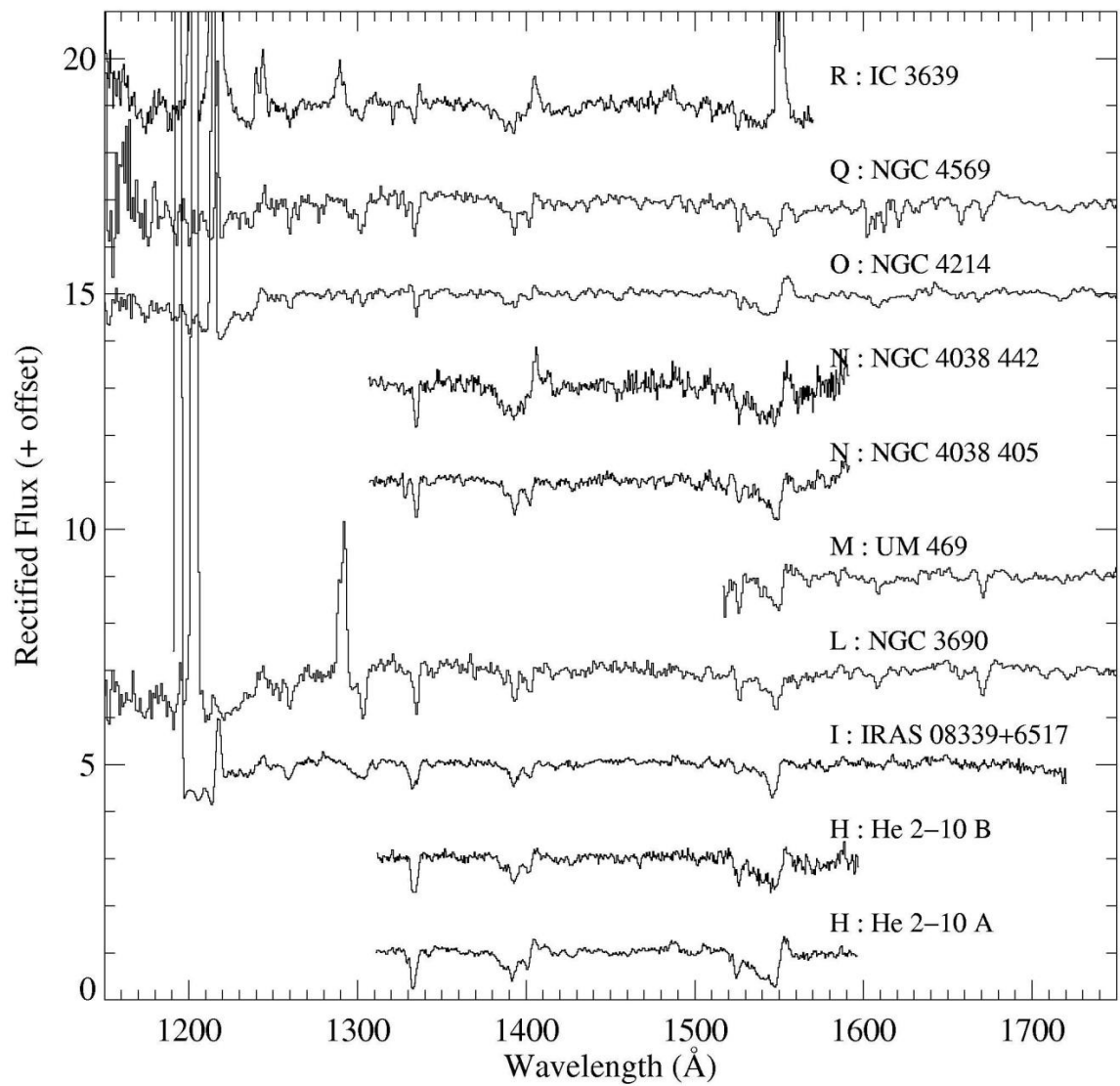

Figure 3. − continued.

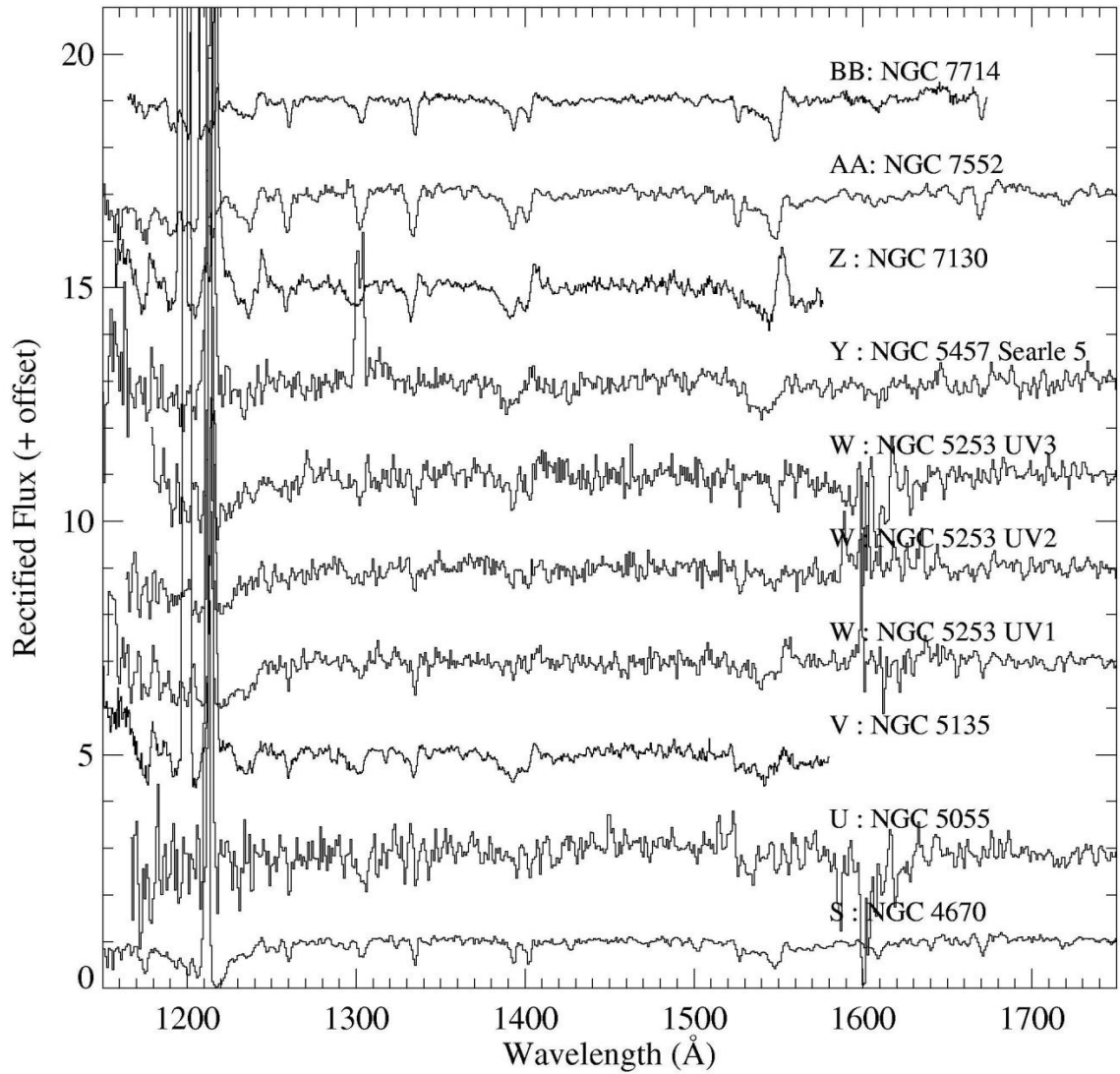

Figure 3. – continued.

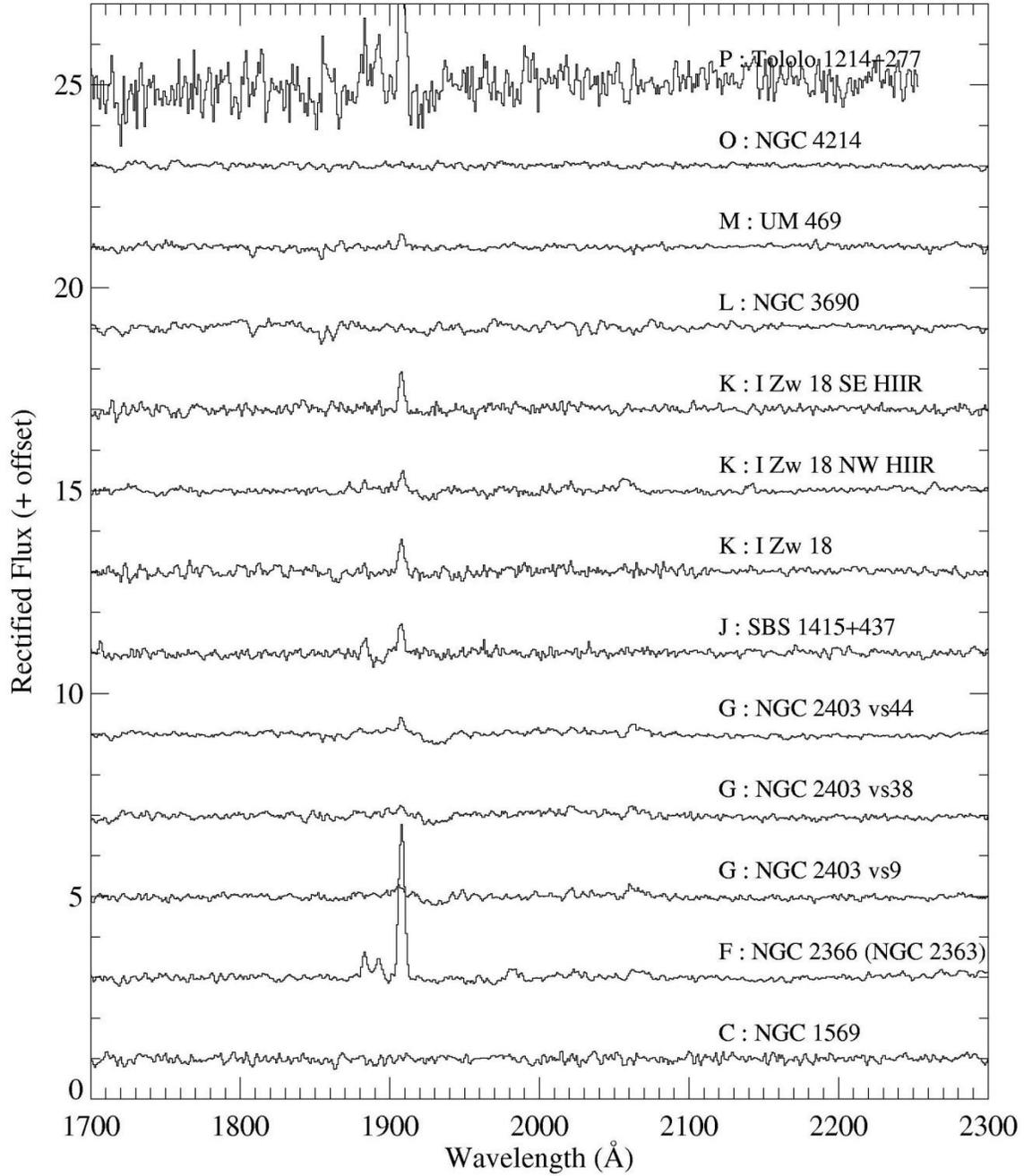

Figure 4. – Galaxies with spectra in the 1700 – 2300 Å region. All of these spectra were obtained with the FOS and the G190H grating. They have been corrected for foreground Milky Way ISM line contamination as described in Section 3.5. The emission features are the nebular lines Si III $\lambda\lambda1882/92$ and C III] $\lambda\lambda1907/09$.

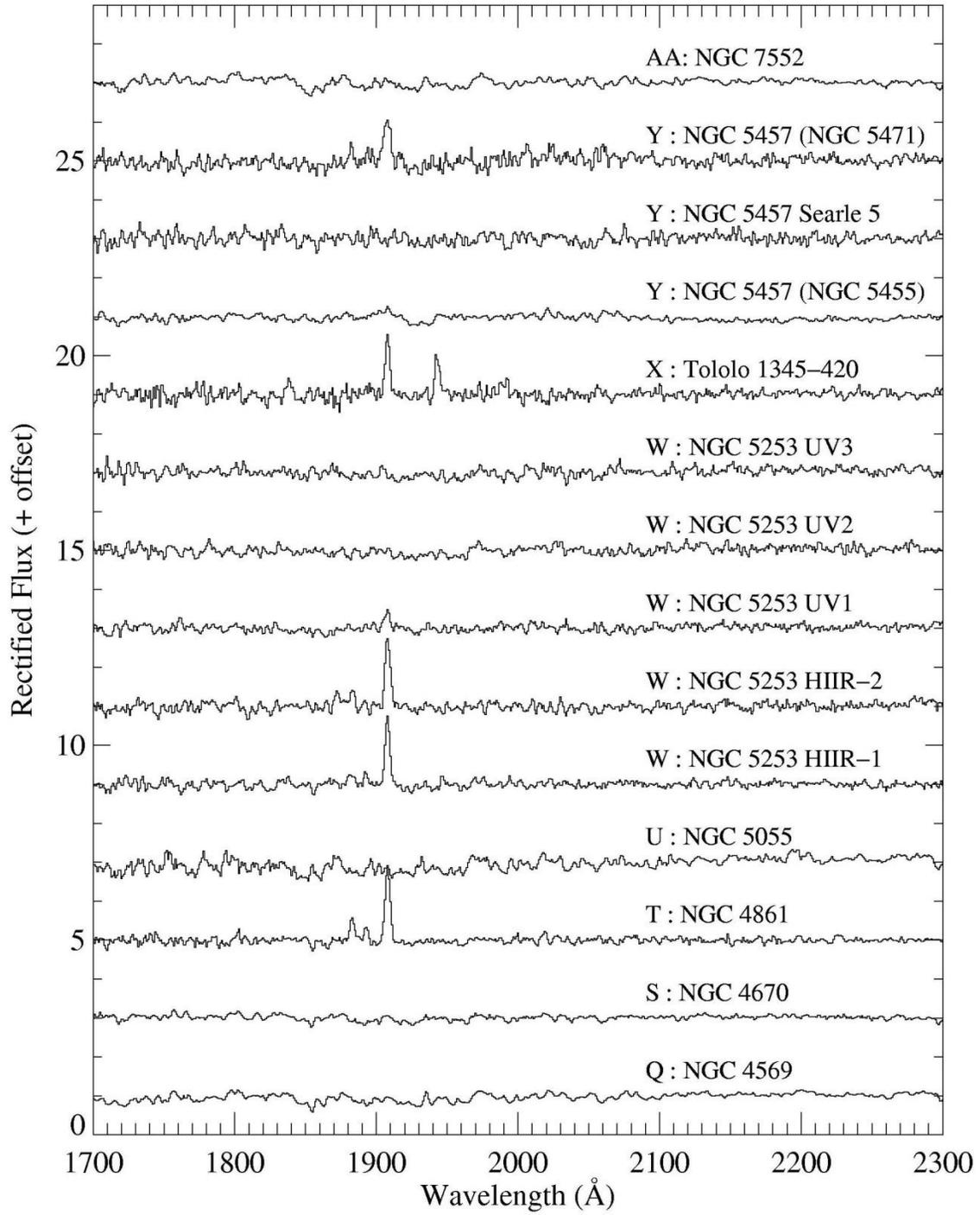

Figure 4. – continued.

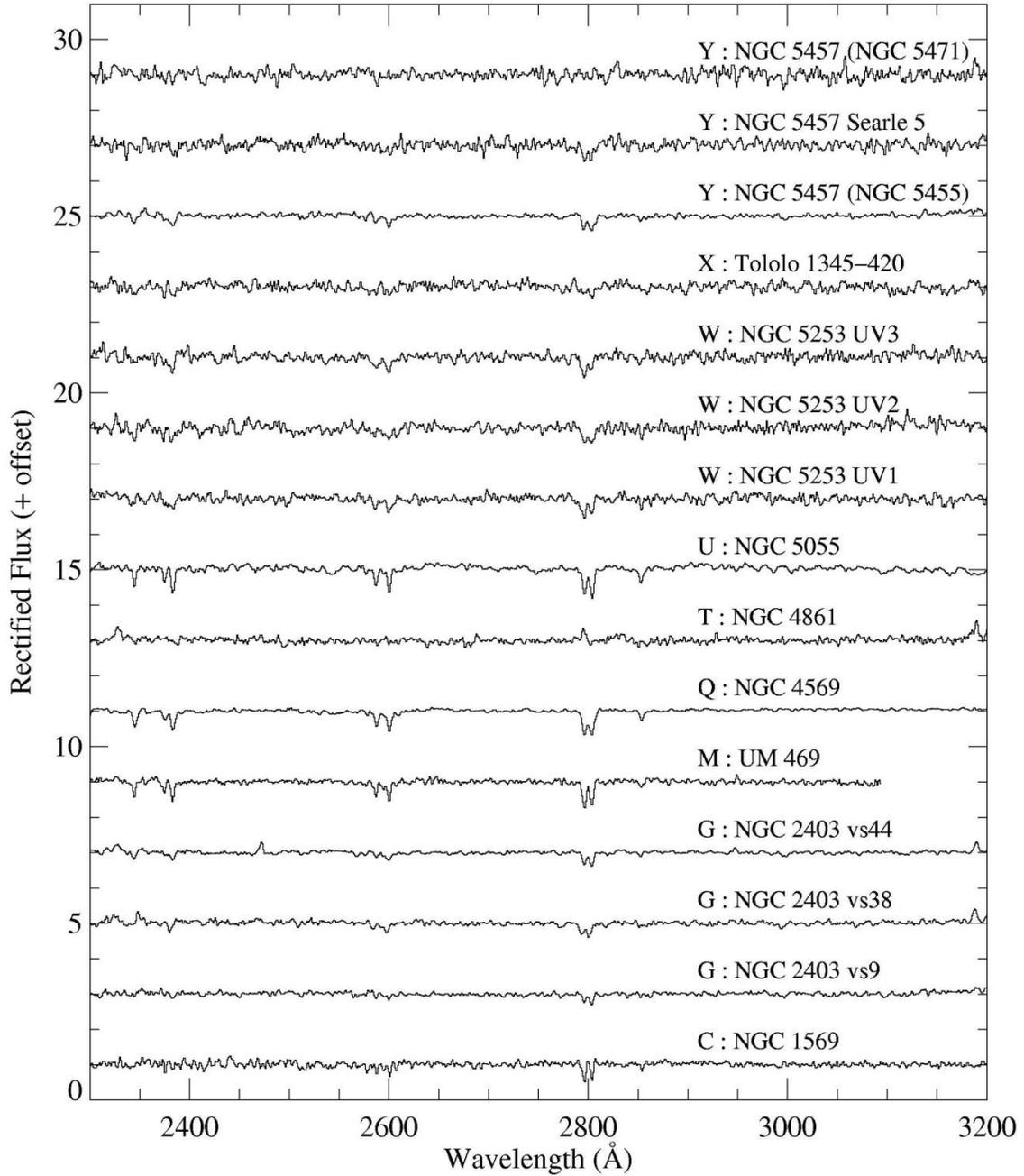

Figure 5. Galaxies with spectra in the 2300 − 3200 Å region. All of these spectra were obtained with the FOS and the G270H grating. They have been corrected for foreground Milky Way ISM line contamination as described in Section 3.5.

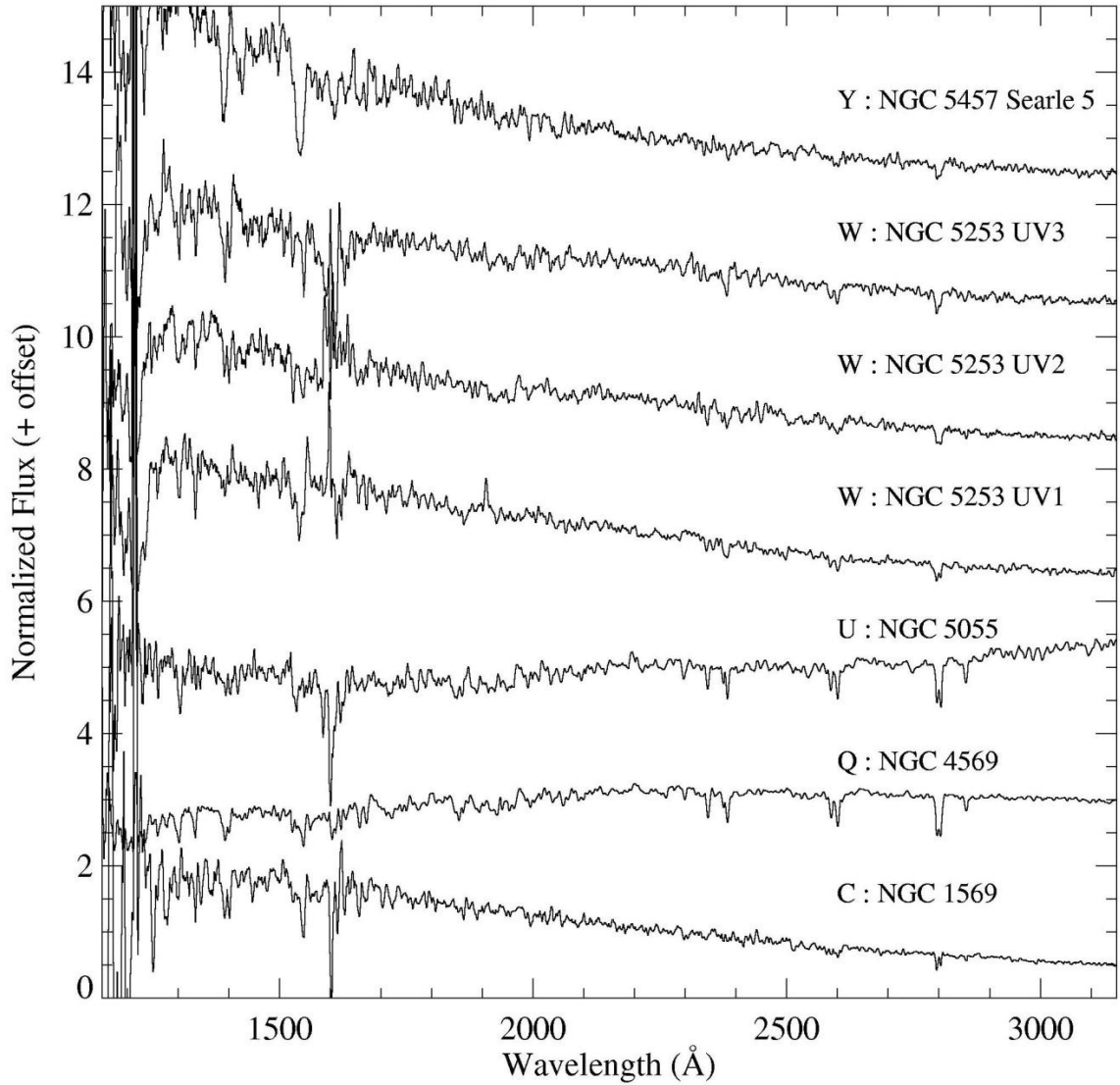

Figure 6. − Galaxies with spectra covering 1150 − 3200 Å. These spectra were obtained with the FOS G130H, G190H, and G270H gratings and subsequently combined. (Note the relatively noisy join region around 1600 Å.) The spectra have been corrected for foreground Milky Way ISM line contamination and reddening as described in Section 3.5. The spectra have been normalized by their median flux, offset from one another, and smoothed by five pixels for clarity.

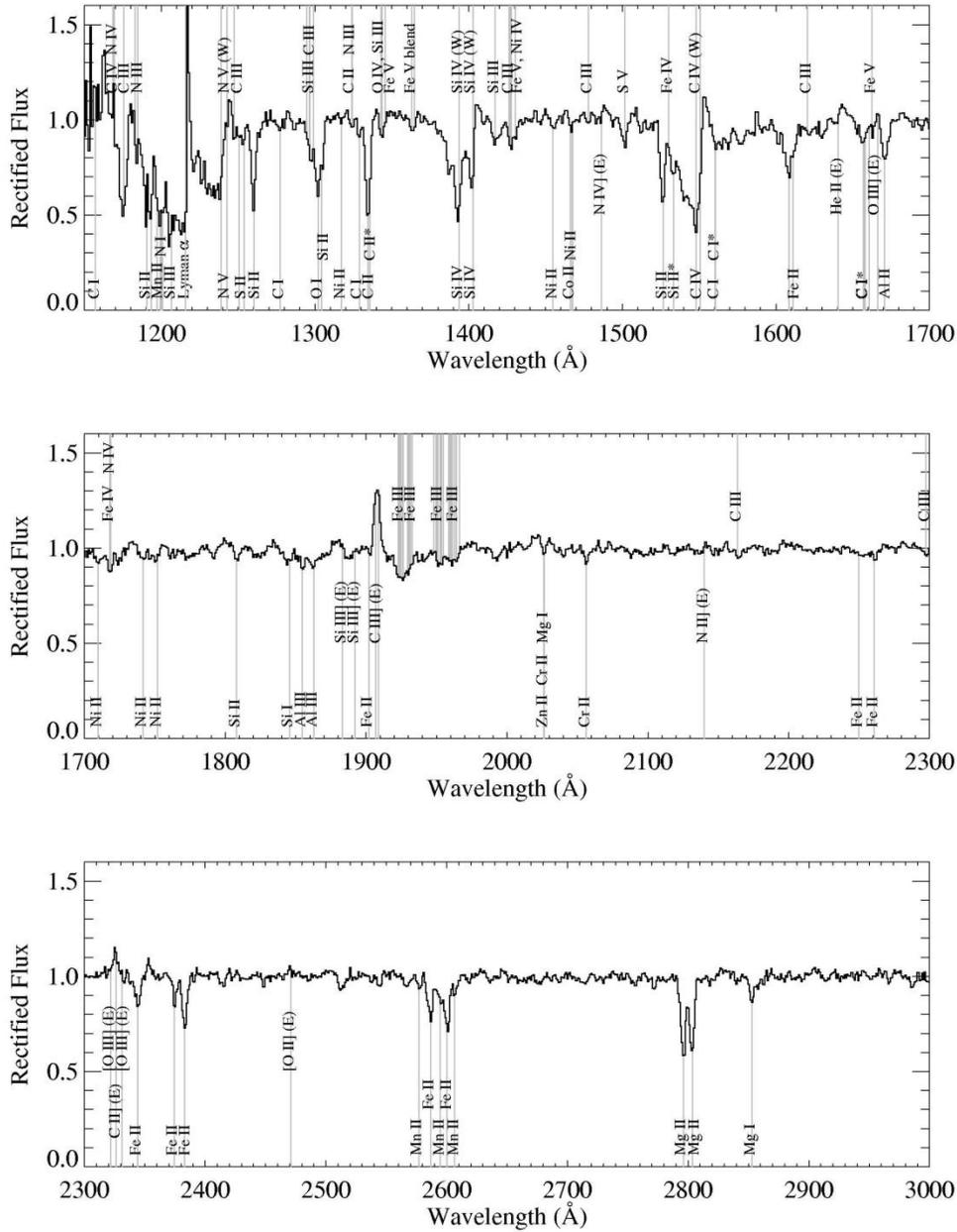

Figure 7. – Composite starburst spectrum constructed from our sample spectra by taking the median of the available spectra at each wavelength. Prominent interstellar lines are labeled below the spectrum; stellar-wind and photospheric features are labeled above. The stellar-wind lines are denoted with a "W". Nebular emission lines are signified with an "E". Laboratory wavelengths for the marked features can be found in Table 1.

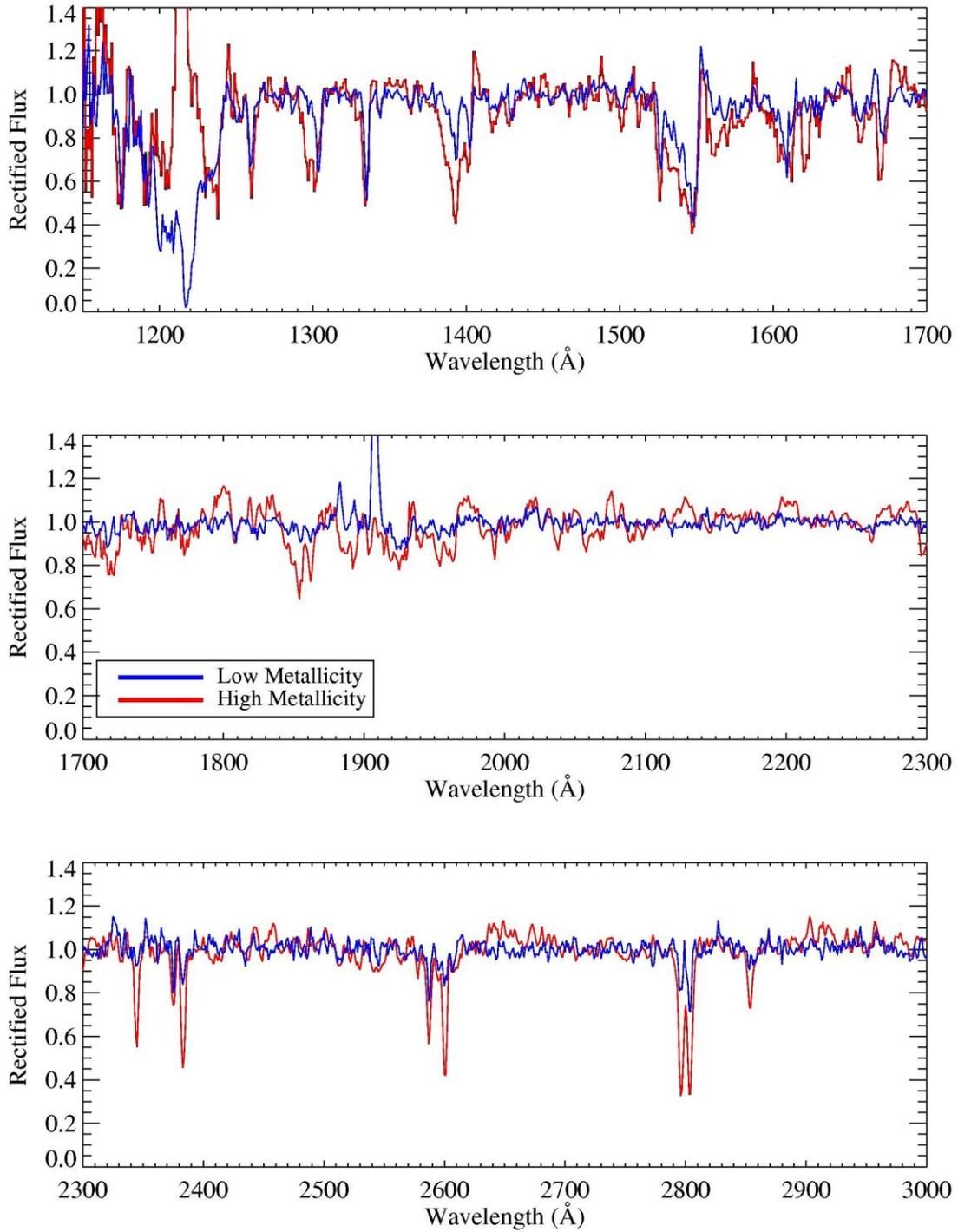

Figure 8. – Same as Figure 7, but for two subsets of galaxy spectra: a high-metallicity set with 12 + log (O/H) > 8.7 (red) and a low-metallicity set with 12 + log (O/H) < 8.2 (blue).

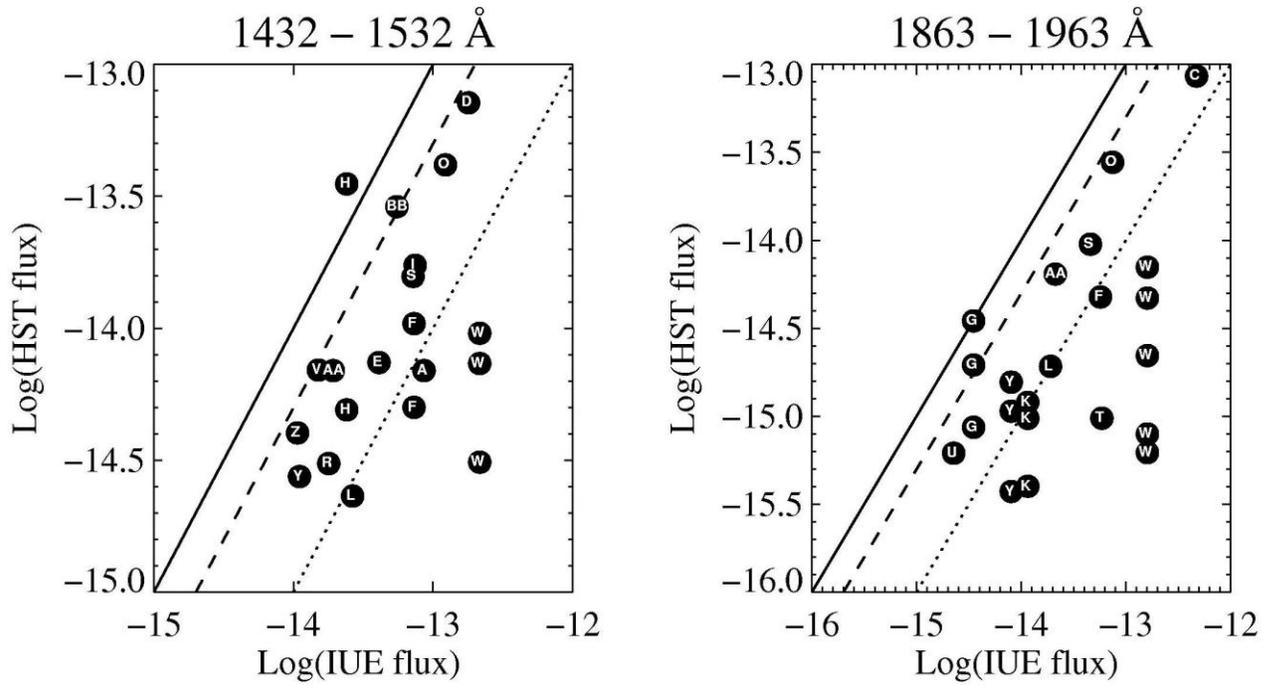

Figure 9. − Comparison of the 1500 (left) and 1900 Å (right) fluxes measured in the HST and IUE apertures. Fluxes are in units of erg s$^{-1}$ cm$^{-2}$ Å$^{-1}$. The solid line is a 1-to-1 correlation; the dashed line a 2-to-1 correlation; and the dotted line a 10-to-1 correlation. Individual galaxies are labeled by their identifiers defined in Table 2.

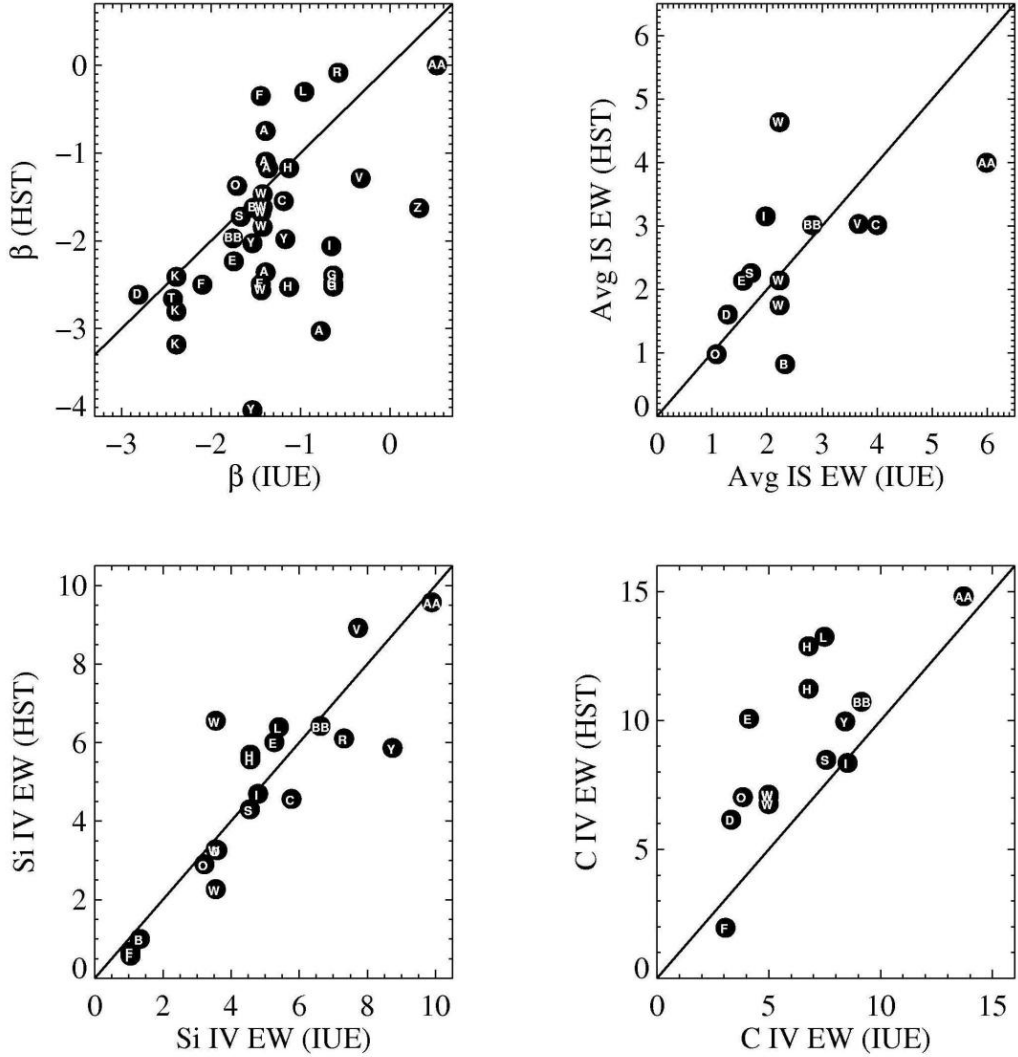

Figure 10. – Upper left: comparison of the UV spectral slopes $\beta_{HST}$ and $\beta_{IUE}$ measured in the HST and IUE data, respectively. The IUE spectra have a spectral slope whose median is redder by 0.6. Upper right: absorption line equivalent widths (Å) measured in the HST and IUE data. The label "Avg IS EW" denotes the average equivalent width of the strongest interstellar features Si II $\lambda1260$, O I $\lambda1303$ + S II $\lambda1304$, and C II $\lambda1335$. Lower: equivalent widths of the stellar-wind lines Si IV $\lambda1400$ (left) and CIV $\lambda1550$ (right). The solid line is a 1-to-1 correlation. Individual galaxies are labeled by their identifiers defined in Table 2.

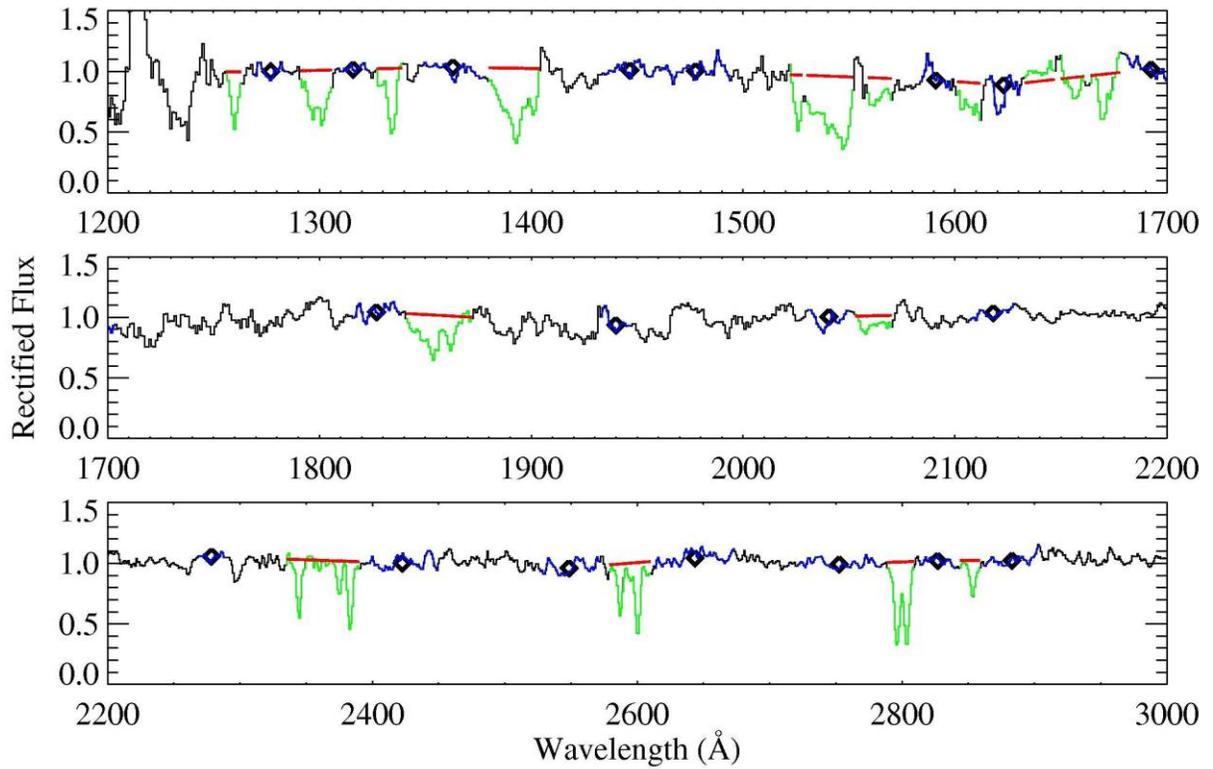

Figure 11. – Composite high-metallicity spectrum with the positions of the UV line indices indicated. The continuum regions are show in blue, with diamonds representing the average value of the bandpass. The central index bandpass is shown in green, with red lines illustrating the continuum.

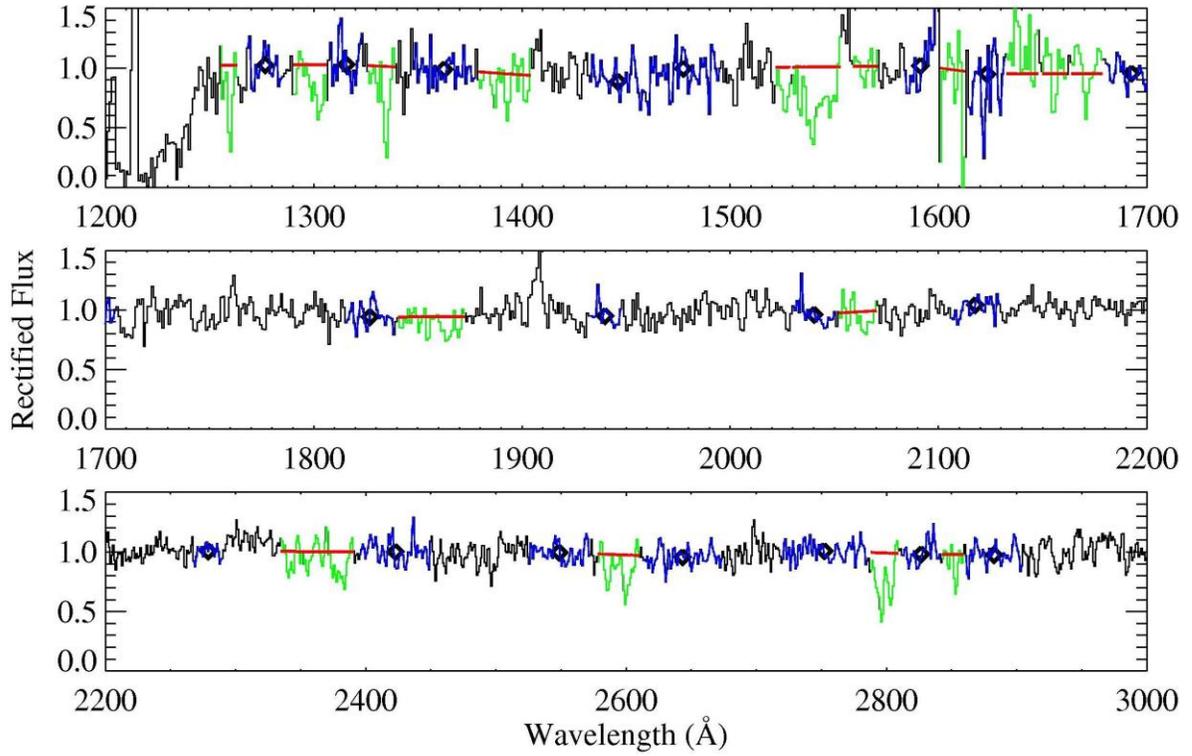

Figure 12. – Example of a low S/N spectrum (NGC 5253-UV1) with the positions of the UV line indices indicated. The color coding is identical to that in Figure 11.

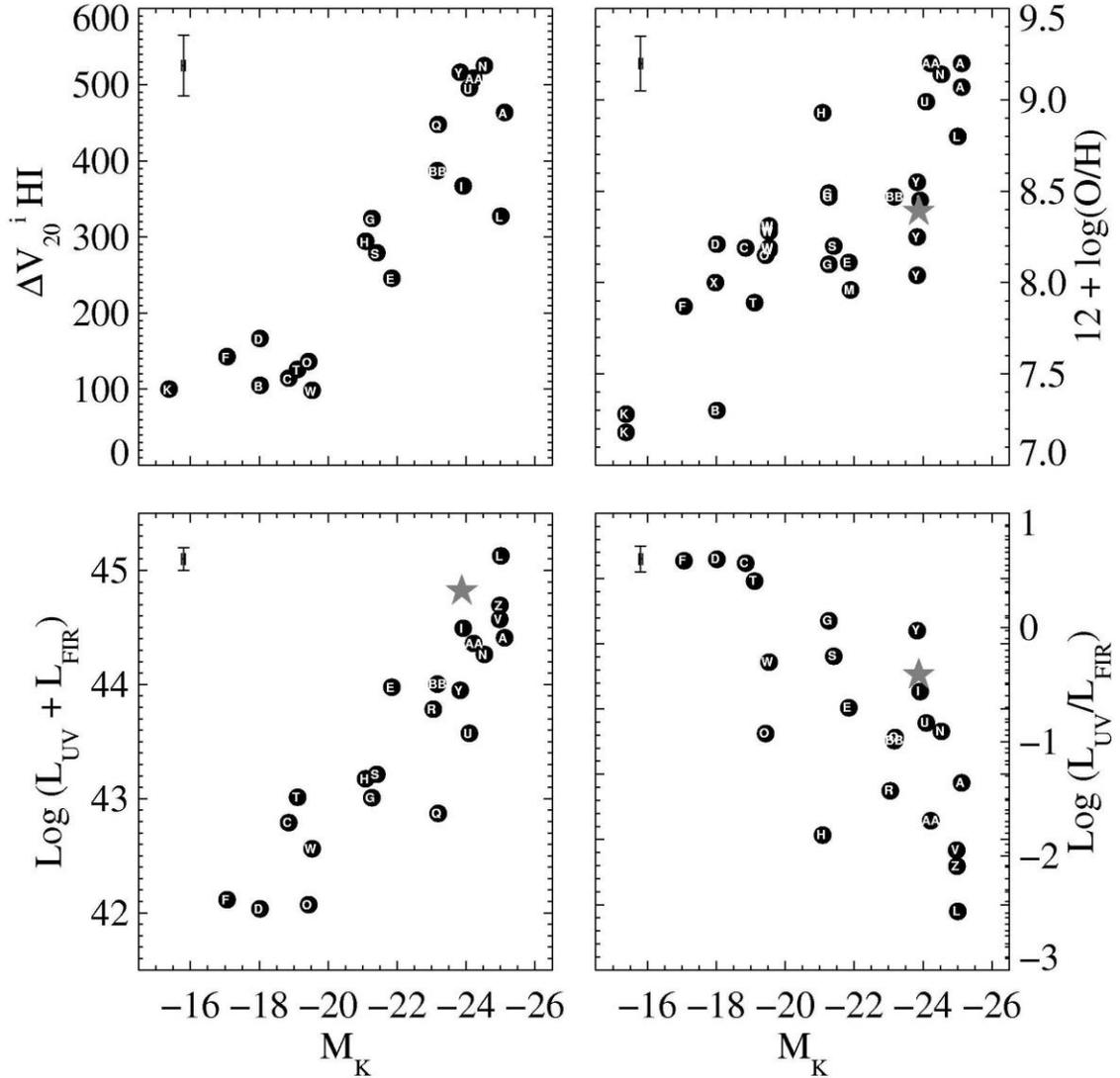

Figure 13. – Correlations between galaxy luminosity $M_K$ and other galaxy properties. Shown are clockwise starting from top left: mass, as traced by the width of the H I line, $\Delta v_{20,i}$; metallicity $12 + \log(O/H)$; starburst strength $L_{UV} + L_{IR}$; dust obscuration $L_{UV}/L_{IR}$. The sample galaxies are represented with solid dots. The asterisk denotes the location of the Lyman-break galaxy MS 1512-cB58. Individual galaxies are labeled by their identifiers defined in Table 2.

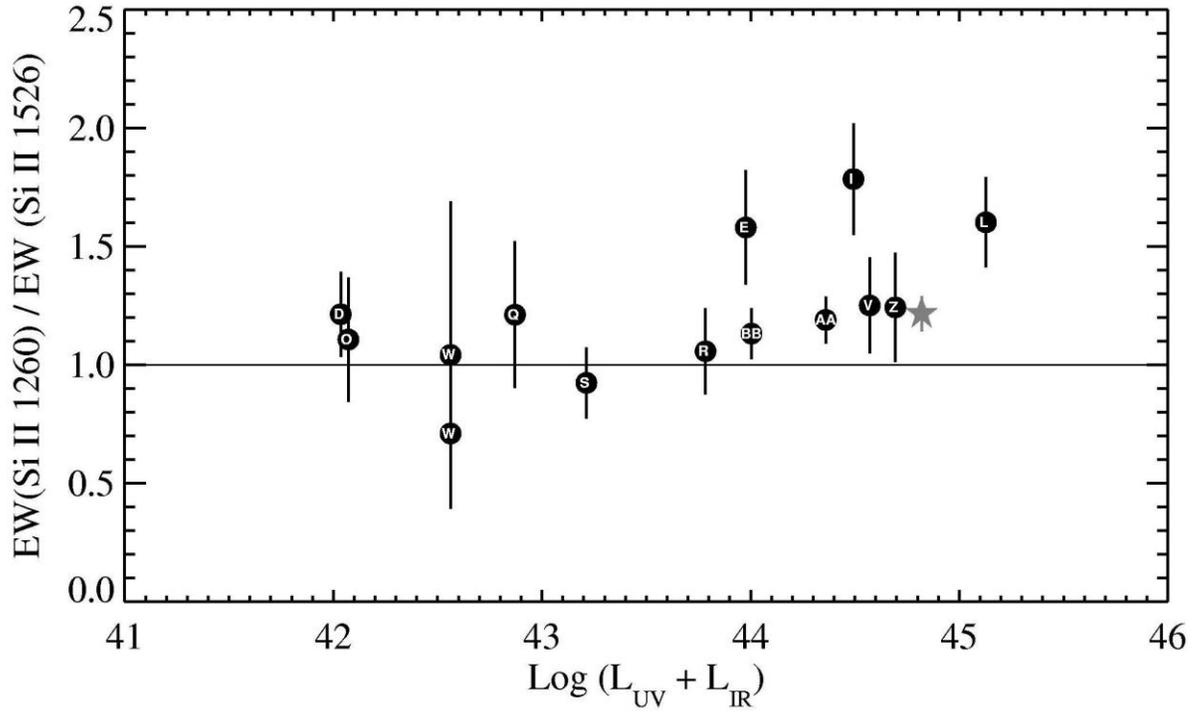

Figure 14. – Ratio of the equivalent widths of the Si II λ1260 and λ1526 lines versus the intrinsic UV luminosity. The line ratio is ~5.2 and ~1.0 in the optically thin and optically thick case, respectively. The asterisk denotes the location of the Lyman-break galaxy MS 1512-cB58. Individual galaxies are labeled by their identifiers defined in Table 2.

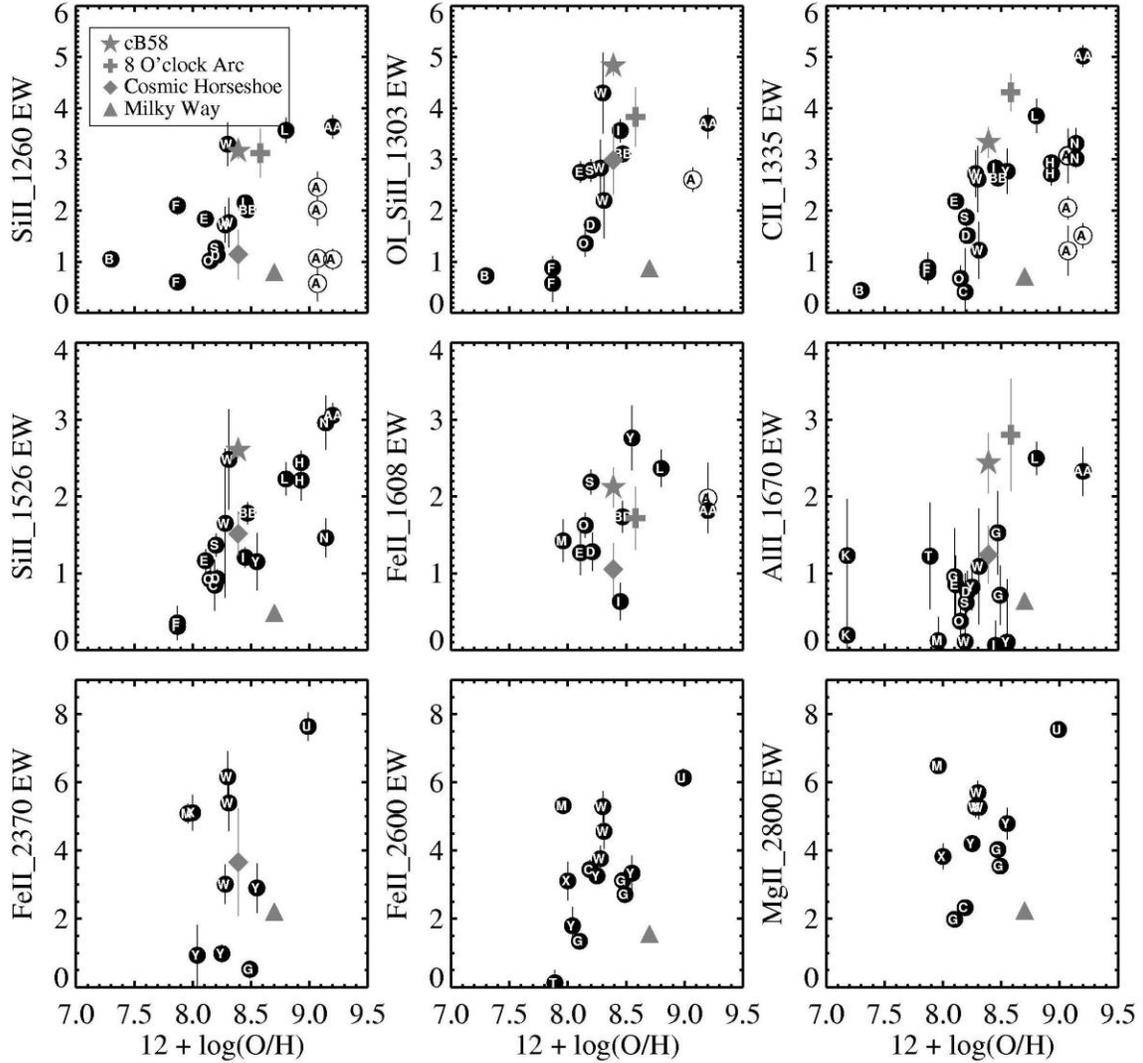

Figure 15. – Equivalent widths of the interstellar lines versus metallicity. The grey triangle indicates the median equivalent width of the Milky Way ISM absorption as reported by Savage et al. (2000). The asterisk denotes MS 1512-cB58. The cross and the square indicate the 8 O'clock Arc and the Cosmic Horseshoe, respectively. AGN are highlighted with open symbols. Individual galaxies are labeled by their identifiers defined in Table 2.

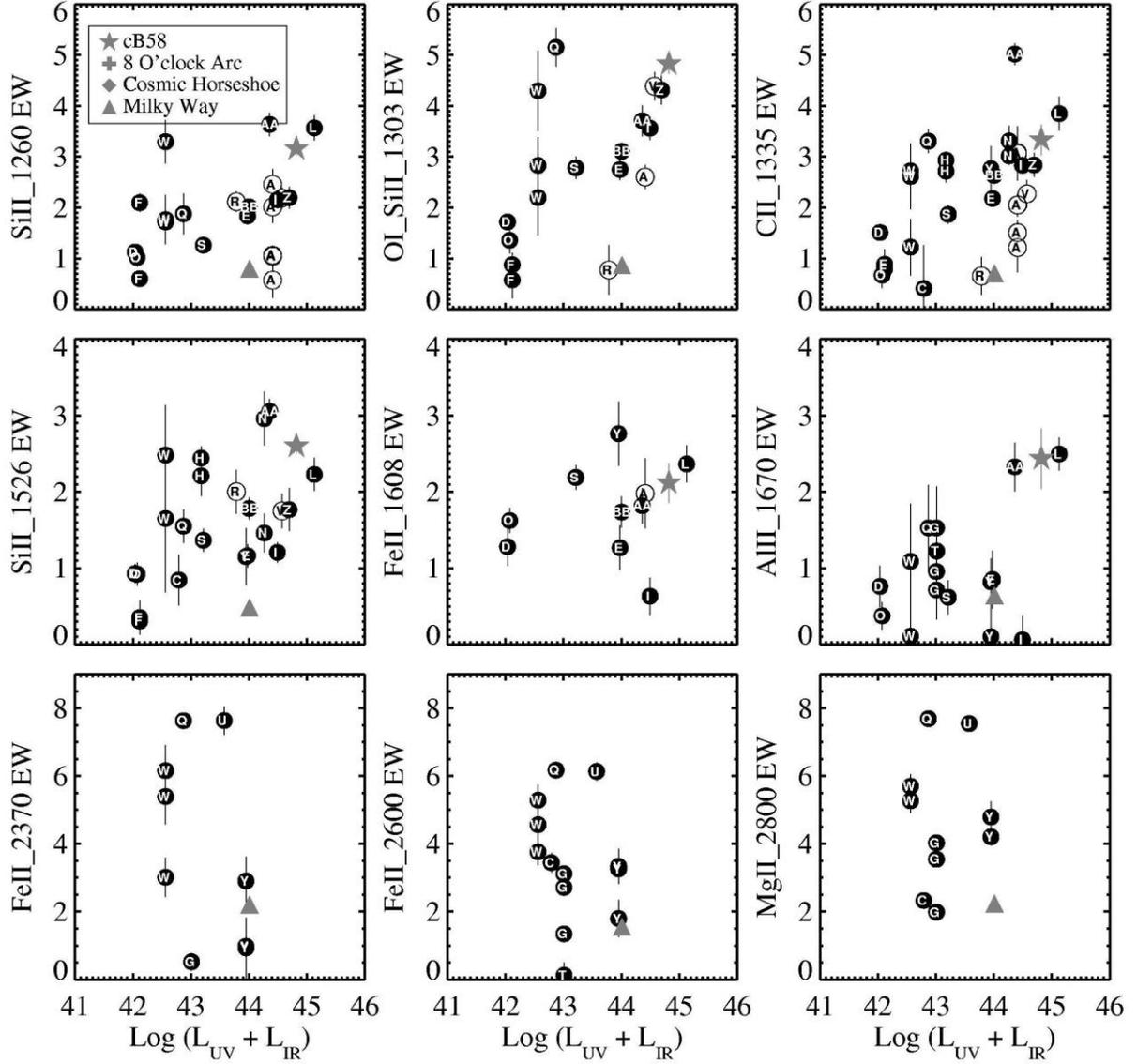

Figure 16. − Equivalent widths of the interstellar lines versus the intrinsic UV luminosity of the starburst, as parameterized by the sum of the far-UV and far-IR luminosities. Symbols and notations are as in Figure 15.

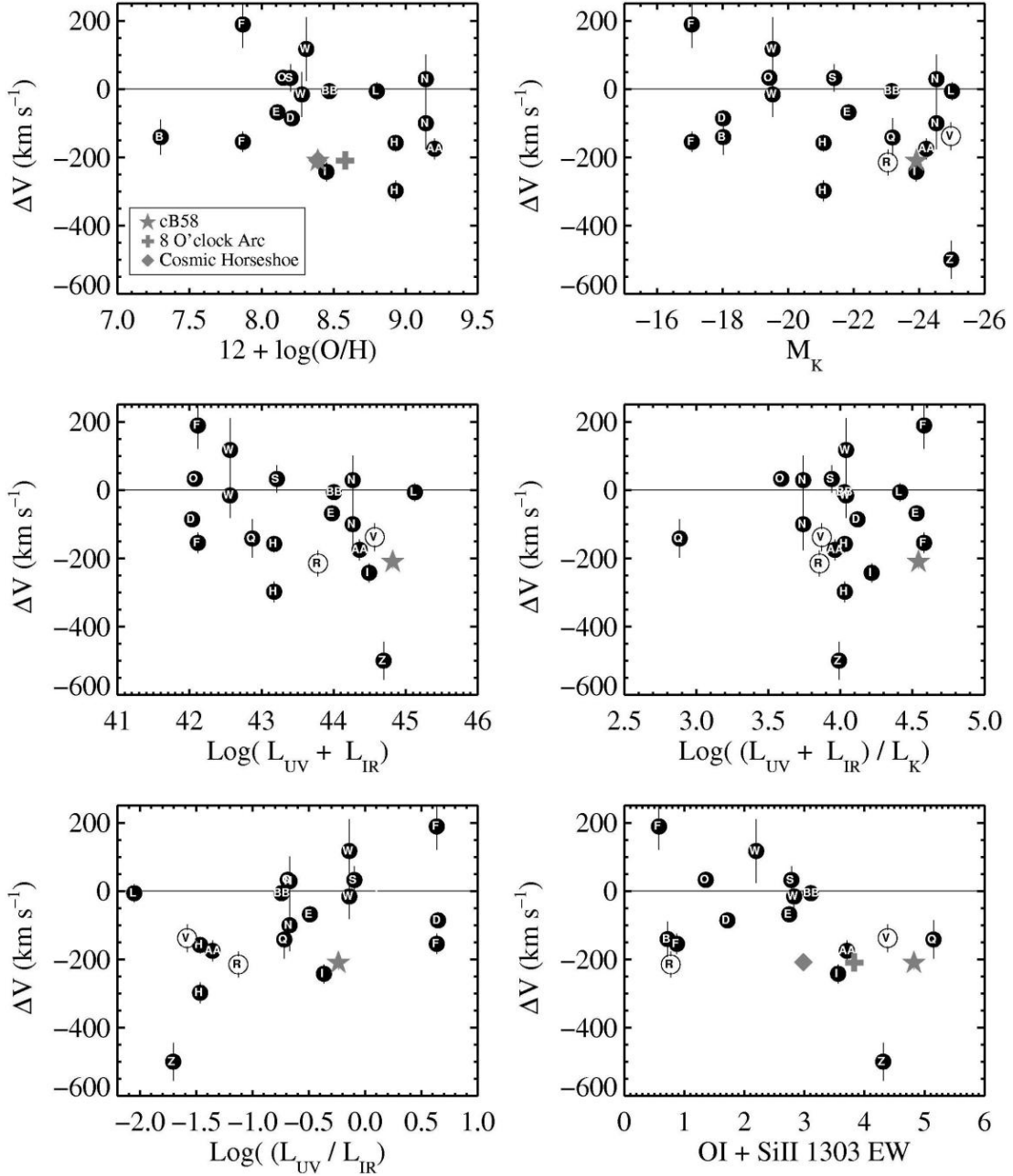

Figure 17. − Correlation of the velocity offset with various galaxy parameters. Top left: oxygen abundance; top right: $M_K$; middle left: intrinsic UV luminosity; middle right: burst strength as expressed by the ratio of young over old stars; bottom left: absorption of UV light by dust; bottom right: EW of the interstellar lines. Symbols and notations are as in Figure 15.

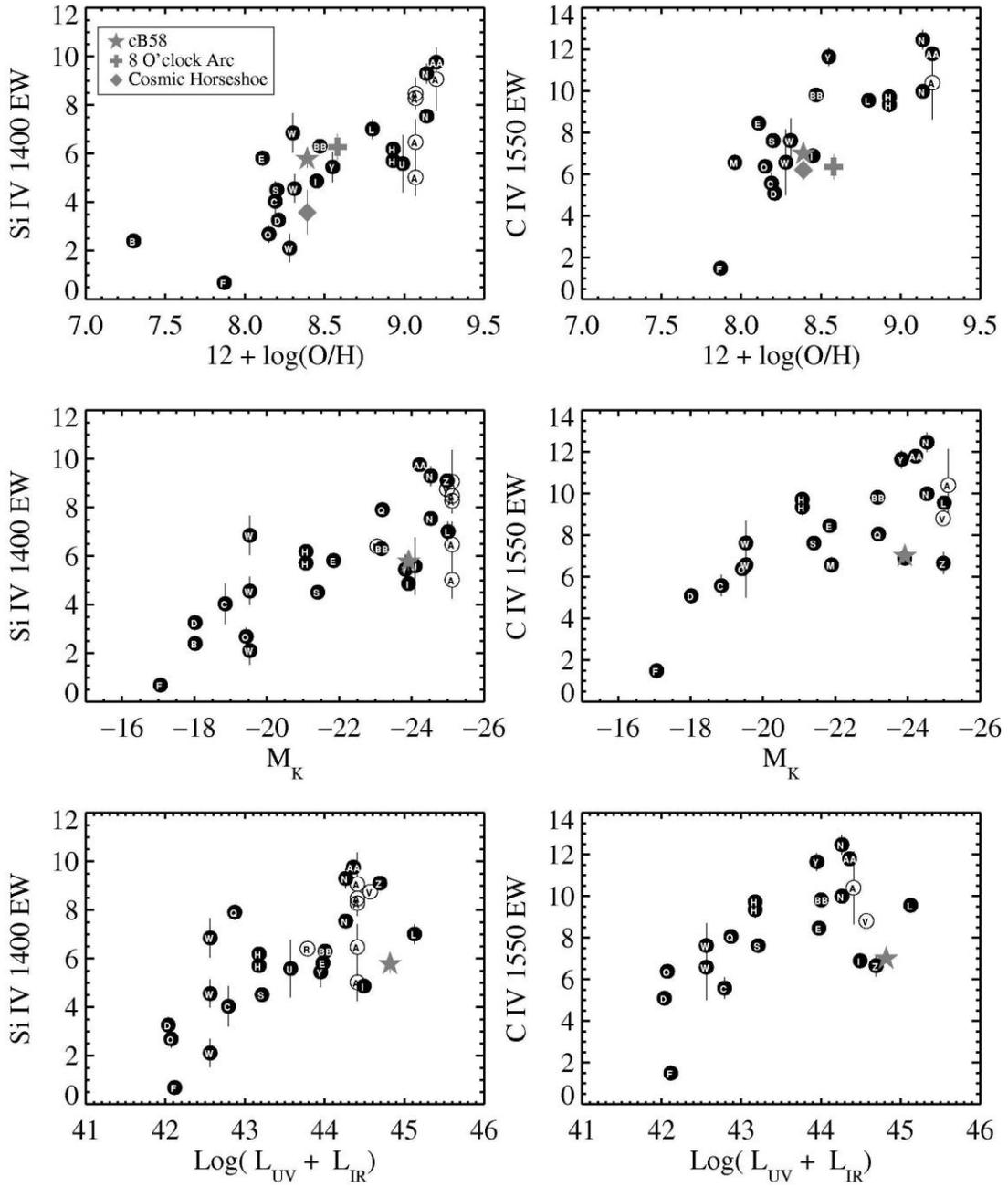

Figure 18. – Correlation of the stellar-wind lines Si IV (left) and C IV (right) with various galaxy parameters. Top: oxygen abundance; middle: $M_K$; bottom: intrinsic UV luminosity. Symbols and notations are as in Figure 15.

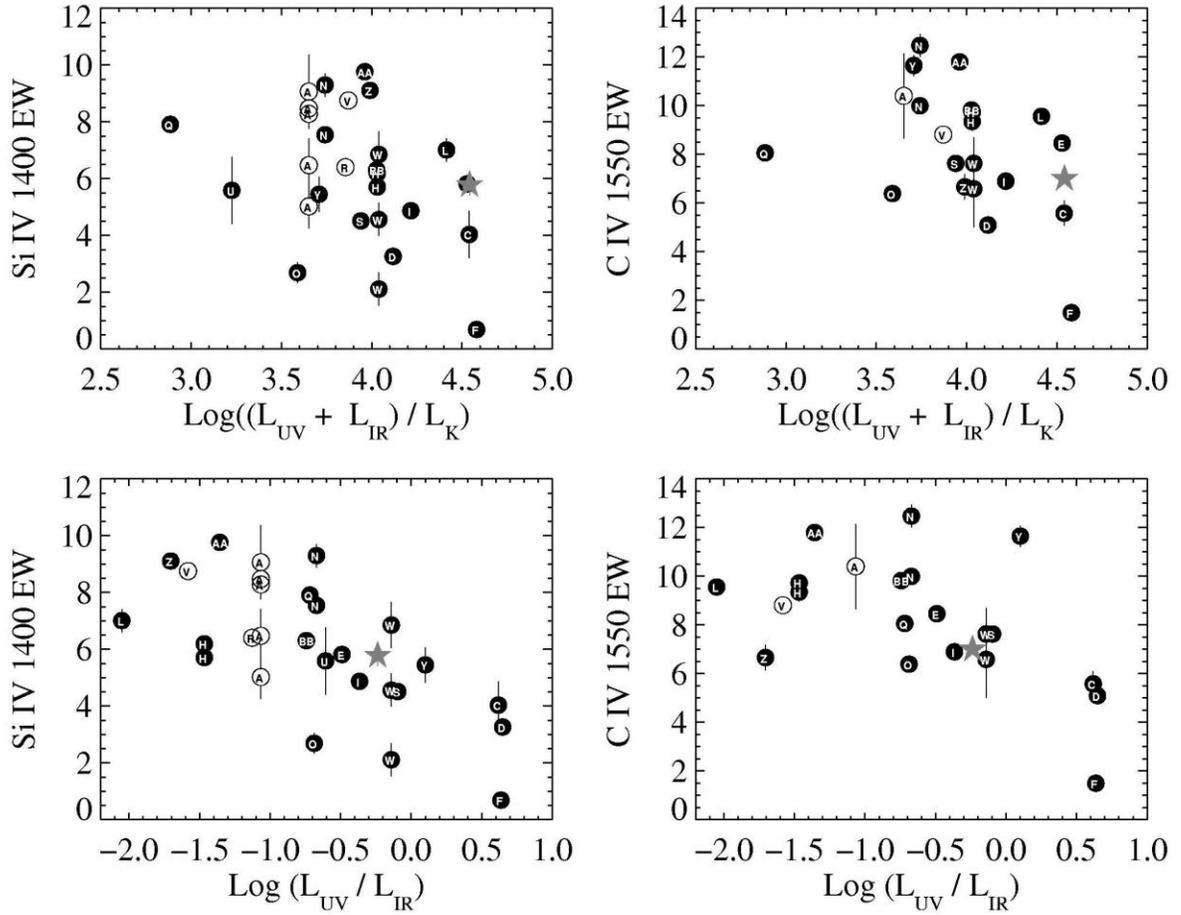

Figure 19. – Same as Figure 18, but for the correlation with burst strength (upper panels) and absorption of UV light by dust (lower panels.)

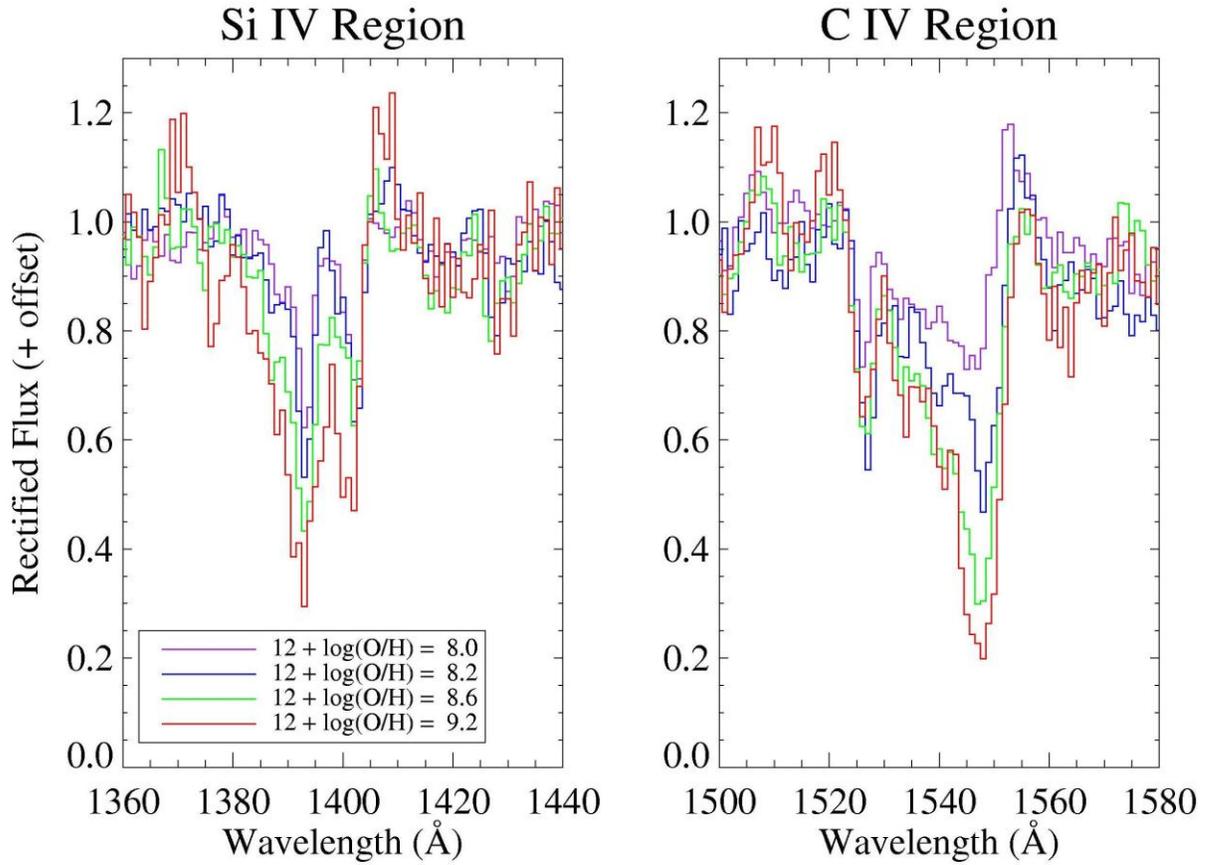

Figure 20. – Illustration of the effect of metallicity on the stellar-wind lines Si IV λ1400 (left) and C IV λ1550 (right). The four sets of spectra shown are averages of four to seven individual galaxies. The mean oxygen abundances of the four sets are listed in the left panel.

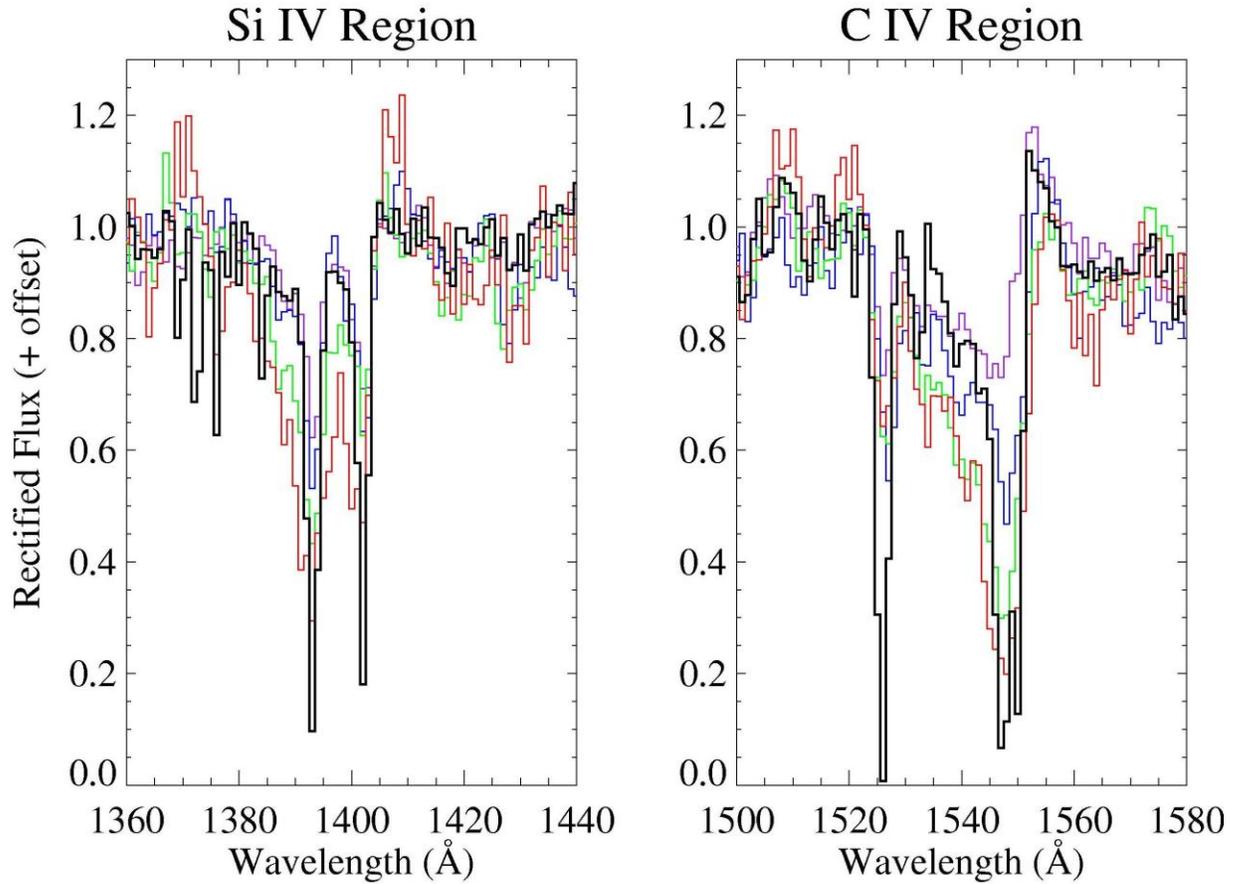

Figure 21. − Comparison of the Si IV (left) and C IV (right) line profiles of MS 1512-cB58 (black) with the template spectra shown in Figure 20. The color coding of the templates is the same as in Figure 20. The metallicity of MS 1512-cB58, as derived from the strong-line method, is $12 + \log(O/H) = 8.4$ (Teplitz et al. 2000).

.

# Tables

Table 1. UV spectral lines observed in star-forming galaxies

| Ion | $\lambda_{lab}$ (Å) | $E_{ion}$ (eV) | $E_{exc}$ (eV) | Wind | Photo | ISM | Nebular | Reference |
|---|---|---|---|---|---|---|---|---|
| C I    | 1157.19 | 0.00  | 0.00  |   |   | + |   | 1 |
| N IV   | 1168.60 | 47.45 | 52.08 |   | + |   |   | 2 |
| C IV   | 1168.87 | 47.89 | 40.28 |   | + |   |   | 2 |
| C IV   | 1168.99 | 47.89 | 40.28 |   | + |   |   | 2 |
| N IV   | 1169.06 | 47.45 | 52.08 |   | + |   |   | 2 |
| N IV   | 1169.48 | 47.45 | 52.08 |   | + |   |   | 2 |
| C III  | 1175.53 | 24.38 | 6.50  | + | + |   |   | 3 |
| N III  | 1183.03 | 29.60 | 18.09 |   | + |   |   | 4 |
| N III  | 1184.54 | 29.60 | 18.10 |   | + |   |   | 4 |
| Mn II  | 1199.39 | 7.44  | 0.00  |   |   | + |   | 1 |
| N I    | 1199.55 | 0.00  | 0.00  |   |   | + |   | 1 |
| N I    | 1200.22 | 0.00  | 0.00  |   |   | + |   | 1 |
| N I    | 1200.71 | 0.00  | 0.00  |   |   | + |   | 1 |
| Si III | 1206.50 | 16.35 | 0.00  |   |   | + |   | 1 |
| H I    | 1215.67 | 0.00  | 0.00  |   |   | + |   | 1 |
| N V    | 1238.82 | 77.47 | 0.00  | + |   | + |   | 1 |
| N V    | 1242.80 | 77.47 | 0.00  | + |   | + |   | 1 |
| C III  | 1247.38 | 24.38 | 12.69 |   | + |   |   | 5 |
| S II   | 1250.58 | 10.36 | 0.00  |   |   | + |   | 1 |
| S II   | 1253.81 | 10.36 | 0.00  |   |   | + |   | 1 |
| Si II  | 1260.42 | 8.15  | 0.00  |   |   | + |   | 1 |
| C I    | 1277.24 | 0.00  | 0.00  |   |   | + |   | 1 |
| Si III | 1294.54 | 16.35 | 6.55  |   | + |   |   | 5 |
| C III  | 1296.33 | 24.38 | 33.48 |   | + |   |   | 6 |
| Si III | 1296.74 | 16.35 | 6.54  |   | + |   |   | 5 |
| Si III | 1298.93 | 16.35 | 6.57  |   | + |   |   | 5 |
| O I    | 1302.17 | 0.00  | 0.00  |   |   | + |   | 1 |
| Si II  | 1304.37 | 8.15  | 0.00  |   |   | + |   | 1 |
| Ni II  | 1317.22 | 7.64  | 0.00  |   |   | + |   | 1 |
| C II   | 1323.93 | 11.26 | 9.29  |   | + |   |   | 6 |
| N III  | 1324.35 | 29.60 | 33.13 |   | + |   |   | 6 |
| C I    | 1328.83 | 0.00  | 0.00  |   |   | + |   | 1 |
| C II   | 1334.53 | 11.26 | 0.00  |   |   | + |   | 1 |
| C II*  | 1335.71 | 11.26 | 0.01  |   |   | + |   | 1 |
| O IV   | 1341.64 | 54.93 | 22.40 |   | + |   |   | 4 |
| Fe V   | 1345.61 | 54.80 | 28.99 |   | + |   |   | 4 |
| Fe V   | 1362.86 | 54.80 | 26.84 |   | + |   |   | 4 |
| Fe V   | 1363.08 | 54.80 | 23.20 |   | + |   |   | 4 |

| Ion | $\lambda_{\text{lab}}$ (Å) | $E_{\text{ion}}$ (eV) | $E_{\text{exc}}$ (eV) | Wind | Photo | ISM | Nebular | Reference |
|---|---|---|---|---|---|---|---|---|
| Fe V   | 1364.82 | 54.80 | 23.15 |   | + |   |   | 4 |
| O V    | 1371.30 | 77.41 | 19.69 |   | + |   |   | 3 |
| Si IV  | 1393.76 | 33.49 | 0.00  | + |   | + |   | 1 |
| Si IV  | 1402.77 | 33.49 | 0.00  | + |   | + |   | 1 |
| Si III | 1417.24 | 16.35 | 10.28 |   | + |   |   | 5 |
| C III  | 1426.45 | 24.38 | 29.53 |   | + |   |   | 5 |
| Fe V   | 1427.27 | 54.80 | 26.49 |   | + |   |   | 4 |
| C III  | 1427.85 | 24.38 | 29.53 |   | + |   |   | 5 |
| Fe V   | 1430.44 | 54.80 | 23.36 |   | + |   |   | 4 |
| Ni II  | 1454.84 | 7.64  | 0.00  |   |   | + |   | 1 |
| Co II  | 1466.21 | 7.86  | 0.00  |   |   | + |   | 1 |
| Ni II  | 1467.26 | 7.64  | 0.00  |   |   | + |   | 1 |
| Ni II  | 1467.76 | 7.64  | 0.00  |   |   | + |   | 1 |
| C III  | 1478.00 | 24.38 | 33.47 |   | + |   |   | 3 |
| N IV]  | 1486.50 | 47.45 | 0.00  | + |   |   | + | 3 |
| S V    | 1501.76 | 47.30 | 15.77 |   | + |   |   | 5 |
| Si II  | 1526.71 | 8.15  | 0.00  |   |   | + |   | 1 |
| Fe IV  | 1530.04 | 30.65 | 15.89 |   | + |   |   | 4 |
| Si II* | 1533.43 | 8.15  | 0.04  |   |   | + |   | 1 |
| C IV   | 1548.19 | 47.89 | 0.00  | + |   | + |   | 1 |
| C IV   | 1550.77 | 47.89 | 0.00  | + |   | + |   | 1 |
| C I    | 1560.31 | 0.00  | 0.00  |   |   | + |   | 1 |
| C I*   | 1560.68 | 0.00  | 0.00  |   |   | + |   | 1 |
| Fe II  | 1608.45 | 7.87  | 0.00  |   |   | + |   | 1 |
| Fe II  | 1611.20 | 7.97  | 0.00  |   |   | + |   | 1 |
| C III  | 1620.40 | 24.38 | 32.20 |   | + |   |   | 3 |
| He II  | 1640.42 | 24.59 | 40.81 | + |   |   | + | 3 |
| C I    | 1656.93 | 0.00  | 0.00  |   |   | + |   | 1 |
| C I*   | 1657.38 | 0.00  | 0.00  |   |   | + |   | 1 |
| C I*   | 1657.91 | 0.00  | 0.00  |   |   | + |   | 1 |
| O III] | 1660.81 | 35.12 | 0.01  |   |   |   | + | 3 |
| Fe V   | 1662.32 | 54.80 | 24.20 |   | + |   |   | 4 |
| O III] | 1666.15 | 35.12 | 0.04  |   |   |   | + | 3 |
| Al II  | 1670.79 | 5.99  | 0.00  |   |   | + |   | 1 |
| Ni II  | 1709.60 | 7.64  | 0.00  |   |   | + |   | 1 |
| Fe IV  | 1717.90 | 30.65 | 17.15 |   | + |   |   | 4 |
| N IV   | 1718.55 | 47.45 | 16.20 | + | + |   |   | 6 |
| Ni II  | 1741.55 | 7.64  | 0.00  |   |   | + |   | 1 |
| Ni II  | 1751.91 | 7.64  | 0.00  |   |   | + |   | 1 |
| Si II  | 1808.01 | 8.15  | 0.00  |   |   | + |   | 1 |
| Si I   | 1845.52 | 0.00  | 0.00  |   |   | + |   | 1 |
| Al III | 1854.72 | 18.83 | 0.00  | + |   | + |   | 1 |
| Al III | 1862.79 | 18.83 | 0.00  | + |   | + |   | 1 |
| Si III]| 1883.00 | 16.35 | 0.00  |   |   |   | + | 3 |

| Ion | $\lambda_{\text{lab}}$ (Å) | $E_{\text{ion}}$ (eV) | $E_{\text{exc}}$ (eV) | Wind | Photo | ISM | Nebular | Reference |
|---|---|---|---|---|---|---|---|---|
| Si III] | 1892.03 | 16.35 | 0.00 | | | | + | 3 |
| Fe II | 1901.77 | 7.87 | 0.00 | | | + | | 1 |
| C III] | 1906.68 | 24.38 | 0.00 | | | | + | 3 |
| C III] | 1908.73 | 24.38 | 0.00 | | | | + | 3 |
| Fe III | 1922.79 | 16.16 | 7.87 | | + | | | 3 |
| Fe III | 1923.01 | 16.16 | 10.28 | | + | | | 3 |
| Fe III | 1923.88 | 16.16 | 8.21 | | + | | | 3 |
| Fe III | 1924.53 | 16.16 | 9.50 | | + | | | 3 |
| Fe III | 1926.04 | 16.16 | 8.21 | | + | | | 3 |
| Fe III | 1926.30 | 16.16 | 3.71 | | + | | | 3 |
| Fe III | 1930.39 | 16.16 | 7.87 | | + | | | 3 |
| Fe III | 1931.51 | 16.16 | 8.64 | | + | | | 3 |
| Fe III | 1932.82 | 16.16 | 10.29 | | + | | | 3 |
| Fe III | 1949.67 | 16.16 | 10.33 | | + | | | 3 |
| Fe III | 1950.33 | 16.16 | 11.03 | | + | | | 3 |
| Fe III | 1951.01 | 16.16 | 8.76 | | + | | | 3 |
| Fe III | 1952.65 | 16.16 | 8.77 | | + | | | 3 |
| Fe III | 1953.33 | 16.16 | 8.77 | | + | | | 3 |
| Fe III | 1954.61 | 16.16 | 8.77 | | + | | | 3 |
| Fe III | 1958.59 | 16.16 | 8.21 | | + | | | 3 |
| Fe III | 1960.32 | 16.16 | 9.86 | | + | | | 3 |
| Fe III | 1962.72 | 16.16 | 8.65 | | + | | | 3 |
| Fe III | 1964.15 | 16.16 | 8.65 | | + | | | 3 |
| Fe III | 1966.41 | 16.16 | 8.66 | | + | | | 3 |
| Zn II | 2026.14 | 9.39 | 0.00 | | | + | | 1 |
| Cr II | 2026.27 | 6.77 | 0.00 | | | + | | 1 |
| Mg I | 2026.48 | 0.00 | 0.00 | | | + | | 1 |
| Cr II | 2056.25 | 6.77 | 0.00 | | | + | | 1 |
| N II] | 2140.00 | 14.53 | 0.02 | | | | + | 3 |
| C III | 2163.61 | 24.38 | 34.28 | | + | | | 3 |
| Fe II | 2249.88 | 7.87 | 0.00 | | | + | | 1 |
| Fe II | 2260.78 | 7.87 | 0.00 | | | + | | 1 |
| C III | 2297.58 | 24.38 | 12.69 | | + | | | 6 |
| [O III] | 2321.66 | 35.12 | 0.01 | | | | + | 3 |
| C II] | 2326.00 | 11.26 | 0.00 | | | | + | 3 |
| [O III] | 2331.00 | 35.12 | 0.04 | | | | + | 3 |
| Fe II | 2344.21 | 7.87 | 0.00 | | | + | | 1 |
| Fe II | 2374.46 | 7.87 | 0.00 | | | + | | 1 |
| Fe II | 2382.76 | 7.87 | 0.00 | | | + | | 1 |
| [O II] | 2470.97 | 13.62 | 0.00 | | | | + | 3 |
| [O II] | 2471.09 | 13.62 | 0.00 | | | | + | 3 |
| Mn II | 2576.88 | 7.44 | 0.00 | | | + | | 1 |
| Fe II | 2586.65 | 7.87 | 0.00 | | | + | | 1 |
| Mn II | 2594.50 | 7.44 | 0.00 | | | + | | 1 |

| Ion | $\lambda_{lab}$ (Å) | $E_{ion}$ (eV) | $E_{exc}$ (eV) | Wind | Photo | ISM | Nebular | Reference |
|---|---|---|---|---|---|---|---|---|
| Fe II | 2600.17 | 7.87 | 0.00 | | | + | | 1 |
| Mn II | 2606.46 | 7.44 | 0.00 | | | + | | 1 |
| Mg II | 2796.35 | 7.65 | 0.00 | | | + | | 1 |
| Mg II | 2803.53 | 7.65 | 0.00 | | | + | | 1 |
| Mg I | 2852.96 | 0.00 | 0.00 | | | + | | 1 |

References: 1 – Prochaska et al. (2001); 2 – Robert et al. (2003); 3 – this work; 4 – Brandt et al. (1998); 5 – de Mello, Leitherer, & Heckman (2000); 6 – Pettini et al. (2000).

Table 2. Summary of HST observations

| ID | Galaxy | Region | R.A. (h:m:s) | Dec (d:m:s) | Instrument | Grating | Aperture | Program | Wavelength (Å) | S/N |
|---|---|---|---|---|---|---|---|---|---|---|
| A | NGC 1068 | Knot2 | 02:42:40.3 | −00:00:38 | GHRS | G140L | LSA | 3024 | 1164 − 1451 | 1 |
| A | NGC 1068 | Knot2 | 02:42:40.3 | −00:00:38 | GHRS | G140L | LSA | 3024 | 1493 − 1780 | 3 |
| A | NGC 1068 | POS1 | 02:42:40.3 | −00:00:38 | GHRS | G140L | LSA | 5942 | 1165 − 1461 | 9 |
| A | NGC 1068 | POS3 | 02:42:39.9 | −00:00:45 | GHRS | G140L | LSA | 5942 | 1165 − 1461 | 2 |
| A | NGC 1068 | POS4 | 02:42:41.4 | −00:00:37 | GHRS | G140L | LSA | 5942 | 1166 − 1462 | 2 |
| A | NGC 1068 | POS8A | 02:42:39.9 | −00:00:58 | GHRS | G140L | LSA | 5942 | 1165 − 1461 | 2 |
| B | SBS 0335−052 | | 03:37:44.0 | −05:02:39 | GHRS | G140L | LSA | 6678 | 1162 − 1448 | 9 |
| C | NGC 1569 | A | 04:30:48.2 | +64:50:59 | FOS | G130H | B-2 | 6408 | 1087 − 1606 | 2 |
| C | NGC 1569 | A | 04:30:48.2 | +64:50:59 | FOS | G190H | B-2 | 6408 | 1572 − 2312 | 5 |
| C | NGC 1569 | A | 04:30:48.2 | +64:50:59 | FOS | G270H | B-2 | 6408 | 2222 − 3277 | 9 |
| D | NGC 1705 | | 04:54:13.4 | −53:21:39 | GHRS | G140L | SSA | 5691 | 1175 − 1462 | 13 |
| D | NGC 1705 | | 04:54:13.4 | −53:21:39 | GHRS | G140L | SSA | 5691 | 1453 − 1740 | 9 |
| E | NGC 1741 | Offset | 05:01:37.7 | −04:15:29 | GHRS | G140L | LSA | 5738 | 1175 − 1461 | 10 |
| E | NGC 1741 | Offset | 05:01:37.7 | −04:15:29 | GHRS | G140L | LSA | 5738 | 1453 −1739 | 6 |
| F | NGC 2366 | NGC 2363 | 07:28:42.6 | +69:11:22 | FOS | G190H | C-1 | 5246 | 1571 − 2311 | 7 |
| F | NGC 2366 | NGC 2363 CP1-A | 07:28:42.6 | +69:11:22 | FOS | G130H | B-3 | 6096 | 1087 − 1606 | 5 |
| F | NGC 2366 | NGC 2363 CP1-B | 07:28:43.5 | +69:11:23 | FOS | G130H | B-3 | 6096 | 1087 − 1606 | 11 |
| G | NGC 2403 | VS9 | 07:36:28.6 | +65:33:49 | FOS | G190H | B-3 | 6044 | 1571 − 2311 | 10 |
| G | NGC 2403 | VS9 | 07:36:28.6 | +65:33:49 | FOS | G270H | B-3 | 6044 | 2221 − 3277 | 14 |
| G | NGC 2403 | VS38 | 07:36:52.1 | +65:36:49 | FOS | G190H | B-3 | 6044 | 1571 − 2311 | 9 |
| G | NGC 2403 | VS38 | 07:36:52.1 | +65:36:49 | FOS | G270H | B-3 | 6044 | 2221 − 3277 | 11 |
| G | NGC 2403 | VS44 | 07:37:06.8 | +65:36:39 | FOS | G190H | B-3 | 6044 | 1571 − 2311 | 13 |
| G | NGC 2403 | VS44 | 07:37:06.8 | +65:36:39 | FOS | G270H | B-3 | 6044 | 2221 −3277 | 19 |
| H | He 2-10 | A | 08:36:15.1 | −26:24:34 | GHRS | G140L | LSA | 6580 | 1314 − 1600 | 14 |
| H | He 2-10 | B | 08:36:15.8 | −26:24:35 | GHRS | G140L | LSA | 6580 | 1314 − 1601 | 6 |
| I | IRAS 08339+6517 | | 08:38:23.0 | +65:07:14 | GHRS | G140L | LSA | 5954 | 1213 − 1500 | 11 |
| I | IRAS 08339+6517 | | 08:38:23.0 | +65:07:14 | GHRS | G140L | LSA | 5954 | 1466 − 1753 | 9 |
| J | SBS 1415+437 | | 14:17:01.4 | +43:30:05 | FOS | G190H | B-3 | 5408 | 1572 − 2312 | 5 |
| K | IZw18 | | 09:34:02.1 | +55:14:27 | FOS | G190H | B-3 | 3840 | 1572 − 2312 | 5 |

| ID | Galaxy | Region | R.A. (h:m:s) | Dec (d:m:s) | Instrument | Grating | Aperture | Program | Wavelength (Å) | S/N |
|---|---|---|---|---|---|---|---|---|---|---|
| K | IZw18 | NW HIIR | 09:34:02.0 | +55:14:28 | FOS | G190H | C-1 | 5434 | 1571 − 2311 | 7 |
| K | IZw18 | SE HIIR | 09:34:02.3 | +55:14:23 | FOS | G190H | B-3 | 6536 | 1572 − 2312 | 5 |
| L | NGC 3690 | | 11:28:30.8 | +58:33:43 | FOS | G130H | B-3 | 4250 | 1087 − 1606 | 4 |
| L | NGC 3690 | | 11:28:30.8 | +58:33:43 | FOS | G190H | B-3 | 4250 | 1573 − 2330 | 9 |
| M | UM 469 | POS1 | 11:57:12.5 | +02:28:29 | FOS | G190H | B-3 | 6801 | 1572 − 2312 | 9 |
| M | UM 469 | POS1 | 11:57:12.5 | +02:28:29 | FOS | G270H | B-3 | 6801 | 2222 − 3277 | 12 |
| N | NGC 4038 | 405 updated | 12:01:50.4 | −18:52:12 | GHRS | G140L | LSA | 5962 | 1314 − 1601 | 7 |
| N | NGC 4038 | 442 updated | 12:01:53.0 | −18:52:08 | GHRS | G140L | LSA | 5962 | 1314 − 1601 | 3 |
| O | NGC 4214 | | 12:15:39.3 | +36:19:36 | FOS | G130H | B-3 | 4122 | 1087 − 1606 | 8 |
| O | NGC 4214 | | 12:15:39.3 | +36:19:36 | FOS | G130H | B-3 | 4122 | 1573 − 2330 | 13 |
| P | Tololo 1214−277 | | 12:17:17.1 | −28:02:32 | GHRS | G140L | LSA | 6678 | 1175 − 1461 | 3 |
| P | Tololo 1214−277 | | 12:17:17.1 | −28:02:32 | FOS | G190H | B-3 | 3840 | 1572 − 2312 | 1 |
| Q | NGC 4569 | | 12:36:49.8 | +13:09:46 | FOS | G130H | B-3 | 6112 | 1087 − 1606 | 6 |
| Q | NGC 4569 | | 12:36:49.8 | +13:09:46 | FOS | G190H | B-3 | 6112 | 1572 − 2312 | 16 |
| Q | NGC 4569 | | 12:36:49.8 | +13:09:46 | FOS | G270H | B-3 | 6112 | 2222 − 3277 | 29 |
| R | IC 3639 | | 12:40:52.9 | −36:45:22 | GHRS | G140L | LSA | 6539 | 1162 − 1449 | 5 |
| R | IC 3639 | | 12:40:52.9 | −36:45:22 | GHRS | G140L | LSA | 6539 | 1301 − 1588 | 6 |
| S | NGC 4670 | | 12:45:17.2 | +27:07:32 | FOS | G130H | B-3 | 4250 | 1087 − 1606 | 9 |
| S | NGC 4670 | | 12:45:17.2 | +27:07:32 | FOS | G190H | B-3 | 4250 | 1573 − 2330 | 12 |
| T | NGC 4861 | POS1 | 12:59:00.3 | +34:50:44 | FOS | G190H | B-3 | 6801 | 1572 − 2312 | 6 |
| T | NGC 4861 | POS1 | 12:59:00.3 | +34:50:44 | FOS | G270H | B-3 | 6801 | 2222 − 3277 | 7 |
| U | NGC 5055 | | 13:15:49.3 | +42:01:46 | FOS | G130H | B-2 | 6112 | 1087 − 1606 | 1 |
| U | NGC 5055 | | 13:15:49.3 | +42:01:46 | FOS | G190H | B-2 | 6112 | 1572 − 2312 | 5 |
| U | NGC 5055 | | 13:15:49.3 | +42:01:46 | FOS | G270H | B-2 | 6112 | 2222 − 3277 | 14 |
| V | NGC 5135 | | 13:25:44.0 | −29:50:01 | GHRS | G140L | LSA | 6539 | 1162 − 1449 | 6 |
| V | NGC 5135 | | 13:25:44.0 | −29:50:01 | GHRS | G140L | LSA | 6539 | 1313 − 1600 | 6 |
| W | NGC 5253 | HII-1 | 13:39:56.0 | −31:38:26 | FOS | G190H | B-3 | 6021 | 1572 − 2312 | 5 |
| W | NGC 5253 | HII-2 | 13:39:56.0 | −31:38:24 | FOS | G190H | B-3 | 6021 | 1572 − 2312 | 4 |
| W | NGC 5253 | UV1 | 13:39:55.8 | −31:38:27 | FOS | G130H | B-3 | 6001 | 1087 − 1606 | 3 |
| W | NGC 5253 | UV1 | 13:39:55.8 | −31:38:27 | FOS | G190H | B-3 | 6001 | 1572 − 2312 | 5 |
| W | NGC 5253 | UV1 | 13:39:55.8 | −31:38:27 | FOS | G270H | B-3 | 6001 | 2222 − 3277 | 5 |
| W | NGC 5253 | UV2 | 13:39:55.9 | −31:38:32 | FOS | G130H | B-3 | 6001 | 1087 − 1606 | 3 |
| W | NGC 5253 | UV2 | 13:39:55.9 | −31:38:32 | FOS | G190H | B-3 | 6001 | 1572 − 2312 | 5 |
| W | NGC 5253 | UV2 | 13:39:55.9 | −31:38:32 | FOS | G270H | B-3 | 6001 | 2222 − 3277 | 5 |
| W | NGC 5253 | UV3 | 13:39:55.4 | −31:38:30 | FOS | G130H | B-3 | 6001 | 1087 − 1606 | 2 |
| W | NGC 5253 | UV3 | 13:39:55.4 | −31:38:30 | FOS | G190H | B-3 | 6001 | 1572 − 2312 | 4 |
| W | NGC 5253 | UV3 | 13:39:55.4 | −31:38:30 | FOS | G270H | B-3 | 6001 | 2222 − 3277 | 5 |
| X | Tololo 1345−420 | POS1 | 13:48:22.3 | −42:21:15 | FOS | G190H | B-3 | 6801 | 1572 − 2312 | 3 |
| X | Tololo 1345−420 | POS1 | 13:48:22.3 | −42:21:15 | FOS | G270H | B-3 | 6801 | 2222 − 3277 | 5 |
| Y | NGC 5457 | NGC 5455 | 14:03:01.2 | +54:24:27 | FOS | G190H | C-1 | 6044 | 1571 − 2311 | 9 |
| Y | NGC 5457 | NGC 5455 | 14:03:01.2 | +54:24:27 | FOS | G270H | C-1 | 6044 | 2221 − 3277 | 13 |
| Y | NGC 5457 | Searle5 | 14:02:55.0 | +54:22:26 | FOS | G130H | B-3 | 3813 | 1087 − 1606 | 3 |
| Y | NGC 5457 | Searle5 | 14:02:55.0 | +54:22:26 | FOS | G190H | B-3 | 3813 | 1573 − 2330 | 3 |
| Y | NGC 5457 | Searle5 | 14:02:55.0 | +54:22:26 | FOS | G270H | B-3 | 3813 | 2221− 3301 | 4 |
| Y | NGC 5457 | NGC 5471 | 14:04:29.0 | +54:23:49 | FOS | G190H | C-1 | 6044 | 1573 − 2330 | 3 |
| Y | NGC 5457 | NGC 5471 | 14:04:29.0 | +54:23:49 | FOS | G270H | C-1 | 6044 | 2221 − 3301 | 4 |
| Z | NGC 7130 | | 21:48:19.5 | −34:57:05 | GHRS | G140L | LSA | 6539 | 1175 − 1462 | 6 |
| Z | NGC 7130 | | 21:48:19.5 | −34:57:05 | GHRS | G140L | LSA | 6539 | 1314 − 1601 | 6 |
| AA | NGC 7552 | | 23:16:10.8 | −42:35:03 | FOS | G130H | B-3 | 4250 | 1087 − 1606 | 7 |
| AA | NGC 7552 | | 23:16:10.8 | −42:35:03 | FOS | G190H | B-3 | 4250 | 1573 − 2330 | 12 |
| BB | NGC 7714 | | 23:36:14.1 | +02:09:18 | GHRS | G140L | LSA | 6672 | 1175 − 1461 | 10 |
| BB | NGC 7714 | | 23:36:14.1 | +02:09:18 | GHRS | G140L | LSA | 6672 | 1402 − 1689 | 10 |

Table 3. Milky Way lines

| Line | Wavelength (Å) | $W_v$ (km s$^{-1}$) | $\sigma$ (km s$^{-1}$) |
|---|---|---|---|
| Si III 1206 | 1206.50 | 180.0 | 25.0 |
| S II 1253 | 1253.81 | 70.0 | 90.0 |
| Si II 1260 | 1260.42 | 190.0 | 65.0 |
| O I 1302 | 1302.17 | 105.0 | 55.0 |
| Si II 1304 | 1304.37 | 95.0 | 35.0 |
| C II 1334 | 1334.53 | 160.0 | 45.0 |
| Si IV 1393 | 1393.76 | 60.0 | 40.0 |
| Si IV 1402 | 1402.77 | 40.0 | 10.0 |
| Si II 1526 | 1526.71 | 95.0 | 25.0 |
| C IV 1548 | 1548.20 | 80.0 | 25.0 |
| C IV 1550 | 1550.77 | 45.0 | 20.0 |
| Al II 1670 | 1670.79 | 115.0 | 30.0 |
| Si II 1808 | 1808.01 | 50.0 | 30.0 |
| Al III 1854 | 1854.72 | 55.0 | 20.0 |
| Zn II 2026 | 2026.14 | 55.0 | 30.0 |
| Zn II 2062 | 2062.66 | 35.0 | 20.0 |
| Fe II 2344 | 2344.21 | 95.0 | 25.0 |
| Fe II 2374 | 2374.46 | 75.0 | 20.0 |
| Fe II 2382 | 2382.77 | 110.0 | 30.0 |
| Fe II 2586 | 2586.65 | 80.0 | 25.0 |
| Fe II 2600 | 2600.17 | 100.0 | 30.0 |
| Mg II 2796 | 2796.35 | 125.0 | 35.0 |
| Mg II 2803 | 2803.53 | 115.0 | 25.0 |
| Mg II 2852 | 2852.96 | 50.0 | 15.0 |

Table 4. Optical properties of the galaxies

| Galaxy | Morphology | $v_{helio}$ (km s$^{-1}$) | $m_B$ | $D_{optical}$ (armin) | $i$ (°) | $E(B-V)$ | Other names |
|---|---|---|---|---|---|---|---|
| NGC1068 | (R)SA(rs)b; Sey2 | 1137 | 9.6 | 7.1 × 6.0 | 40 | 0.03 | M77, UGC 02188, IRAS 02401-0013, Arp 037 |
| SBS0335-052 | BCG | 4043 | 16.6 | 0.23 × 0.20 | ….. | 0.05 | SBS 0335-052E |
| NGC1569 | IBm | −104 | 11.9 | 3.6 × 1.8 | 63 | 0.55 | UGC 03056, ARP 210, VII Zw 016, IRAS 04260+6444 |
| NGC1705 | SAO- pec, HII | 628 | 12.8 | 1.9 × 1.4 | 78 | 0.01 | ESO158- G 013, IRAS 04531-5326 |
| NGC1741 | SB(s)m pec, HII | 4107 | 13.3 | 0.61 × 0.23 | 62 | 0.05 | MRK 1089, ARP 259, IRAS 04591-0419 |
| NGC2366 | IB(s)m | 100 | 11.4 | 8.1 × 3.3 | 62 | 0.04 | UGC 03851, IRAS 07233+6917 |
| NGC2403 | SAB(s)cd, HII | 131 | 8.9 | 21.9 × 12.3 | 57 | 0.04 | UGC 03918, IRAS 07321+6543 |
| He2-10 | I0 pec | 873 | 12.4 | 1.93 × 1.35 | 32 | 0.11 | ESO 495- G 021, IRAS 08341-2614 |
| IRAS08339+6517 | Pec HII | 5730 | 14.2 | 0.3 × 0.3 | ….. | 0.09 | PGC 024283 |
| SBS1415+437 | BCG | 609 | 15.6 | 0.75 × 0.15 | ….. | 0.01 | CGCG 219-066, CG 0389 |
| I Zw 18 | Compact | 751 | 16.1 | 0.3 × 0.1 | 57 | 0.04 | MKN 116, UGCA 116, SBS 0930+554 |
| NGC3690 | Ibm pec+Sbm Pair, HII | 3121 | 12.0 | 2.0 × 1.5 | 42 | 0.02 | MRK 0171, IRAS 11257+5850, ARP 299 |
| UM469 | HII | 17388 | 18.0 | 0.29 × 0.15 | ….. | 0.03 | IRAS F11546+0245, Tololo 1154+027 |
| NGC4038 | SB(s)m pec | 1642 | 10.9 | 5.2 × 3.1 | 51 | 0.05 | UGCA 264, ESO 572-IG 047, ARP 244, Antennae |
| NGC4214 | IAB(s)m, HII | 291 | 10.2 | 8.5 × 6.6 | 37 | 0.02 | NGC 4228, UGC 07278, IRAS 12131+3636 |
| Tololo1214–277 | HII | 7795 | 17.5 | 0.08 × 0.08 | ….. | 0.06 | Tololo 21, SCHG 1214-277 |
| NGC4569 | SAB(rs) ab, LINER | −235 | 10.3 | 9.5 × 4.4 | 63 | 0.05 | M90, UGC 07786, ARP 076, IRAS 12343+1326 |
| IC3639 | Sb(rs)bc, Sey2 | 3275 | 13.0 | 1.2 × 1.2 | ….. | 0.07 | ESO 381-G08, Fairall 0312, IRAS 12381-3638, Tololo 74 |
| NGC4670 | SB(s)0/a pec, BCDG | 1069 | 13.1 | 1.4 × 1.1 | 37 | 0.01 | UGC 07930, ARP 163, HARO9, IRAS 12428+2724 |
| NGC4861 | SB(s)m | 847 | 12.9 | 4.0 × 1.5 | 67 | 0.01 | IC 3961, UGC 08098, ARP 266, I Zw 049, MRK 59 |
| NGC5055 | SA(rs)bc, HII, LINER | 504 | 9.3 | 12.6 × 7.2 | 55 | 0.02 | M63, UGC 08334, IRAS 13135+4217, Sunflower |
| NGC5135 | SB(l)ab, Sey2 | 4112 | 12.9 | 2.6 × 1.8 | 67 | 0.06 | ESO 444- G 032, Tololo 34, IRAS 13229-2934 |
| NGC5253 | Im pec, HII | 404 | 11.2 | 5.0 × 1.9 | 77 | 0.06 | UGC 369, ESO 445- G 004, Haro 10, IRAS 13370-3123 |
| Tololo1345–420 | HII | 2398 | 16.3 | 0.22 × 0.18 | ….. | 0.10 | Tololo 111, SCHG 1345-421 |
| NGC5457 | SAB(rs)cd | 241 | 8.3 | 28.8 × 26.9 | 24 | 0.01 | M101, UGC 08981, ARP 026, IRAS 14013+5435 |
| NGC7130 | Sa pec, LINER | 4842 | 12.9 | 1.5 × 1.4 | 25 | 0.03 | IC 5135, ESO 403- G 032, IRAS 21453-3511 |
| NGC7552 | SA(s)c pec, HII, LINER | 1585 | 11.2 | 3.4 ×2.7 | 46 | 0.01 | IC 5294, ESO 291- G 012, IRAS 23134-4251 |
| NGC7714 | SB(s)b:pec, HII, LINER | 2798 | 13.0 | 1.9 × 1.4 | 43 | 0.05 | UGC 12699, ARP 284, MRK 538, IRAS 23336+0152 |

Table 5. Multi-wavelength properties of the galaxies

| Galaxy | IRAS 60 μm | IRAS 100 μm | IUE 1500 Å | IUE 1900 Å | $\beta_{IUE}$ | H I $\Delta v_{20}$ (km s$^{-1}$) | Ref | $m_K$ (mag) | Ref | $m_{FUV}$ (mag) | $\beta_{GALEX}$ |
|---|---|---|---|---|---|---|---|---|---|---|---|
| NGC1068 | 176.2 | 224.0 | 8.61 | 9.68 | −0.57 | 298 | 1 | 5.79 | 5 | 12.52 | −0.58 |
| SBS0335−052 | ..... | ..... | 1.13 | 0.64 | −2.19 | 105 | 2 | 15.63 | 6 | ….. | ….. |
| NGC1569 | 45.4 | 47.3 | 72.69 | 47.29 | −1.19 | 102 | 3 | 7.86 | 5 | 9.85 | −0.44 |
| NGC1705 | 1.0 | 2.6 | 17.95 | 8.98 | −2.44 | 163 | 1 | 10.52 | 5 | ….. | ….. |
| NGC1741 | 3.9 | 5.8 | 4.10 | 2.76 | −1.55 | 217 | 1 | 11.84 | 7 | ….. | ….. |
| NGC2366 | 3.5 | 4.7 | 7.30 | 5.71 | −1.80 | 126 | 4 | 10.62 | 5 | 12.50 | −1.69 |
| NGC2403 | 51.5 | 148.5 | 0.59 | 0.35 | −0.66 | 272 | 4 | 6.19 | 5 | 10.37 | −1.44 |
| He2-10 | 24.1 | 26.4 | 2.41 | 2.02 | −0.68 | 156 | 1 | 9.00 | 7 | ….. | ….. |
| IRAS08339+6517 | 5.9 | 6.5 | 7.43 | 4.76 | −1.48 | 367 | 4 | 10.64 | 7 | ….. | ….. |
| SBS1415+437 | ..... | ..... | ..... | ..... | ..... | 74 | 2 | ….. | ….. | ….. | ….. |
| I Zw 18 | ..... | ..... | 2.10 | 1.15 | −2.38 | 84 | 2 | 15.92 | 8 | 15.82 | −1.98 |
| NGC3690 | 103.7 | 107.4 | 2.65 | 1.90 | −1.21 | 219 | 4 | 8.42 | 5 | ….. | ….. |
| UM469 | 0.3 | 0.6 | ..... | ..... | ..... | ..... | ..... | 15.06 | 6 | ….. | ….. |
| NGC4038 | 48.7 | 82.0 | ..... | ..... | ..... | 408 | 4 | 7.18 | 5 | 12.82 | −1.07 |
| NGC4214 | 17.9 | 29.0 | 12.26 | 7.46 | −1.74 | 82 | 1 | 7.91 | 5 | ….. | ….. |
| Tololo1214−277 | ..... | ..... | ..... | ..... | ..... | ..... | ..... | ….. | ….. | ….. | ….. |
| NGC4569 | 9.2 | 27.3 | ..... | ..... | 1.50 | 399 | 4 | 6.58 | 5 | 14.46 | 1.08 |
| IC3639 | 7.5 | 10.7 | 1.79 | 1.27 | −0.91 | ..... | ..... | 9.69 | 7 | ….. | ….. |
| NGC4670 | 2.6 | 4.5 | 7.18 | 4.58 | −1.68 | 168 | 4 | 10.41 | 7 | ….. | ….. |
| NGC4861 | 2.0 | 2.5 | 10.74 | 5.89 | −2.48 | 116 | 4 | 11.77 | 5 | 13.54 | −1.92 |
| NGC5055 | 40.0 | 157.7 | ..... | 0.23 | ..... | 406 | 4 | 5.61 | 5 | 12.41 | −0.68 |
| NGC5135 | 16.9 | 28.6 | 1.51 | 1.85 | −0.19 | ..... | ..... | 8.83 | 5 | ….. | ….. |
| NGC5253 | 30.0 | 30.9 | 21.56 | 16.03 | −1.42 | 96 | 1 | 8.25 | 5 | 12.20 | −0.89 |
| Tololo1345-420 | ..... | ..... | ..... | ..... | ..... | ..... | ..... | 14.11 | 7 | ….. | ….. |
| NGC5457 | 88.0 | 252.8 | 1.10 | 0.80 | −1.17 | 210 | 4 | 5.51 | 5 | 9.98 | −1.49 |
| NGC7130 | 16.5 | 25.6 | 1.07 | 0.84 | −0.20 | ..... | ..... | 9.20 | 7 | ….. | ….. |
| NGC7552 | 72.0 | 101.5 | 1.92 | 2.12 | 0.40 | 366 | 4 | 7.54 | 5 | 14.18 | −0.20 |
| NGC7714 | 10.4 | 11.5 | 5.49 | 3.75 | −1.24 | 264 | 4 | 9.76 | 5 | ….. | ….. |

References: 1 – Kobulnicky & Gebhardt (2000); 2 – Thuan et al. (1999); 3 – Meurer, Staveley-Smith, & Killeen (1998); 4 – Huchtmeier & Richter (1989); 5 – Jarrett et al. (2003); 6 – 2MASS Point Source Catalog; 7 – 2MASS Extended Source Catalog; 8 – Thuan (1983)

Note: IRAS fluxes are in Jansky and IUE fluxes are in $10^{-14}$ erg s$^{-1}$ cm$^{-2}$ Å$^{-1}$.

Table 6. Physical properties of the galaxies

| Galaxy | $d$ (Mpc) | Ref | $M_B$ (mag) | $M_K$ (mag) | log $L_{UV}$ | log $L_{IR}$ | H I $\Delta v_{20,i}$ (km s$^{-1}$) | 12+log(O/H) | Ref | $A_{IUE}$ (pc) | $A_{HST}$ (pc) |
|---|---|---|---|---|---|---|---|---|---|---|---|
| NGC 1068 knot2 | 15.2 | 1 | −21.4 | −25.1 | 43.3 | 44.4 | 463 | 9.2 | 12 | 1105 | 147 |
| NGC 1068 pos1 | | | | | | | | 9.1 | 12 | 1105 | 128 |
| NGC 1068 pos3 | | | | | | | | 9.1 | 12 | 1105 | 128 |
| NGC 1068 pos4 | | | | | | | | 9.1 | 12 | 1105 | 128 |
| NGC 1068 pos8a | | | | | | | | 9.1 | 12 | 1105 | 128 |
| SBS 0335−052 | 53.7 | 1 | −17.1 | −18.0 | 42.8 | ..... | 105 | 7.3 | 13 | 3905 | 452 |
| NGC 1569 | 2.2 | 2 | −16.6 | −18.9 | 42.7 | 42.1 | 114 | 8.2 | 14 | 159 | 2 |
| NGC 1705 | 5.1 | 3 | −15.8 | −18.0 | 41.9 | 41.3 | 166 | 8.2 | 15 | 370 | 5 |
| NGC 1741 | 54.6 | 1 | −20.5 | −21.8 | 43.4 | 43.9 | 245 | 8.1 | 16 | 3970 | 460 |
| NGC 2366 (NGC 2363) | 3.4 | 4 | −16.4 | −17.1 | 42.0 | 41.4 | 142 | 7.9 | 17 | 250 | 16 |
| NGC 2366 (NGC 2363-a) | | | | | | | | 7.9 | 17 | 250 | 14 |
| NGC 2366 (NGC 2363-b) | | | | | | | | 7.9 | 17 | 250 | 14 |
| NGC 2403 vs9 | 3.1 | 5 | −18.7 | −21.3 | 42.8 | 42.6 | 324 | 8.1 | 18 | 225 | 12 |
| NGC 2403 vs38 | | | | | | | | 8.5 | 18 | 225 | 12 |
| NGC 2403 vs44 | | | | | | | | 8.5 | 18 | 225 | 12 |
| He 2-10 A | 10.4 | 1 | −18.0 | −21.1 | 41.7 | 43.2 | 294 | 8.9 | 19 | 756 | 87 |
| He 2-10 B | | | | | | | | 8.9 | 19 | 756 | 87 |
| IRAS 08339+6517 | 81.5 | 1 | −20.7 | −23.9 | 44.0 | 44.3 | 367 | 8.4 | 20 | 5926 | 687 |
| SBS 1415+437 | 13.6 | 6 | −15.1 | ..... | ..... | ..... | 74 | 7.4 | 13 | ..... | 56 |
| I Zw 18 | 18.2 | 7 | −15.3 | −15.4 | 42.1 | ..... | 100 | 7.2 | 13 | 1323 | 88 |
| I Zw 18 NW HIIR | | | | | | | | 7.2 | 13 | 1323 | 75 |
| I Zw 18 SE HIIR | | | | | | | | 7.3 | 13 | 1323 | 75 |
| NGC 3690 | 48.5 | 1 | −21.5 | −25.0 | 43.1 | 45.1 | 327 | 8.8 | 21 | 3527 | 235 |
| UM 469 | 246.1 | 1 | −19.1 | −21.9 | ..... | 44.1 | ..... | 8.0 | 13 | ..... | 1026 |
| NGC 4038 405 | 22.0 | 8 | −21.0 | −24.5 | 43.5 | 44.2 | 524 | 9.1 | 22 | ..... | 185 |
| NGC 4038 442 | | | | | | | | 9.1 | 22 | ..... | 185 |
| NGC 4214 | 2.9 | 9 | −17.2 | −19.4 | 41.3 | 42.0 | 136 | 8.1 | 19 | 213 | 14 |
| Tololo 1214−277 | 115.6 | 1 | −18.0 | ..... | ..... | ..... | ..... | 7.5 | 13 | ..... | 560 |
| NGC 4569 | 9.0 | 10 | −19.7 | −23.2 | 42.1 | 42.8 | 447 | ..... | ..... | ..... | 37 |
| IC 3639 | 35.3 | 1 | −19.9 | −23.0 | 42.6 | 43.8 | ..... | ..... | ..... | 2567 | 297 |
| NGC 4670 | 23.1 | 1 | −18.8 | −21.4 | 42.9 | 43.0 | 279 | 8.2 | 21 | 1679 | 111 |
| NGC 4861 | 15.0 | 1 | −18.0 | −19.1 | 42.9 | 42.4 | 126 | 7.9 | 13 | 1090 | 62 |
| NGC 5055 | 8.7 | 1 | −20.5 | −24.1 | 42.9 | 43.5 | 495 | 9.0 | 12 | ..... | 36 |
| NGC 5135 | 57.7 | 1 | −21.1 | −25.0 | 43.0 | 44.6 | ..... | ..... | ..... | 4196 | 486 |
| NGC 5253 HIIR-1 | 3.6 | 5 | −16.8 | −19.5 | 42.2 | 42.3 | 98 | 8.2 | 23 | 261 | 15 |
| NGC 5253 HIIR-2 | | | | | | | | 8.2 | 23 | 261 | 15 |
| NGC 5253 UV1 | | | | | | | | 8.3 | 23 | 261 | 15 |
| NGC 5253 UV2 | | | | | | | | 8.3 | 23 | 261 | 15 |
| NGC 5253 UV3 | | | | | | | | 8.3 | 23 | 261 | 15 |
| Tololo 1345−420 | 26.1 | 1 | −16.1 | −18.0 | ..... | ..... | ..... | 8.0 | 13 | ..... | 108 |
| NGC 5457 (NGC 5455) | 7.4 | 11 | −21.1 | −23.8 | 43.7 | 43.6 | 516 | 8.2 | 24 | 536 | 30 |
| NGC 5457 Searle 5 | | | | | | | | 8.6 | 24 | 536 | 35 |
| NGC 5457 (NGC 5471) | | | | | | | | 8.0 | 24 | 536 | 30 |
| NGC 7130 | 68.7 | 1 | −21.4 | −25.0 | 43.0 | 44.7 | ..... | ..... | ..... | 4996 | 579 |
| NGC 7552 | 22.5 | 1 | −20.6 | −24.2 | 43.0 | 44.3 | 508 | 9.2 | 21 | 1636 | 109 |
| NGC 7714 | 38.5 | 1 | −20.1 | −23.2 | 43.2 | 43.9 | 387 | 8.5 | 25 | 2799 | 324 |

References: 1 − NASA Extragalactic Database (NED); 2 − Angeretti et al. (2005); 3 − Tosi et al. (2001); 4 − Karachentsev et al. (2003); 5 − Sakai et al. (2004); 6 − Aloisi et al. (2005); 7 − Aloisi et al. (2007); 8 − Schweizer et al. (2008); 9 − Maíz-Apellániz, Cieza, & MacKenty (2002); 10 − Shapley, Fabbiano, & Eskridge (2001); 11 − Ferrarese et al. (2000); 12 − Garnett (2002); 13 − Nava et al. (2006); 14 − Kobulnicky & Skillman (1997); 15 − Lee & Skillman (2004); 16 − Izotov, Thuan, & Stasińska (2007); 17 − Esteban et al. (2002); 18 − Garnett et al. (1997); 19 − Kobulnicky, Kennicutt, & Pizagno (1999); 20 − López-Sánchez, Esteban, & García-Rojas (2006); 21 − H98; 22 − Bastian et al. (2006); 23 − López-Sánchez et al. (2007); 24 − Kennicutt, Bresolin, & Garnett (2003); 25 − González-Delgado et al. (1995)

Table 7. UV line indices

| Index Name | Index Bandpass | Blue Continuum | Red Continuum | Features included |
|---|---|---|---|---|
| SiII_1260 | 1255 – 1264 | 1268 – 1286 | 1268 – 1286 | Si II 1260, S II 1259 |
| OI_SiII_1303 | 1290 – 1307 | 1268 – 1286 | 1308 – 1324 | O I 1302, Si II 1304, Si III 1295 |
| CII_1335 | 1326 – 1340 | 1308 – 1324 | 1348 – 1378 | C II 1334, C II* 1335 |
| SiIV_1400 | 1379 – 1405 | 1348 – 1378 | 1433 – 1460 | Si IV 1393, 1402 |
| SiII_1526 | 1521 – 1529 | 1460 – 1495 | 1572 – 1599 | Si II 1526 |
| CIV_1550 | 1529 – 1553 | 1460 – 1495 | 1583 – 1599 | C IV 1548, 1550, Si II* 1533 |
| FeII _1608 | 1600 – 1613 | 1583 – 1599 | 1614 – 1632 | Fe II 1608 |
| AlII_1670 | 1663 – 1679 | 1614 – 1632 | 1680 – 1705 | Al II 1670 |
| AlIII_1860 | 1840 – 1873 | 1815 – 1839 | 1932 – 1948 | Al III 1854, 1862 |
| FeII_2370 | 2334 – 2391 | 2267 – 2290 | 2395 – 2450 | Fe II 2344, 2374, 2382 |
| FeII_2600 | 2578 – 2611 | 2525 – 2572 | 2613 – 2674 | Mn II 2576, Fe II 2586, Mn II 2594, Fe II 2600, Mn II 2606 |
| MgII_2800 | 2788 – 2810 | 2720 – 2785 | 2812 – 2842 | Mg II  2796, 2803 |

Table 8. Line index measurements

| Galaxy | SiII_1260 | OI_SiII_1303 | CII_1335 | SiIV_1400 | SiII_1526 | CIV_1550 | FeII_1608 | AlII_1670 | AlIII_1860 | FeII_2370 | FeII_2600 | MgII_2800 | $v_{off}$ (km s$^{-1}$) |
|---|---|---|---|---|---|---|---|---|---|---|---|---|---|
| NGC 1068 knot2 | 1.05 ± 0.20 | −3.43 ± 0.37 | 1.51 ± 0.25 | 9.07 ± 1.31 | 0.38 ± 1.60 | 10.41 ± 1.75 | 1.98 ± 0.46 | −1.41 ± 0.79 | ..... | ..... | ..... | ..... | ..... |
| NGC 1068 pos1 | 1.07 ± 0.14 | 2.60 ± 0.23 | 2.05 ± 0.22 | 8.28 ± 0.24 | ..... | ..... | ..... | ..... | ..... | ..... | ..... | ..... | ..... |
| NGC 1068 pos3 | 2.01 ± 0.31 | −1.21 ± 0.60 | 3.08 ± 0.40 | 8.48 ± 0.65 | ..... | ..... | ..... | ..... | ..... | ..... | ..... | ..... | ..... |
| NGC 1068 pos4 | 0.57 ± 0.35 | −5.92 ± 0.80 | 1.22 ± 0.48 | 5.03 ± 0.77 | ..... | ..... | ..... | ..... | ..... | ..... | ..... | ..... | ..... |
| NGC 1068 pos8a | 2.46 ± 0.30 | −4.73 ± 0.85 | 3.06 ± 0.54 | 6.48 ± 0.94 | ..... | ..... | ..... | ..... | ..... | ..... | ..... | ..... | ..... |
| SBS 0335−052 | 1.05 ± 0.12 | 0.72 ± 0.17 | 0.44 ± 0.17 | 2.41 ± 0.27 | ..... | ..... | ..... | ..... | ..... | ..... | ..... | ..... | −140 ± 51 |
| NGC 1569 | −0.11 ± 1.26 | −1.50 ± 1.55 | 0.41 ± 0.85 | 4.04 ± 0.82 | 0.84 ± 0.33 | 5.58 ± 0.51 | 5.61 ± 1.19 | 0.80 ± 1.18 | 0.95 ± 0.40 | −0.18 ± 0.69 | 3.45 ± 0.29 | 2.33 ± 0.21 | ..... |
| NGC 1705 | 1.13 ± 0.08 | 1.72 ± 0.14 | 1.51 ± 0.13 | 3.27 ± 0.17 | 0.93 ± 0.12 | 5.10 ± 0.19 | 1.28 ± 0.25 | 0.76 ± 0.27 | ..... | ..... | ..... | ..... | −85 ± 16 |
| NGC 1741 | 1.84 ± 0.15 | 2.75 ± 0.21 | 2.18 ± 0.15 | 5.82 ± 0.17 | 1.16 ± 0.15 | 8.47 ± 0.21 | 1.27 ± 0.29 | 0.85 ± 0.38 | ..... | ..... | ..... | ..... | −67 ± 21 |
| NGC 2366 (NGC 2363) | ..... | ..... | ..... | ..... | ..... | ..... | ..... | −1.36 ± 0.74 | 0.27 ± 0.35 | ..... | ..... | ..... | ..... |
| NGC 2366 (NGC 2363-a) | 2.10 ± 0.18 | 0.58 ± 0.36 | 0.89 ± 0.30 | −0.21 ± 0.37 | 0.35 ± 0.22 | −3.16 ± 0.48 | ..... | ..... | ..... | ..... | ..... | ..... | 189 ± 68 |
| NGC 2366 (NGC 2363-b) | 0.60 ± 0.12 | 0.88 ± 0.24 | 0.80 ± 0.23 | 0.69 ± 0.30 | 0.31 ± 0.12 | 1.50 ± 0.20 | ..... | ..... | ..... | ..... | ..... | ..... | −154 ± 29 |
| NGC 2403 vs9 | ..... | ..... | ..... | ..... | ..... | ..... | ..... | 0.95 ± 0.63 | 1.32 ± 0.38 | −0.54 ± 0.32 | 1.34 ± 0.22 | 1.99 ± 0.19 | ..... |
| NGC 2403 vs38 | ..... | ..... | ..... | ..... | ..... | ..... | ..... | 1.53 ± 0.54 | 1.64 ± 0.41 | −0.30 ± 0.34 | 3.12 ± 0.21 | 4.03 ± 0.18 | ..... |
| NGC 2403 vs44 | ..... | ..... | ..... | ..... | ..... | ..... | ..... | 0.71 ± 0.39 | 1.60 ± 0.41 | 0.52 ± 0.26 | 2.72 ± 0.20 | 3.55 ± 0.15 | ..... |
| He 2-10 A | ..... | ..... | 2.93 ± 0.17 | 6.19 ± 0.19 | 2.44 ± 0.15 | 9.73 ± 0.25 | ..... | ..... | ..... | ..... | ..... | ..... | −297 ± 30 |
| He 2-10 B | ..... | ..... | 2.72 ± 0.23 | 5.71 ± 0.28 | 2.21 ± 0.26 | 9.35 ± 0.39 | ..... | ..... | ..... | ..... | ..... | ..... | −157 ± 24 |
| IRAS 08339+6517 | 2.15 ± 0.15 | 3.56 ± 0.22 | 2.83 ± 0.17 | 4.87 ± 0.19 | 1.21 ± 0.13 | 6.90 ± 0.21 | 0.63 ± 0.24 | 0.06 ± 0.32 | ..... | ..... | ..... | ..... | −241 ± 27 |
| SBS 1415+437 | ..... | ..... | ..... | ..... | ..... | ..... | ..... | −0.10 ± 0.90 | 0.02 ± 0.45 | ..... | ..... | ..... | ..... |
| I Zw 18 | ..... | ..... | ..... | ..... | ..... | ..... | ..... | 1.23 ± 0.74 | 1.81 ± 0.59 | ..... | ..... | ..... | ..... |
| I Zw 18 NW HIIR | ..... | ..... | ..... | ..... | ..... | ..... | ..... | 0.19 ± 0.78 | 1.07 ± 0.28 | ..... | ..... | ..... | ..... |
| I Zw 18 SE HIIR | ..... | ..... | ..... | ..... | ..... | ..... | ..... | −1.64 ± 1.05 | −2.30 ± 0.53 | ..... | ..... | ..... | ..... |
| NGC 3690 | 3.57 ± 0.25 | −4.07 ± 0.94 | 3.85 ± 0.33 | 7.01 ± 0.41 | 2.23 ± 0.22 | 9.56 ± 0.30 | 2.36 ± 0.24 | 2.50 ± 0.22 | 4.38 ± 0.32 | ..... | ..... | ..... | −6 ± 26 |
| UM 469 | ..... | ..... | ..... | ..... | ..... | 6.58 ± 0.38 | 1.42 ± 0.27 | 0.12 ± 0.31 | 1.05 ± 0.32 | 5.09 ± 0.29 | 5.32 ± 0.26 | 6.49 ± 0.16 | ..... |
| NGC 4038 405 | ..... | ..... | 3.02 ± 0.18 | 7.55 ± 0.27 | 1.46 ± 0.26 | 10.00 ± 0.34 | ..... | ..... | ..... | ..... | ..... | ..... | 29 ± 72 |
| NGC 4038 442 | ..... | ..... | 3.31 ± 0.29 | 9.30 ± 0.41 | 2.96 ± 0.35 | 12.48 ± 0.47 | ..... | ..... | ..... | ..... | ..... | ..... | −100 ± 75 |
| NGC 4214 | 1.02 ± 0.18 | 1.36 ± 0.26 | 0.67 ± 0.25 | 2.69 ± 0.36 | 0.92 ± 0.15 | 6.39 ± 0.22 | 1.62 ± 0.17 | 0.37 ± 0.18 | 1.64 ± 0.28 | ..... | ..... | ..... | 33 ± 20 |
| Tololo 1214−277 | ..... | ..... | ..... | ..... | ..... | ..... | ..... | −2.88 ± 5.58 | 4.77 ± 2.91 | ..... | ..... | ..... | ..... |
| NGC 4569 | 1.88 ± 0.40 | 5.15 ± 0.38 | 3.31 ± 0.23 | 7.91 ± 0.29 | 1.55 ± 0.22 | 8.06 ± 0.31 | 4.73 ± 0.46 | 1.53 ± 0.56 | 4.94 ± 0.39 | 7.63 ± 0.22 | 6.17 ± 0.15 | 7.69 ± 0.10 | −141 ± 55 |
| IC 3639 | 2.12 ± 0.20 | 0.77 ± 0.49 | 0.66 ± 0.37 | 6.41 ± 0.33 | 2.00 ± 0.29 | −6.96 ± 1.18 | ..... | ..... | ..... | ..... | ..... | ..... | −214 ± 37 |
| NGC 4670 | 1.26 ± 0.15 | 2.78 ± 0.22 | 1.87 ± 0.18 | 4.52 ± 0.19 | 1.37 ± 0.15 | 7.63 ± 0.21 | 2.19 ± 0.16 | 0.62 ± 0.22 | 2.70 ± 0.27 | ..... | ..... | ..... | 32 ± 39 |
| NGC 4861 | ..... | ..... | ..... | ..... | ..... | ..... | ..... | 1.22 ± 0.69 | 1.66 ± 0.30 | −0.32 ± 0.37 | 0.12 ± 0.37 | −0.66 ± 0.29 | ..... |
| NGC 5055 | 1.41 ± 1.13 | 5.70 ± 1.20 | 1.41 ± 1.17 | 5.59 ± 1.18 | −1.63 ± 1.75 | 1.58 ± 2.65 | 20.82 ± 6.71 | −2.02 ± 1.85 | 3.34 ± 0.58 | 7.64 ± 0.41 | 6.14 ± 0.27 | 7.55 ± 0.17 | ..... |
| NGC 5135 | 2.19 ± 0.21 | 4.39 ± 0.28 | 2.27 ± 0.27 | 8.76 ± 0.29 | 1.75 ± 0.23 | 8.81 ± 0.32 | ..... | ..... | ..... | ..... | ..... | ..... | −137 ± 40 |
| NGC 5253 HIIR-1 | ..... | ..... | ..... | ..... | ..... | ..... | ..... | 0.11 ± 0.88 | −0.10 ± 0.76 | ..... | ..... | ..... | ..... |
| NGC 5253 HIIR-2 | ..... | ..... | ..... | ..... | ..... | ..... | ..... | 2.89 ± 1.17 | −1.65 ± 0.71 | ..... | ..... | ..... | ..... |
| NGC 5253 UV1 | 1.72 ± 0.35 | 2.83 ± 0.56 | 2.72 ± 0.46 | 2.11 ± 0.59 | 1.65 ± 0.97 | 6.59 ± 1.58 | 1.98 ± 2.00 | −0.72 ± 0.87 | 1.84 ± 0.52 | 3.02 ± 0.58 | 3.76 ± 0.39 | 5.28 ± 0.30 | −15 ± 65 |
| NGC 5253 UV2 | 1.76 ± 0.48 | 2.20 ± 0.74 | 1.22 ± 0.55 | 4.56 ± 0.58 | 2.48 ± 0.65 | 7.63 ± 1.07 | −2.39 ± 1.88 | 1.09 ± 0.76 | 0.61 ± 0.59 | 5.40 ± 0.82 | 4.56 ± 0.51 | 5.27 ± 0.35 | 117 ± 93 |
| NGC 5253 UV3 | 3.30 ± 0.43 | 4.30 ± 0.79 | 2.61 ± 0.64 | 6.86 ± 0.82 | −0.49 ± 1.15 | −2.89 ± 2.31 | 11.12 ± 3.63 | −0.56 ± 1.61 | 0.17 ± 0.51 | 6.16 ± 0.75 | 5.29 ± 0.46 | 5.70 ± 0.35 | ..... |
| Tololo 1345−420 | ..... | ..... | ..... | ..... | ..... | ..... | ..... | 0.40 ± 1.71 | 2.23 ± 1.57 | 5.11 ± 0.53 | 3.11 ± 0.56 | 3.83 ± 0.38 | ..... |
| NGC 5457 (NGC 5455) | ..... | ..... | ..... | ..... | ..... | ..... | ..... | 0.82 ± 0.30 | 1.14 ± 0.52 | 0.98 ± 0.26 | 3.26 ± 0.21 | 4.21 ± 0.15 | ..... |
| NGC 5457 Searle 5 | −0.90 ± 0.67 | −11.59 ± 1.53 | 2.77 ± 0.44 | 5.45 ± 0.62 | 1.15 ± 0.38 | 11.65 ± 0.43 | 2.76 ± 0.42 | 0.10 ± 0.82 | 3.07 ± 0.89 | 2.90 ± 0.72 | 3.34 ± 0.52 | 4.79 ± 0.46 | ..... |
| NGC 5457 (NGC 5471) | ..... | ..... | ..... | ..... | ..... | ..... | ..... | 1.79 ± 1.43 | 2.01 ± 0.82 | 0.93 ± 0.89 | 1.80 ± 0.55 | −0.00 ± 0.72 | ..... |
| NGC 7130 | 2.20 ± 0.21 | 4.31 ± 0.28 | 2.84 ± 0.23 | 9.11 ± 0.26 | 1.77 ± 0.28 | 6.66 ± 0.52 | ..... | ..... | ..... | ..... | ..... | ..... | −499 ± 55 |
| NGC 7552 | 3.64 ± 0.23 | 3.71 ± 0.30 | 5.02 ± 0.21 | 9.77 ± 0.27 | 3.06 ± 0.16 | 11.80 ± 0.24 | 1.82 ± 0.23 | 2.33 ± 0.32 | 5.87 ± 0.32 | ..... | ..... | ..... | −174 ± 30 |
| NGC 7714 | 2.02 ± 0.10 | 3.11 ± 0.16 | 2.63 ± 0.17 | 6.31 ± 0.18 | 1.78 ± 0.14 | 9.82 ± 0.20 | 1.74 ± 0.21 | ..... | ..... | ..... | ..... | ..... | −6 ± 16 |

# Appendix

For reference we show images of our sample galaxies with the spectroscopic apertures overlaid. The images were obtained from the HST Archive and from the NASA/IPAC Extragalactic Database (NED). Where multi-color data were available, we selected observations near 3000 Å to best display the UV-optical morphology of the galaxies. All images were obtained with HST's WFPC2 camera, except for NGC 2403 and NGC 5457, for which ground-based images were substituted due to their large apparent diameter. No useful images were available for UM 469 and Tololo 1345−420, both of which are highly compact sources. The absolute astrometry of the images is uncertain by at least $\pm 1''$. The position of the spectroscopic apertures shown in Appendix Figure 1 represents the initial target coordinates specified in the HST proposal. The spectroscopic targets are centered in the science apertures using an on-board acquisition procedure Sections 3.2 and 3.3, hence final placement of the aperture is difficult to ascertain.

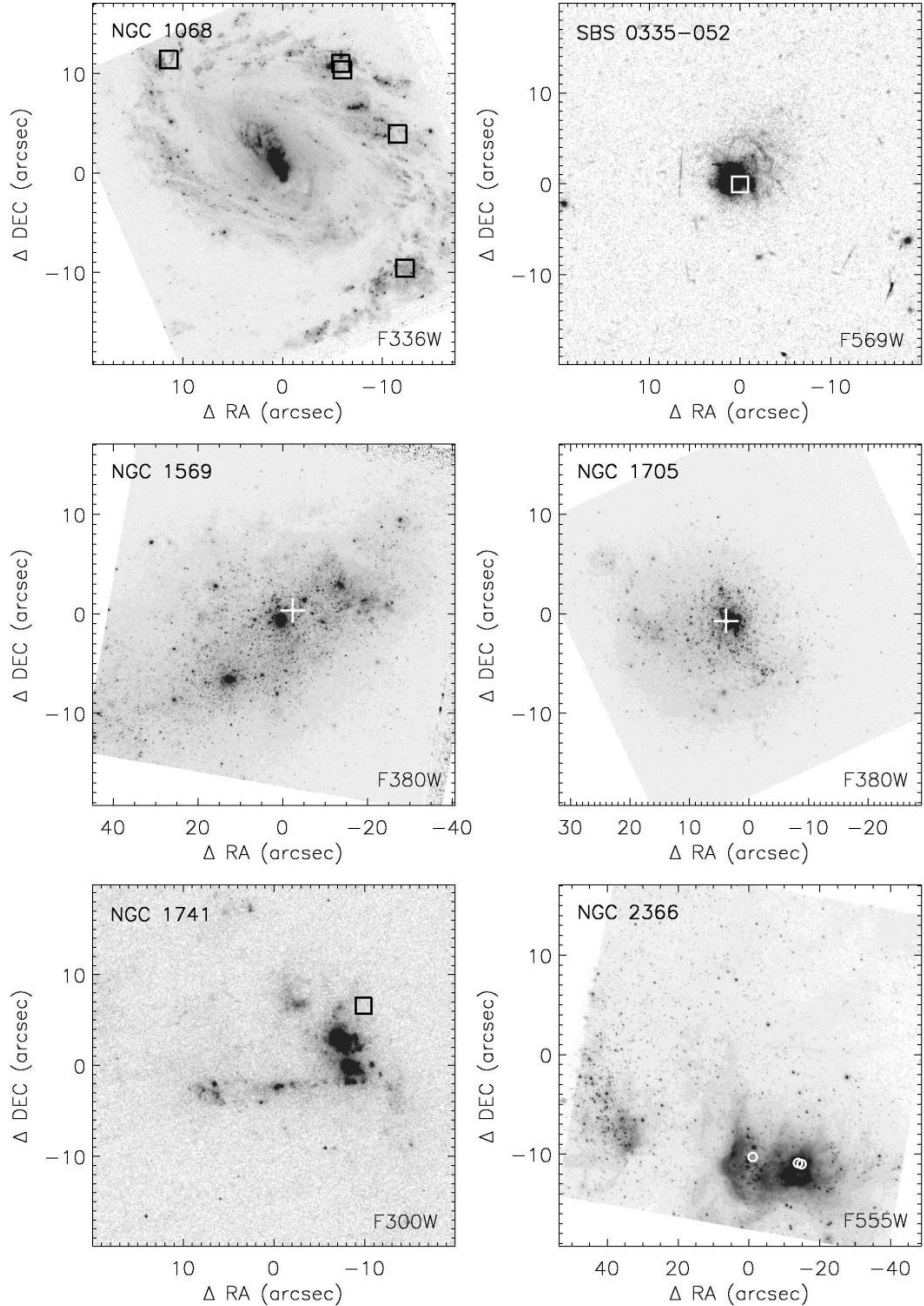

Appendix Figure 1. – Images of the sample galaxies with the spectroscopic apertures overlaid. The aperture positions represent the initial target coordinates. A peak-up is used to center the aperture on the target. Apertures are shown to scale (black or white circles and squares). In some cases cross-hairs have been added to make the apertures more visible. North is up and East to the left. Filters are listed in the lower right hand corner.

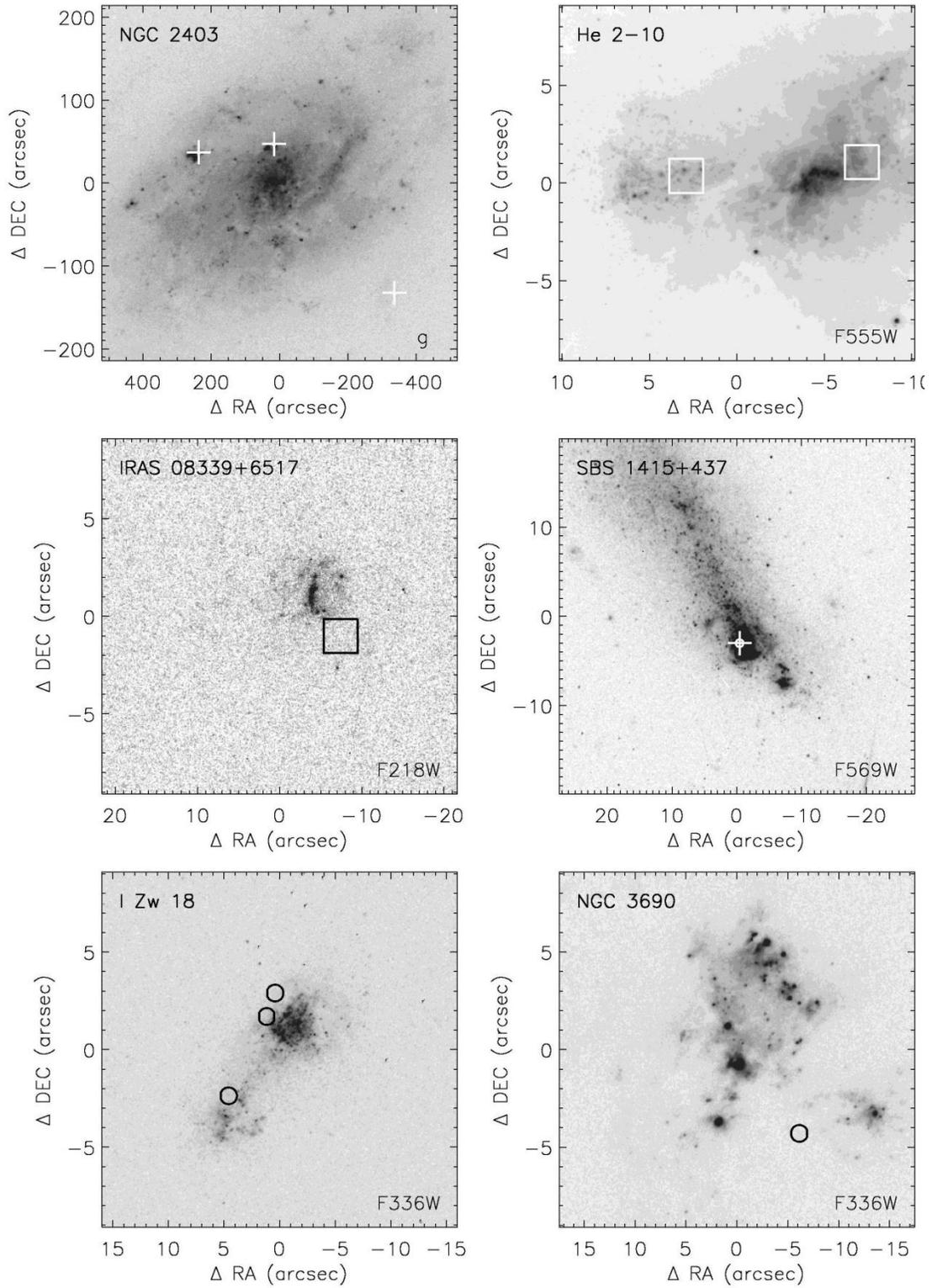

Appendix Figure 1. – continued.

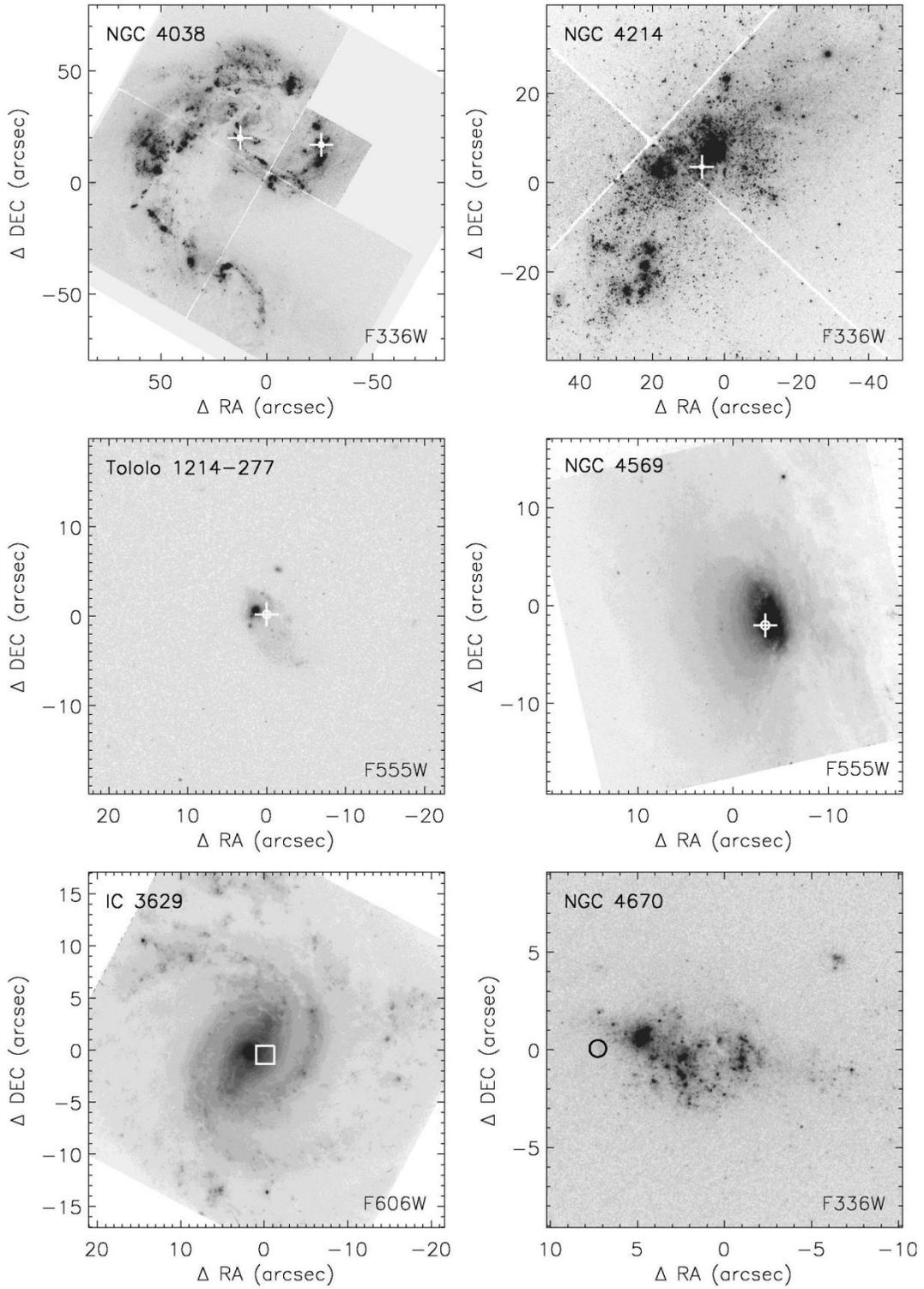

Appendix Figure 1. – continued.

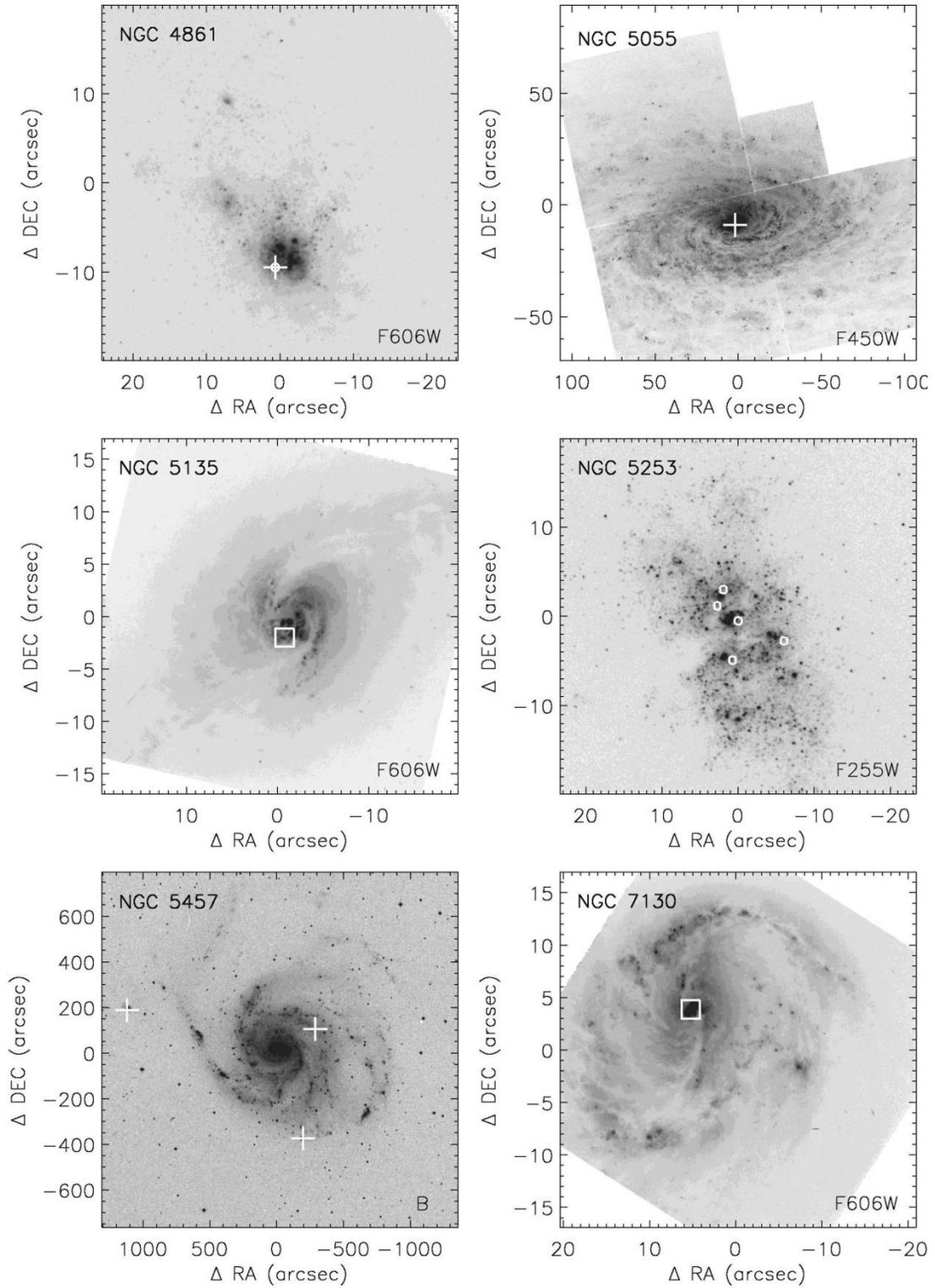

Appendix Figure 1. – continued.

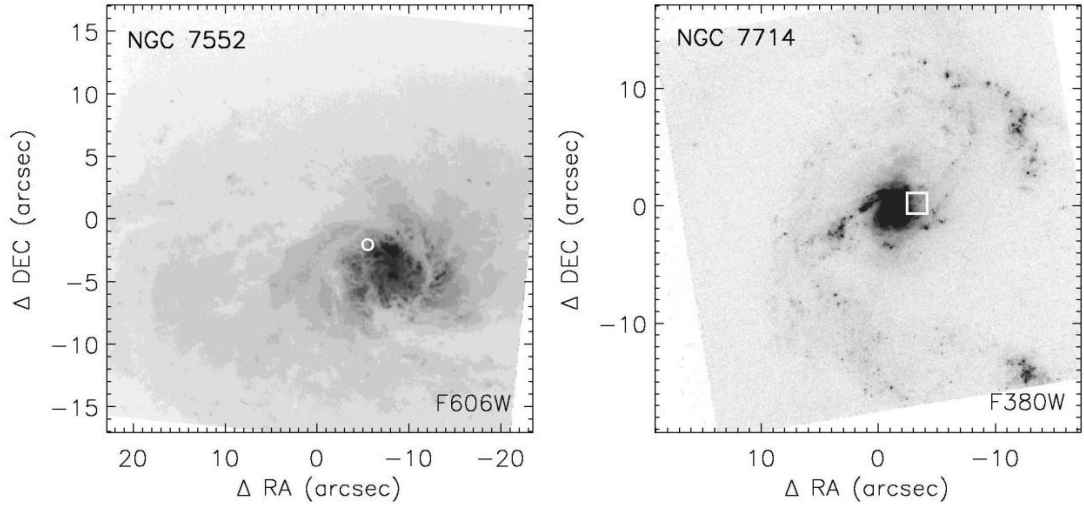

Appendix Figure 1. – continued.